\documentclass[preprint]{aastex631}
\shorttitle{Formation of Low-${}^{26}{\rm Al}/{}^{27}{\rm Al}$ Corundum/Hibonite Inclusions}
\shortauthors{Desch et al.}

\usepackage{graphics}
\graphicspath{{./}{figures/}}
\usepackage{subfigure}
\usepackage{xcolor}

\newcommand{\altwosix}{\mbox{${}^{26}{\rm Al}$}}
\newcommand{\mgtwosix}{\mbox{${}^{26}{\rm Mg}$}}
\newcommand{\alratio}{\mbox{${}^{26}{\rm Al}/{}^{27}{\rm Al}$}}
\newcommand{\mgratio}{\mbox{${}^{26}{\rm Mg}/{}^{24}{\rm Mg}$}}
\newcommand{\beten}{\mbox{${}^{10}{\rm Be}$}}
\newcommand{\beratio}{\mbox{${}^{10}{\rm Be}/{}^{9}{\rm Be}$}}
\newcommand{\efiftyti}{\mbox{$\epsilon^{50}{\rm Ti}$}}
\newcommand{\permil}{\mbox{\text{\textperthousand}}}

\newenvironment{Steve}{\color{black}}{\ignorespacesafterend}

\newenvironment{EDIT}{\color{black}}{\ignorespacesafterend}

\begin{document}

\title{Origin of Low-${}^{26}{\rm Al}/{}^{27}{\rm Al}$ Corundum/Hibonite Inclusions in Meteorites}

\correspondingauthor{Steve Desch}
\email{steve.desch@asu.edu}

\author[0000-0002-1571-0836]{Steven J. Desch}
\affiliation{
School of Earth and Space Exploration, Arizona State University, \\
PO Box 871404, Tempe, AZ 85287-1404}

\author[0000-0003-3695-3295]{Emilie T. Dunham}
\affiliation{Department of Earth, Planetary, and Space Sciences, University of California, Los Angeles, Los Angeles, CA 90095, and Department of Earth and Planetary Sciences, University of California, Santa Cruz, Santa Cruz CA 95064}
%\affiliation{Arizona State University
%School of Earth and Space Exploration \\
%Tempe, AZ 85287}

\author[0000-0002-6051-9002]{Ashley K. Herbst}
\affiliation{
School of Earth and Space Exploration, Arizona State University, \\
PO Box 871404, Tempe, AZ 85287-1404}

\author[0000-0001-8991-3110]{Cayman T. Unterborn}
\affiliation{Southwest Research Institute, 6220 Culebra Road, \\
San Antonio, TX 78238}

\author[0000-0003-2287-9370]{Thomas Sharp}
\affiliation{School of Earth and Space Exploration, Arizona State University, \\
PO Box 871404, Tempe, AZ 85287-1404}

\author[0000-0002-7978-6370]{Maitrayee Bose}
\affiliation{School of Earth and Space Exploration, Arizona State University, \\
PO Box 871404, Tempe, AZ 85287-1404}

\author[0000-0002-6918-2653]{Prajkta Mane}
\affiliation{Lunar and Planetary Institute, \\
3600 Bay Area Blvd.,
Houston TX 77058}

\author[0000-0003-0496-5734]{Curtis D. Williams}
\affiliation{Arctic Slope Regional Corporation Federal, 11091 Sunset Hills Road \#800, Reston VA 20190}
%\affiliation{Earth and Planetary Sciences Department, University of California, Davis \\ Davis, CA 95616}

%\author[0000-0002-5246-4400]{Zachary A. Torrano}
%\affiliation{Earth and Planets Laboratory,Carnegie Institution for Science,\\ Washington, DC 20015}

\begin{abstract}
Most meteoritic calcium-rich, aluminum-rich inclusions (CAIs) formed from a reservoir with $\alratio \approx 5 \times 10^{-5}$, but some record lower $(\alratio)_0$, demanding they sampled a reservoir without live $\altwosix$. This has been interpreted as evidence for ``late injection" of supernova material into our protoplanetary disk. We instead interpret the heterogeneity as chemical, demonstrating that these inclusions are strongly associated with the refractory phases corundum or hibonite. We name them ``Low-$\alratio$ Corundum/Hibonite Inclusions" (LAACHIs). We present a detailed astrophysical model for LAACHI formation in which they derive their Al from presolar corundum, spinel or hibonite grains $0.5 - 2 \, \mu{\rm m}$  in size with no live $\altwosix$; live $\altwosix$ is carried on smaller ($<$50 nm) presolar chromium spinel grains from recent nearby Wolf-Rayet stars or supernovae. In hot ($\approx$ 1350-1425 K) regions of the disk these grains, and perovskite grains, would be the only survivors. These negatively charged grains would grow to sizes $1 - 10^3 \, \mu{\rm m}$,  even incorporating positively charged perovskite grains, but not the small, negatively charged $\altwosix$-bearing grains. Chemical and isotopic fractionations due to grain charging was a significant process in hot regions of the disk. Our model explains the sizes, compositions, oxygen isotopic signatures, and the large, correlated ${}^{48}{\rm Ca}$ and ${}^{50}{\rm Ti}$ anomalies (if carried by presolar perovskite) of LAACHIs, and especially how they incorporated no $\altwosix$ in a solar nebula with uniform, canonical $\alratio$. A late injection of supernova material is obviated, although formation of the Sun in a high-mass star-forming region is demanded. 
\end{abstract}


\section{Introduction}
\label{sec:intro}

\subsection{Low ${}^{26}{\rm Al}/{}^{27}{\rm Al}$ in Some Inclusions}
\label{sec:lowratio}

One of the enduring mysteries of Solar System formation is the origin of the short-lived radionuclides (SLRs), in particular the important isotope $\altwosix$, which decays to $\mgtwosix$ with a half-life of 0.72 Myr \citep{AuerEtal2009,Kondev2021}.
When this isotope was discovered, it was immediately recognized for its potential to be the cause of asteroid melting \citep{Urey1955}.
Its existence in the solar nebula at a level $\alratio \approx 5 \times 10^{-5}$ was established later \citep{LeeEtal1976}, by measurements of Al and Mg isotopes in calcium-rich, aluminum-rich inclusions (CAIs).
A linear correlation, or ``isochron", between $\mgratio$ and ${}^{27}{\rm Al}/{}^{24}{\rm Mg}$ measurements in minerals in an inclusion is a signature that $\mgtwosix$ formed by decay of $\altwosix$, and the slope of the linear correlation is the initial $(\alratio)_0$ ratio in the solid when it formed, or when it last achieved isotopic closure.
The differences in $(\alratio)_0$ between two inclusions (e.g., CAIs and chondrules) can be used to determine the time difference between their formations, assuming they formed from the same reservoir with common initial $(\alratio)_{\rm SS}$.
After measurements of thousands of CAIs, a ``canonical" Solar System value has been established \citep{MacPhersonEtal1995}, and refined today to be $(\alratio)_{\rm SS} = (5.23 \pm 0.13)\times 10^{-5}$ \citep{JacobsenEtal2008}.
Although debate continues whether some CAIs record slightly higher values \citep{SimonYoung2011}, the vast majority of CAIs formed with very similar ratios, apparently recording a spatially uniform ratio $(\alratio)_{\rm SS}$  in the solar nebula, with the slight variations (among those not melted later) interpreted as small differences in formation time, within about $5 \times 10^5$ years \citep[e.g.,][]{MacPhersonEtal2012,LiuEtal2019,KawasakiEtal2020}.

The canonical $(\alratio)_{\rm SS}$ ratio has been inferred from the majority of CAIs, which primarily contain such minerals as melilite,
%[${\rm Ca}_{2}{\rm Al}_{2}{\rm SiO}_{7}$-${\rm Ca}_{2}{\rm Mg}{\rm Si}_{2}{\rm O}_{7}$ mixture],
anorthite,
% [${\rm CaAl}_{2}{\rm Si}_{2}{\rm O}_{8}$],
Ti-rich clinopyroxene,
%[${\rm Ca}({\rm Mg,Al,Ti})({\rm Si,Al})_{2}{\rm O}_{6}$], 
and spinel.
%[${\rm MgAl}_{2}{\rm O}_{4}$]; 
Depending on their exact mineralogy and thermal histories, these may be categorized as fluffy type A (FTA), compact type A (CTA), type B, or type C CAIs
\citep[e.g.,][]{Grossman1975,GrossmanEtal2000,MacPhersonEtal2022}.
These can be considered ``normal" CAIs, and the community's intuition about CAIs and $\altwosix$ in the solar nebula was initially shaped by measurements in large (up to cm-sized) type B CAIs from the Allende (CV3) carbonaceous chondrite in particular.  

Some types of CAIs, not falling into the categories above, record much lower $(\alratio)_0$ ratios \citep[see the review by][]{KrotEtal2012}.
In particular, PLAty Crystals of hibonite (PLACs) are $10-100 \, \mu{\rm m}$-sized fragments or aggregates of hibonite laths that tend to record low values $(\alratio)_0 < 3 \times 10^{-6}$ \citep{Ireland1988,Ireland1990,LiuEtal2009,KoopEtal2016a}.
%\citep{Ireland1988,Ireland1990,ViragEtal1991,MacPhersonEtal2005,Liu2008thesis,LiuEtal2009,MakideEtal2011,KrotEtal2012,KoopEtal2016a,KrotEtal2019}.
Numerous corundum grains tens of microns in size exhibit similarly small $(\alratio)_0$ ratios \citep{ViragEtal1991,MakideEtal2011,MakideEtal2013}.
A large fraction of grossite-dominated CAIs from CH chondrites have been shown to record $(\alratio)_0 \approx 4 \times 10^{-7}$ \citep{Krot2019}.
There are also a few examples of large, corundum- and/or hibonite-dominated inclusions: {\it M98-8} {\Steve (from the CM chondrite Murchison)}, {\it DOM 31-2} {\Steve (from the CO chondrite Dominion Range 08006)}, and {\it A-COR-01} {\Steve (from the CO chondrite Allan Hills 77307)} record values from $(\alratio)_0 < 1.6 \times 10^{-6}$, down to $(\alratio)_0 \approx (1.63 \pm 0.31) \times 10^{-7}$ \citep{SimonEtal2002,SimonEtal2019b,BodenanEtal2020}.
The most {\Steve difficult to explain} of these may be the hibonite-dominated inclusion {\it HAL}, which apparently records $(\alratio)_0 \approx 5 \times 10^{-8}$ \citep{FaheyEtal1987a}.

These low values of $(\alratio)_0 \sim 10^{-7}$ cannot be explained as a late formation or thermal resetting of the Al-Mg chronometer of materials that sampled the reservoir with canonical $(\alratio)_{\rm SS}$.
Most of these inclusions would have been incorporated into their parent body asteroids within 3 Myr, when $\alratio > 3 \times 10^{-6}$ still, and remained too cold to alter the Al-Mg isochron.
The cause of the heterogeneity in $(\alratio)_0$ ratios has been an ongoing mystery for decades.

\subsection{Causes of the ${}^{26}{\rm Al}/{}^{27}{\rm Al}$  Heterogeneities?}
\label{sec:why}

While many papers have identified this clear heterogeneity of $\altwosix$, there is no consensus in the field for its origin.
The term ``heterogeneity" is agnostic about cause, but often it is implied that the heterogeneities are erased by physical mixing between materials at different parts of the disk \citep[e.g.,][]{KrotEtal2012,MishraChaussidon2014}, by processes such as those described by \citet{Boss2013}.
Such {\it spatial} heterogeneities might be expected if $\altwosix$ were produced by irradiation by solar energetic particles in the solar nebula \citep[e.g.,][]{GounelleEtal2001,GachesEtal2020,Jacquet2019}, but there are severe limitations to production of $\altwosix$ in ${\rm H}_{2}$ gas \citep{ClaytonJin1995}, or in material devoid of ${\rm H}_{2}$ \citep{DeschEtal2010}.
\citet{Wood1998} suggested that the molecular cloud was heterogeneous in its $\altwosix$ distribution, leading to $\altwosix$-free gas during ``clumpy infall"; likewise, spatial variations in isotopic compositions of disk materials, because of variations in isotopic compositions of gas infalling from the molecular cloud, are the underlying assumptions of several contemporary models \citep{NanneEtal2019,PignataleEtal2019,LichtenbergEtal2021,LiuBEtal2022}.
However, these assumptions are untested and such models are counter to the expectation that the molecular cloud was spatially homogeneous \citep{PanEtal2012} and well mixed during infall \citep{KuffmeierEtal2017}, even before mixing within the disk. %\citep{Boss2013}.

The evidence for large-scale spatial heterogeneities is weak in any case.
\citet{BollardEtal2019} argued that chondrules formed in a region with lower (by a factor of 2) $\alratio$ than CAIs did, although this interpretation was based on the untested  assumption that CAIs achieved isotopic closure in the Pb-Pb system at the same time they closed in the Al-Mg system \citep{DeschEtal2023a}.
Similarly, \citet{LarsenEtal2011} argued for the solar nebula generally having lower (factor of 2) $\alratio$ ratios than the CAI-forming region, based on measured $\mgtwosix$ deficits in CAIs; but recently \citet{GregoryEtal2020} measured greater deficits in forsteritic inclusions, again favoring a uniform distribution of $\altwosix$.
{\Steve These variations, if they existed, would not explain}
%These putative factors-of-two variations are attributable to prosaic causes, and would not explain 
$\alratio$ ratios as low as $\sim 10^{-7}$, anyway.

Currently, the interpretation of a {\it temporal} heterogeneity, and the ``late injection" model \citep{SahijpalGoswami1998,SahijpalEtal2000}, is most commonly accepted \citep[e.g.,][]{LiuEtal2009,MakideEtal2011,KrotEtal2012,vanKootenEtal2016,Krot2019,KuEtal2022}.
In this model, the solar nebula formed with little $\altwosix$, which was instead acquired later, just prior to the formation of most CAIs, either due to some late input to the Sun's molecular cloud \citep{FosterBoss1997}, or injected into the Sun's protoplanetary disk \citep{OuelletteEtal2007} by a nearby core collapse supernova injecting gas or, more likely, dust grains.
While this scenario cannot be ruled out, the likelihood of a supernova exploding at the right distance from the Sun's protoplanetary disk, during its first $< 10^5$ years, is very small $\sim 0.1 - 1\%$ \citep{OuelletteEtal2010}.

Although spatial and temporal heterogeneities are popular explanations,  there is a high likelihood %(up to 80\%) 
that a forming solar system would simply acquire $\altwosix$ from its molecular cloud at near-canonical values \citep{JuraEtal2013,DeschEtal2023c}.
Any ${}^{26}{\rm Al}$ inherited from the molecular cloud would have been present and well-mixed throughout the solar nebula, from the earliest times, which would argue against any spatial or temporal heterogeneities.

Therefore, we instead favor the interpretation of a {\it chemical} heterogeneity.
We note that all of the inclusions that sampled a subcanonical $\alratio$ reservoir are dominated by the refractory minerals of corundum [${\rm Al}_{2}{\rm O}_{3}$], the calcium aluminate hibonite [${\rm CaAl}_{12}{\rm O}_{19}$], or more rarely the calcium aluminate grossite [${\rm CaAl}_{4}{\rm O}_{7}$].
These Al-dominated minerals are some of the most refractory minerals found in meteorites, and hibonite and grossite are both formed by reaction of Ca vapor with corundum \citep[e.g.,][]{EbelGrossman2000,UnterbornPanero2017}.
The association of low-$\alratio$ ratios with these minerals, especially corundum and hibonite, is so strong that we collectively call such meteoritic components Low-$\alratio$ Corundum/Hibonite Inclusions (LAACHIs). 

\citet{Ireland1990} and \citet{LarsenEtal2020} offered a chemical heterogeneity interpretation of ${}^{26}{\rm Al}$ distributions, suggesting that inclusions {\Steve low in $\altwosix$} may have formed from material that was predominantly from presolar grains that had spent enough time in the interstellar medium (ISM) that effectively all their $\altwosix$ had decayed.
% {\Steve A presolar origin for PLACs specifically also has been previously suggested  \citep{Ireland1990}.}
In the model of \citet{LarsenEtal2020},
these grains would be preferentially destroyed at high temperatures, and their $\altwosix$ somehow not incorporated into inclusions.
%It has been suggested that live $\altwosix$ was instead carried into the solar nebula on grains that are easily destroyed at high temperatures \citep{TrinquierEtal2009,LarsenEtal2020}. 
Given the wide spread of isotopic ratios in presolar grains, this would be consistent with  the observation that low-$\alratio$ inclusions often show large anomalies in $\efiftyti$, whereas inclusions with canonical $\alratio$ ratios show a small spread in $\efiftyti$ values \citep{FaheyEtal1987b,KoopEtal2016a}. 
Indeed, there are known examples of aluminum oxide presolar grains, including: corundum \citep[e.g.,][]{HutcheonEtal1994,NguyenEtal2007}; spinel \citep[e.g.,][]{NguyenEtal2003}; and hibonite \citep[e.g.,][]{ChoiEtal1999,KrestinaEtal2002lpsc,NittlerEtal2008,ZegaEtal2011} grains.
These grains likely formed in the outflows of asymptotic giant branch (AGB) stars, supernova explosions, or other stellar nucleosynthetic events, throughout Galactic history, likely surviving in the interstellar medium (ISM) for $10^8 - 10^9$ yr \citep{JonesEtal1994,HeckEtal2020}.
Even if they formed with such a high ratio as $(\alratio)_0 \sim 1$, they would have $\alratio < 4 \times 10^{-9}$ after just 20 Myr, so that the vast majority of presolar grains would contain essentially no live $\altwosix$.

Objections to the idea that LAACHIs are predominantly formed from presolar grains include the fact that they have oxygen isotopic signatures that exactly match other known solar nebula reservoirs, not the wide range of values seen in presolar grains, and despite the observed heterogeneity in ${}^{50}{\rm Ti}$.
Most importantly, there has not been an astrophysical model for how these Al-rich presolar grains---and only these Al-rich presolar grains---could have coagulated with each other into $10-100 \, \mu{\rm m}$ objects like PLACs, and yet avoided accreting $\altwosix$ from the gas, or any carriers of $\altwosix$.

\subsection{Outline of this Paper}
\label{sec:outline}

The goal of this paper is to present a quantitative astrophysical model for how inclusions, in particular those dominated by corundum and hibonite (or grossite) could have formed without {\Steve significant} $\altwosix$, in a solar nebula with uniform $(\alratio)_{\rm SS} \approx 5 \times 10^{-5}$. 
Once such inclusions formed, isotopic exchange with other materials in the solar nebula would allow these inclusions to acquire more canonical $\alratio$ ratios.
This paper addresses the greater challenge, which is to explain how any inclusions could have formed without $\altwosix$, in a solar nebula with $\altwosix$ uniformly distributed even in the molecular cloud.

The paper is organized as follows.
In \S 2 we review the meteoritic data that shows that some inclusions record $(\alratio)_0 < 3 \times 10^{-6}$, which cannot be explained by late resetting.
We show that these comprise only inclusions dominated by corundum or hibonite (or grossite), which we call LAACHIs; and that conversely, inclusions dominated by these phases mostly formed with low $(\alratio)_0$.
In \S 3 we present an astrophysical model for formation of LAACHIs.
We argue these formed from partially processed presolar grains of corundum and spinel and hibonite, in a very early stage of nebular evolution.
A central, novel aspect of our model is that chemical and isotopic fractionations occur because of electrical charging of coagulating dust grains in hot regions of the solar nebula.
Our model explains the growth of LAACHIs up to tens to hundreds of microns in size, devoid of $\altwosix$, but with correlated excesses/deficits in ${}^{48}{\rm Ca}$ and ${}^{50}{\rm Ti}$, as well as their solar oxygen isotopic compositions and other features.
In \S 4 we discuss the implications for observable features of these inclusions, including refractory metal abundances, stable isotope anomalies, short-lived radionuclide abundances, and mineralogy and microstructures.
In \S 5 we draw conclusions and discuss implications for Solar System chronometry and formation environment.

%
% SECTION 2: METEORITIC DATA 
%
\section{Meteoritic Evidence for Low ${}^{26}{\rm Al}/{}^{27}{\rm Al}$}

\subsection{Background}

The definition of ``CAIs" as objects rich in Ca and Al encompasses many different types of inclusions with different mineralogies, sizes and morphologies, and $(\alratio)_0$ ratios.
Our goal {\Steve in this section} is to review these different types and show that the $(\alratio)$ ratio in the solar nebula reservoir sampled by all CAIs is uniform and canonical, {\Steve except for some CAIs} dominated by the refractory phases corundum and/or hibonite or grossite. 
Conversely, we show that among the types of CAIs dominated by corundum and/or hibonite or grossite, the majority sample a reservoir with very subcanonical $\alratio$.
{\Steve It is because the}
association between low $\alratio$ ratios and corundum and/or hibonite (and less commonly grossite) mineralogies is so strong {\Steve that} we use the term ``LAACHIs" to collectively describe these inclusions.
It is likely LAACHIs have a common origin, separate from most CAIs.

Most CAIs are characterized by mineralogies rich in melilite [a solid {\Steve solution} of ${\rm Ca}_{2}{\rm Al}_{2}{\rm SiO}_{7}$ and ${\rm Ca}_{2}{\rm Mg}{\rm Si}_{2}{\rm O}_{7}$], 
anorthite [${\rm CaAl}_{2}{\rm Si}_{2}{\rm O}_{8}$], 
Ti,Al-rich pyroxene [${\rm Ca}({\rm Mg,Al,Ti})({\rm Si,Al})_{2}{\rm O}_{6}$], 
and spinel [${\rm MgAl}_{2}{\rm O}_{4}$], with most (types CTA, B, C) showing igneous textures, and FTA showing signatures of having grown from aggregates of condensates, plus reaction with gas \citep{KrotEtal2009,MacPherson2014}.
Most of these CAIs record nearly identical ratios $(\alratio)_0 \approx 5.23 \times 10^{-5}$ \citep{JacobsenEtal2008}, indicating formation from a common reservoir at a similar time, and defining the canonical $(\alratio)_{\rm SS}$ ratio.

Other, rarer types of CAIs have been identified as recording $(\alratio)_0$ too low to have formed from the same reservoir \citep{KrotEtal2012}.
Although it is always possible to thermally reset the Al-Mg system after $\altwosix$ has decayed, and record a lower $(\alratio)_0$, the temperatures in the parent bodies of carbonaceous chondrites generally were too low for that, and any resetting had to take place in the solar nebula.
Because carbonaceous chondrites generally, and the CV3 chondrite Allende in particular, assembled between 2 and 3 Myr after most CAIs \citep[e.g.,][]{DeschEtal2018}, CAIs would have to record $(\alratio)_0 > 3 \times 10^{-6}$.
The CAI {\it H 030/L} from an R chondrite of petrologic type 4 \citep{RoutEtal2009} is the rare exception where low $(\alratio)_0$ might be explained by resetting on the parent body.
In general, CAIs with $(\alratio)_0 < 3 \times 10^{-6}$ must have sampled a different reservoir, with $\alratio < (\alratio)_{\rm SS}$.

In the following, we compare many properties of LAACHIs and non-LAACHI CAIs, including: their sizes, morphologies, mineralogies, MgO and TiO$_2$ contents, initial $(\alratio)_0$ ratios, oxygen isotopic compositions ($\Delta^{17}{\rm O}$), and titanium isotopic compositions ($\epsilon^{50}{\rm Ti}$).
Here $\Delta^{17}{\rm O} = \delta^{17}{\rm O} - 0.517 \, \delta^{18}{\rm O}$, 
where $\delta^{17}{\rm O} = \left[ ({}^{17}{\rm O}/{}^{16}{\rm O})_{\rm CAI} / ({}^{17}{\rm O}/{}^{16}{\rm O})_{\rm std} - 1 \right]$ and 
$\delta^{18}{\rm O} = \left[ ({}^{18}{\rm O}/{}^{16}{\rm O})_{\rm CAI} / ({}^{18}{\rm O}/{}^{16}{\rm O})_{\rm std} - 1 \right]$ are the deviations in molar ratios of oxygen isotopes from a terrestrial standard, {\Steve expressed} in parts per thousand {\Steve (per mil, or $\permil$)}.
Whereas both $\delta^{17}{\rm O}$ and $\delta^{18}{\rm O}$ can be altered during chemical reactions {\Steve or evaporation}, changes in  $\Delta^{17}{\rm O}$ generally require mixing with another isotopic reservoir.
Likewise, $\epsilon^{50}{\rm Ti} = \left[ ({}^{50}{\rm Ti}/{}^{47}{\rm Ti})_{\rm CAI} / ({}^{50}{\rm Ti}/{}^{47}{\rm Ti})_{\rm std} - 1 \right]$, measured in parts per ten thousand, {\Steve or epsilon units}.
In the following, the ratios have been corrected for mass-dependent fractionation {\Steve fixing the ${}^{49}{\rm Ti}/{}^{47}{\rm Ti}$ ratio}.
In \S 2.2 we review the properties of `normal' CAIs and the intermediate objects known as {\Steve Spinel-HIBonite inclusions (SHIBs)}. 
In \S 2.3 we review the same properties for LAACHIs, types of CAIs that commonly record $\alratio < 3 \times 10^{-6}$.
These include: hibonite-dominated Fractionation and Unknown Nuclear Effects (FUN) CAIs; corundum grains; PLAty Crystals of hibonite (PLACs), and other PLAC-like CAIs; Blue AGgregates of hibonites (BAGs);  grossite-dominated CAIs; and other unusual inclusions dominated by corundum and hibonite.
In \S 2.4 we summarize the data to show that the dichotomy in $(\alratio_0$ values is clearly associated with corundum or hibonite, justifying the classification of some CAIs as LAACHIs.

\subsection{Non-LAACHI CAIS}

\subsubsection{Normal CAIs}

As a baseline for comparison, we review `normal' CAIs, which we define as the multi-minerallic CAIs of types A, B and C.
While normal CAIs are found in all chondrite types, most descriptions are of CAIs from CV chondrites \citep[e.g.,][]{MacPhersonKrot2014}. 
Type A CAIs are primarily melilite, with some Ti-rich clinopyroxene and spinel. 
CTA CAIs have igneous textures, while FTA CAIs appear to be aggregates of grains that condensed from the gas or at least reacted with the gas. 
Type B and C CAIs have igneous textures, and contain melilite, Ti-rich clinopyroxene, anorthite, and spinel. Normal CAIs have distinct populations in each carbonaceous chondrite type and show a wide range of mineralogies, textures, shapes, and sizes. 
For example, the most common type of CAIs in CM, Ordinary, and Enstatite chondrites are spinel-rich and hibonite-rich; CK CAIs are commonly rich in secondary minerals; CO CAIs commonly contain hibonite and grossite; while CR CAIs tend to be melilite-rich \citep{RussellEtal1998,AleonEtal2002,ChaumardEtal2014,FendrichEbel2021,LinEtal2006}.
While many CAIs contain hibonite or grossite, the vast majority of them are not dominated by these minerals. 
% [Russell1998: CO, Aleon2002: CR, Charmaud2014: CK, Fendrich\&Ebel: CM, Lin2006: NC].
CAIs can be small (tens of $\mu{\rm m}$), or up to centimeter-sized in the case of type B CAIs from CV chondrites like Allende. 

Oxygen isotopic compositions of normal CAIs can extend from $\Delta^{17}{\rm O} \approx -26$ to $-22 \permil$, with an average of $\approx -24 \permil$ \citep{KrotEtal2010}.
Titanium isotopic compositions cluster very tightly around $\epsilon^{50}{\rm Ti} \approx +9$, varying from about $+2$ to $+15$ epsilon units \citep{WilliamsEtal2016,DavisEtal2018,TorranoEtal2019}.

The first measurements of $(\alratio)_0$ ratios were undertaken in large type B CAIs from Allende. 
\citet{MacPhersonEtal1995} and \citet{MacPhersonEtal2005} created histograms of $(\alratio_0$ ratios in all CAIs, including many CAIs we do not consider `normal'.
These distributions appear to be bimodal, with the majority of CAIs near the canonical value $\alratio \approx 5 \times 10^{-5}$ and a tail of inclusions extending to $(\alratio)_0 \approx 0$
\citep{Davis2022}. 

\subsubsection{SHIBs}

Spinel-HIBonite inclusions (SHIBs) are a type of CAI we consider intermediate between normal CAIs and LAACHIs. 
They are composed predominantly of spinel and hibonite, with accessory perovskite.
Voids and cracks are commonly observed in both untreated and acid-treated SHIBs, and the former generally contains alteration phases while untreated SHIBs sustain partial rims of secondary, or sometimes primary silicates. 
Grain sizes of SHIBs are quite variable, ranging from 30 $\mu{\rm m}$ to 160 $\mu{\rm m}$ in their longest dimension 
%as seen in Table \ref{table:meteoriticdata}
\citep{KoopEtal2016b}. 
The MgO content for {\Steve the hibonite in} SHIBs ranges from 0.4 to 3.1 wt\% while TiO$_2$ ranges from 0.2 to 6.2 wt\%. 
% (Table~\ref{table:meteoriticdata}).
\citet{KoopEtal2016a} reported that hibonite in SHIBs often is very zoned in MgO and TiO$_2$. 

SHIBs generally vary in their oxygen isotopic ratios from $\Delta^{17}{\rm O} \approx -25$ to $-16 \permil$, with an average $\approx -23 \permil$, almost identical to normal CAIs \citep{KoopEtal2016b}.
Their titanium isotopic ratios vary anywhere from $\epsilon^{50}{\rm Ti} \approx -100$ to $+100$ epsilon units \citep{KoopEtal2016b}.

The distribution of initial $(\alratio)_0$ values for SHIBs {\Steve tends to skew strongly to the canonical value $\approx 5 \times 10^{-5}$,}
%possibly with two peaks near $4.9 \times 10^{-5}$ and $5.4 \times 10^{-5}$,} 
with none found with $(\alratio)_0 < 3 \times 10^{-6}$ \citep{LiuEtal2009,LiuEtal2012,LiuEtal2019,KoopEtal2016b}. 
% (Fig.~\ref{fig:alratios}). 
% (Table \ref{table:meteoriticdata}). 
There is a preponderence of nearly canonical $\alratio$ ratios in both `normal' CAIs (defined by their mineralogy and lacking in corundum or hibonite) and SHIBs (containing hibonite but also significant spinel)
These are {\it not} examples of classes of CAIs that are generally $\altwosix$-poor. 

\subsection{LAACHIs}

In contrast to normal CAIs and SHIBs, other classes of CAIs as a rule exhibit low $(\alratio)_0$ ratios and appear to sample a different reservoir. 
These include FUN CAIs, corundum grains, PLACs, grossite-bearing CAIs, and other hibonite- or corundum-bearing inclusions.
We review these here.

\subsubsection{FUN CAIs}

FUN CAIs are Ca-rich, Al-rich inclusions with both Unknown Nucleosynthetic (UN) mass-independent isotopic anomalies and mass-dependent Fractionation (F) of isotopes due presumably to evaporation \citep{MendybaevEtal2013,MendybaevEtal2017}.
%\citep{WilliamsEtal2017}
This definition permits a variety of mineralogies, and many types of CAIs can be classified as FUN.
To date, about 21 inclusions have been identified as FUN CAIs \citep{DunhamEtal2021metsoc}.

{\EDIT The $\Delta^{17}{\rm O}$ values}
%Oxygen isotopic ratios 
of FUN CAIs {\Steve can range from $-25\permil$ to $-16\permil$ \citep{KrotEtal2014}, but} are generally $\Delta^{17}{\rm O} \approx -23\permil$, indistinguishable from normal CAIs, {\Steve although they show strong mass-dependent fractionation in borth $\delta^{17}{\rm O}$ and $\delta^{18}{\rm O}$} \citep{KrotEtal2010}.
Their titanium isotopic ratios vary, but appear correlated with whether or not they are dominated by hibonite:
{\Steve 
hibonite-rich FUN CAIs can see $\epsilon^{50}{\rm Ti}$ vary from about -200 to +200; FUN CAIs not dominated by hibonite tend to cluster strongly around a single value $\approx -40$ 
}
\citep{TorranoEtal2021metsoc,DunhamEtal2021metsoc}.

FUN CAIs have been traditionally thought of as forming from an $\altwosix$-poor reservoir \citep[e.g.,][]{KrotEtal2012}, but this is true only of the subset of FUN CAIs containing significant hibonite.
As reviewed by \citet{DunhamEtal2021metsoc}, only the examples of {\it H030/L}, {\it HAL}, {\it 2-6-6}, and {\it 1-9-1} definitely formed with $(\alratio)_0 < 3 \times 10^{-6}$,
and these all contain significant hibonite, or are dominated by it.
In contrast, only {\EDIT seven} of the 21 known FUN CAIs ({\it CMS-1}, {\it EK1-4-1}, {\it GG3}, {\it CG14}, {\it STP-1}, {\it C1}, {\EDIT and {\it Vig 1623-5}}) do not contain hibonite. 
These are all large, multi-minerallic CAIs identifiable as type A or B, and all record $(\alratio)_0$ ratios $> 3 \times 10^{-6}$
\citep{DunhamEtal2021metsoc,MarinCarboneEtal2012lpsc}.
These FUN CAIs without hibonite could have formed from the same canonical reservoir as other CAIs, later in the lifetime of the solar nebula, but before parent body accretion.
FUN CAIs strongly suggest a correlation between hibonite and $(\alratio)_0$.

\subsubsection{Corundum Grains}

%Corundum is the first refractory phase to condense in the solar nebula and yet this phase is not very abundant in CAIs \citep{ViragEtal1991}. 
The corundum grains that have been analyzed are single, isolated grains with exceptionally small sizes, ranging from 0.5 $\mu{\rm m}$ to 15 $\mu{\rm m}$ \citep{ViragEtal1991,MakideEtal2011,NakamuraEtal2007}. 
The corundum is very low in TiO$_2$ and generally MgO is below detection limits, while any hibonite included in these CAIs tends to be low in both MgO and TiO$_2$ \citep{MakideEtal2013,ViragEtal1991}. 
Corundum grains often have sharp features, suggesting they are fragments of larger corundum grains \citep{ViragEtal1991}. 
Some of the corundum grains within the matrices of their host samples analyzed by \citet{NakamuraEtal2007} were found as aggregates of two to six similarly-sized grains.
Other corundum inclusions are seen rimmed in hibonite \citep{NeedhamEtal2017}.

The oxygen isotopic compositions of corundum grains range from $\Delta^{17}{\rm O} \approx -31$ to $-14\permil$, with an average near that of other CAIs, $\Delta^{17}{\rm O} \approx -23\permil$ \citep{MakideEtal2011}.
Corundum grains have not, to our knowledge, been analyzed for $\epsilon^{50}{\rm Ti}$.

Over 110 corundum grains have been analyzed for $(\alratio)_0$ \citep{ViragEtal1991,MakideEtal2011,NakamuraEtal2007}. 
Because the corundum grains are so small, it is difficult to measure more than one spot on most grains, so most of the $(\alratio)_0$ ratios are inferred from ``model" isochrons in which the intercept is fixed to the value of bulk chondrites. 
Given that many PLACs show deficits in ${}^{26}{\rm Mg}$ (\S 4.2.2), model isochrons may underestimate the slope.
Despite this, it is clear that these grains were generally deficient in $\altwosix$. 
% In Table \ref{table:meteoriticdata}, 
Although almost half of all analyzed corundum grains exhibit $(\alratio)_0$ values approaching the canonical value, we calculate a modal value for the corundum grains of $< 1 \times 10^{-6}$.

\subsubsection{PLACs and PLAC-like CAIs}

PLAty hibonite Crystals (PLACs) typically range in size from 30 $\mu{\rm m}$ to 150 $\mu{\rm m}$  \citep{KoopEtal2016a}.
They are traditionally characterized as having large, platy or plate-like hibonite crystals, but may range in morphology from platy to equant or stubby, and sometimes appear to be aggregates \citep{KoopEtal2016a,LiuEtal2012}.
%\citet{KoopEtal2016} performed a study on PLACs from a sample of 32 hibonite-rich CAIs with sizes ranging 
%These PLACs are described as single hibonite crystals having a range of morphologies from platy, equant, and stubby, or sometimes part of hibonite aggregates. 
%Void spaces filled in by iron oxide-rich alteration phases have been found within these hibonite aggregates. 
%These hibonite aggregates suggest a possibility that
Many PLACs appear to have fragmented from larger objects or aggregates with sizes $> 150 \, \mu{\rm m}$ \citep{KoopEtal2016a}.

PLACs exhibit a wider range of oxygen isotopic ratios than normal CAIs, $\Delta^{17}{\rm O} \approx -32$ to $-19 \permil$, with average $\Delta^{17}{\rm O} \approx -22\permil$, but similar to the range seen in corundum grains \citep{KoopEtal2016a}.
PLACs are known for their large positive and negative anomalies in stable isotope ratios such as ${}^{50}{\rm Ti}$, which ranges from -700 to +2700 epsilon units.

PLACs generally exhibit low, subcanonical $(\alratio)_0$ ratios \citep{Ireland1988,Ireland1990,KoopEtal2016a,LiuEtal2012,Liu2008thesis}. 
Although a minority ({\Steve about a third}) of PLACs record the canonical value of $5 \times 10^{-5}$, we calculate a modal $(\alratio)_0$ value for PLACs of $2 \times 10^{-6}$, more than an order of magnitude lower than the canonical value.
%However, a minority ($\approx 32\%$) of PLACs, record the canonical value of $5 \times 10^{-5}$, including CB-H2 with a $\alratio$ value of $\sim 5.5 \times 10^{-5}$ \citep{FaheyEtal1987a}. 

Blue AGregates (BAGs) are another type of hibonite-rich inclusion, similar to but not as common as PLACs. 
\citet{Ireland1988} analyzed three BAGs and described them as less than 80 $\mu{\rm m}$ in diameter and and as a conglomerate of hibonite plates and fragments ranging from 2 $\mu{\rm m}$ to 20 $\mu{\rm m}$.
Perovskite grains are common, but they are no larger than 2 $\mu{\rm m}$ in diameter \citet{Ireland1988}. 
Determination of $(\alratio)_0$ has been attempted for five BAGs, and is generally low.
\citet{Ireland1988} {\Steve and \citet{Ireland1990}} found upper limits $(\alratio)_0 < 1 \times 10^{-5}$ in some;
%, \citet{Ireland1990} measured one unresolved from 0 ($(0.14 \pm 1)\times 10^{-5}$) 
\citet{LiuEtal2009} detected no excesses of ${}^{26}{\rm Mg}$ in others. 

\subsubsection{Grossite-bearing CAIs}

Although not as common as corundum or hibonite, grossite is a closely related refractory calcium aluminate.
Examples of CAIs dominated by grossite {\Steve have been identified}. 
These CAIs have a large range in sizes, from 60 $\mu{\rm m}$ to 200 $\mu{\rm m}$ 
and have been found mostly in CO \citep{SimonEtal2019a} and CH chondrites \citep[e.g.,][]{Krot2019}.
Aside from grossite, these CAIs can exhibit a wide mineralogical range containing perovskite, spinel, and/or hibonite enclosed within the grossite cores. 
The hibonite is lath-like in some grossite-enriched CAIs, and layers of melilite have been observed enclosing the cores. 
Rims, or partial rims, of gehlenite or spinel have been observed. 
These grossite-rich CAIs also have been observed as aggregates with gehlenite rims and pore space and matrix between the nodules \citep{SimonEtal2019a}. 
%\citet{SimonEtal2019} measured TiO$_2$ of about 2.1 wt\% or greater, but MgO $< 1$ wt\% in the hibonite within these grossite-enriched CAIs.
%When including grossite and hibonite in these CAIs, the MgO range is 0.03 wt\% to 1.5 wt\% and the TiO$_2$ range is 0.05 wt\% to 1.6 wt\%. 

Like corundum grains and PLACs, grossite-bearing CAIs show a wide range in oxygen isotopic ratios, from $\Delta^{17}{\rm O} \approx -34$ to $-9 \permil$, with an average $\Delta^{17}{\rm O} \approx -19 \permil$, slightly more ${}^{16}{\rm O}$-poor than other CAIs \citep{KrotEtal2019}.
%Steve
{We are not aware of titanium isotopic measurements of grossite-rich CAIs.}

\citet{SimonEtal2019a} reported a bimodal distribution of initial $(\alratio)_0$ values, with 38\% near the canonical value of $5 \times 10^{-5}$; but the majority are low, and we calculate the modal value of  $(\alratio)_0$ for grossite-rich CAIs is $< 1 \times 10^{-6}$.
Complicating the issue {\Steve are two facts: the strongest evidence for low $(\alratio)_0$ ratios among grossite CAIs comes from those in CH chondrites, and although some grossite-dominated CAIs from CH chondrites show low $(\alratio)_0$, as low as  $(3.7 \pm 3.1) \times 10^{-7}$ \citep{KrotEtal2017}, these} may have been subject to a late resetting.
Many CAIs in CH chondrites appear to have been reprocessed, or at least thermally reset, in the impact plume created by the collision between asteroids that led to the CB/CH chondrites \citep{KrotEtal2005}. 
This event occurred at {\Steve 5.8} Myr after CAIs \citep{DeschEtal2023b}, when $\alratio$ would have been $\approx 2 \times 10^{-7}$. 
It is expected that the oxygen isotopic composition of the CAIs also would have been reset in the impact plume.
The fact that $\Delta^{17}{\rm O}$ in grossite-rich CAIs extends to very ${}^{16}{\rm O}$-poor compositions suggests this may have occurred to some CAIs, but perhaps not to others with more typical ${}^{16}{\rm O}$-rich compositions. 
At any rate, {\Steve at least some} grossite-bearing CAIs in general appear to have formed from a low-$\alratio$ reservoir.

\subsubsection{Other Inclusions}

In addition to the examples above, there are unusual, larger, multi-minerallic inclusions dominated by corundum and/or hibonite.

{\it M98-8} has been described by \citet{SimonEtal2002}.
It is a rounded, $90 \times 75 \, \mu{\rm m}$ refractory inclusion from the Murchison (CM) carbonaceous chondrite.
It consists of rounded and anhedral 5 $\mu{\rm m}$ to 15 $\mu{\rm m}$ corundum grains enveloped in hibonite laths that are 10 $\mu{\rm m}$ wide and 30 $\mu{\rm m}$ to 40 $\mu{\rm m}$ long. 
Perovskite grains, a few microns in size, are observed towards the inclusion's edge and at the edge of the hibonite laths. 
The inclusion has a bit of a fluffy texture due to the triangular- or trapezoidal-shaped void spaces between the hibonite crystals. 
%\Steve 
{Oxygen isotopic ratios within {\it M98-8} appear consistent with a single value $\Delta^{17}{\rm O} \approx -24\permil$. 
We are not aware of titanium isotopic measurements of this inclusion.}
This corundum- and hibonite-bearing inclusion records a low $(\alratio)_0$ value of $(-1.9 \pm 3.5) \times 10^{-6}$. %\citep{SimonEtal2002}.

{\it DOM 31-2} is a $230 \times 220 \, \mu{\rm m}$ inclusion found in the CO3.00 carbonaceous chondrite Dominion Range (DOM) 08006, described by \citet{SimonEtal2019b}.
It consists of a core of hibonite laths contained in a grossite and perovskite mantle with anhedral spinel in between, which is then contained in melilite approximately 10 $\mu{\rm m}$ to 15 $\mu{\rm m}$ wide. 
Void space is observed within the grossite and perovskite mantle as well as in between the hibonite laths in the core. 
Outside the melilite layer is a 10 $\mu{\rm m}$- to 15 $\mu{\rm m}$-sized thick layer of diopside enclosing FeNi metal within.
Lastly, a region of unstructured enstatite and forsterite grains and metal occur. 
%\Steve 
{Oxygen isotopic ratios in all minerals, including the hibonite and grossite, cluster tightly around $\Delta^{17}{\rm O} \approx -23\permil$.
Its titanium isotopic composition is $\epsilon^{50}{\rm Ti} = +88 \pm 20$ \citep{SimonEtal2019b}.}
Hibonite in this UN inclusion records an $\alratio$ initial value of $(-3 \pm 8) \times 10^{-7}$.

{\it A-COR-01}, from ALHA 77307 (CO3.0), has been described by \citet{BodenanEtal2020}.
It is a sub-rounded, $175 \times 140 \, \mu{\rm m}$ -sized inclusion with lath-like hibonite crystals with 5 $\mu{\rm m}$- to 10 $\mu{\rm m}$-sized, anhedral corundum grains in the core. 
This core is layered with spinel, which is then layered with a 5 $\mu{\rm m}$- to 10 $\mu{\rm m}$-sized rim of diopside. %\citep{BodenanEtal2020}. 
There is a 1 to $4 \, \mu{\rm m}$-sized crack that separated the spinel from the diopside layer, which \citet{BodenanEtal2020} attribute to a mechanical separation during an energetic event, e.g., an impact. 
%\Steve 
{Oxygen isotopic ratios in the core minerals of corundum and hibonite are $\Delta^{17}{\rm O} \approx -31.2 \pm 3.0\permil$ and $-33.5 \pm 2.3\permil$, in contrast to the ratio in the diopside rim, $\Delta^{17}{\rm O} \approx -24.8 \pm 0.5\permil$, which suggests the ratios in the hibonite and corundum are primary.
We are not aware of titanium isotopic measurements for this inclusion.}
This hibonite- and corundum-bearing inclusion records an $\alratio$ initial value $(1.67 \pm 0.31) \times 10^{-7}$. %\citep{BodenanEtal2020}.

{\it HAL} (Hibonite from ALlende) is a $100 \times 200 \, \mu{\rm m}$-sized inclusion from Allende composed of a hibonite core with exsolved fine needles of a Ti-rich phase, and a $120 \times 220 \, \mu{\rm m}$-sized black rim \citep{FaheyEtal1987a,AllenEtal1980}. 
This black, compact rim looks similar to devitrified glass and is composed of an anisotropic Al-Fe oxide with a feathery-like texture. 
The outer portion is composed of a fine-grained, very friable rim with thickness up to $\approx 2000 \, \mu{\rm m}$. 
This friable rim is made up of five distinct layers with an aggregate thickness of 90 $\mu{\rm m}$ to 450 $\mu{\rm m}$ and composed of a variable mineralogy {that probably formed on the parent body.}
%, with the innermost layer consisting of nepheline and sodalite with minor Ti-Fe oxide and fine, long grains of Al-Fe-oxide, and anorthite.
%The second layer contains the minerals from the innermost layer, as well as abundant perovskite, grossular, andradite, and pyroxene. 
%The third layer is predominatly Ca-phosphate perovskite, and hydroxy-apatite while the fourth layer is composed of mostly hibonite, Al-Fe oxide, garnets, and nepheline. 
%The fifth and final layer is composed of olivine, pentlandite, and Ni-Fe metal amongst other minerals found in the other layers \citep{AllenEtal1980}. 
%\Steve 
{The oxygen isotopic ratios have been altered by mass-dependent fractionation due to evaporation, but are mostly consistent with $\Delta^{17}{\rm O} \approx -25\permil$ \citep{LeeEtal1980}.}
{\it HAL} is isotopically anomalous in titanium, with $\epsilon^{50}{\rm Ti} \approx +160$ epsilon units \citep{FaheyEtal1987a}.
\citet{FaheyEtal1987a} reported $(\alratio)_0 = (5.2 \pm 1.7) \times 10^{-8}$ for this inclusion.
%Among LAACHIs, the very largest of these may be the singular object {\it HAL} (radius $1000 \, \mu{\rm m}$); but more typically, low-$\alratio$ inclusions are not larger than {\it DOM 31-2} (apparent radius $\approx 110 \, \mu{\rm m}$) and {\it A-COR-01} (equivalent radius $\approx 80 \, \mu{\rm m}$).

{\it SHAL} (Son of HAL) is a HAL-like Allende inclusion made up of a $500 \times 200 \, \mu{\rm m}$-sized single hibonite grain along with blocky-like perovskite roughly 200 $\mu{\rm m}$ in size \citep{LiuKeller2017lpsc}. 
%{\Steve {\bf Oxygen Isotopes?}}
%{\Steve {\bf Titanium Isotopes?}}
{\it SHAL} perovskite and hibonite appear devoid of $\mgtwosix$ excesses attributed to the $\altwosix$ decay \citep{LiuEtal2016metsoc}. 

{\it HIDALGO} (HIbonite in Dar AL Gani cO3) is a HAL-like, nearly stoichiometrically pure single hibonite inclusion with a size of $300 \times 300 \, \mu{\rm m}$. 
It contains 0.2 wt\% TiO$_2$ and MgO below the detection limit, making it the most Mg-depleted hibonite found so far \citep{LiuEtal2021metsoc}. 
%{\Steve {\bf Oxygen Isotopes?}}
%{\Steve {\bf Titanium Isotopes?}}
This inclusion records an $(\alratio)_0$ value of $(1.5 \pm 0.02) \times 10^{-5}$ \citep{LiuEtal2021metsoc}.

\subsection{Summary}

While normal CAIs and SHIBs attest to a widespread uniform and canonical value $\alratio \approx 5 \times 10^{-5}$ in the solar nebula, some inclusions in meteorites formed with $(\alratio)_0$ as low as $\sim 10^{-7}$, which cannot be explained by late resetting, and must have formed from a reservoir with subcanonical $\alratio$.
As outlined above, these ``LAACHIs" include: some hibonite-bearing FUN CAIs, most corundum grains, PLACs and PLAC-like CAIs, BAGs, grossite-bearing CAIs, as well as some other unusual inclusions dominated by hibonite and/or corundum.
The properties of LAACHIs, SHIBs, and normal CAIs are summarized in {\bf Table~\ref{table:meteoriticdata}}. 
We list their typical sizes; their mineralogies and morphologies, which suggest origins as aggregates or fragments;  their MgO and TiO$_2$ contents; their typical oxygen isotopic compositions; their typical ${}^{50}{\rm Ti}$ values; and their typical initial $(\alratio)_0$ ratios.

\begin{figure}
\centering
\begin{subfigure}
  \centering
  \includegraphics[width=.5\linewidth]{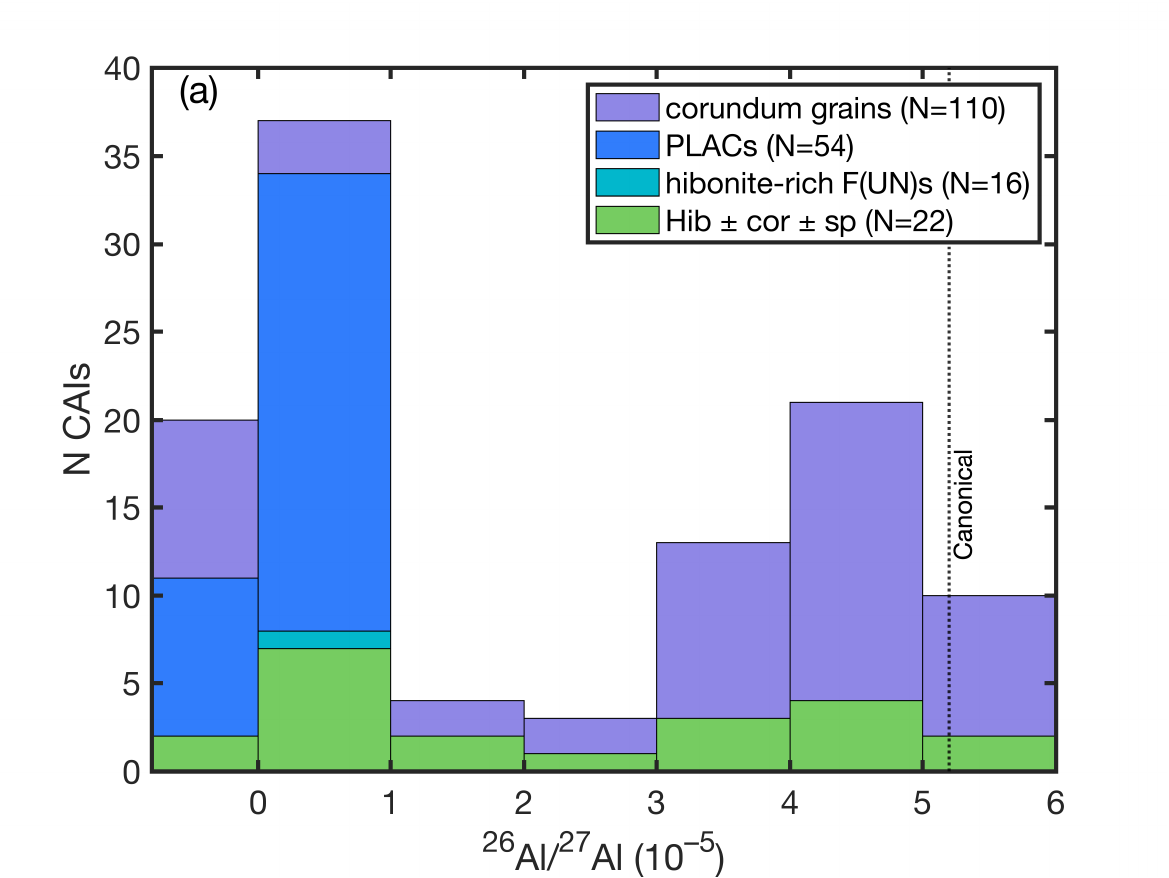}
  %\caption{A subfigure}
  \label{fig:sub1}
\end{subfigure}%
\begin{subfigure}
  \centering
  \includegraphics[width=.41\linewidth]{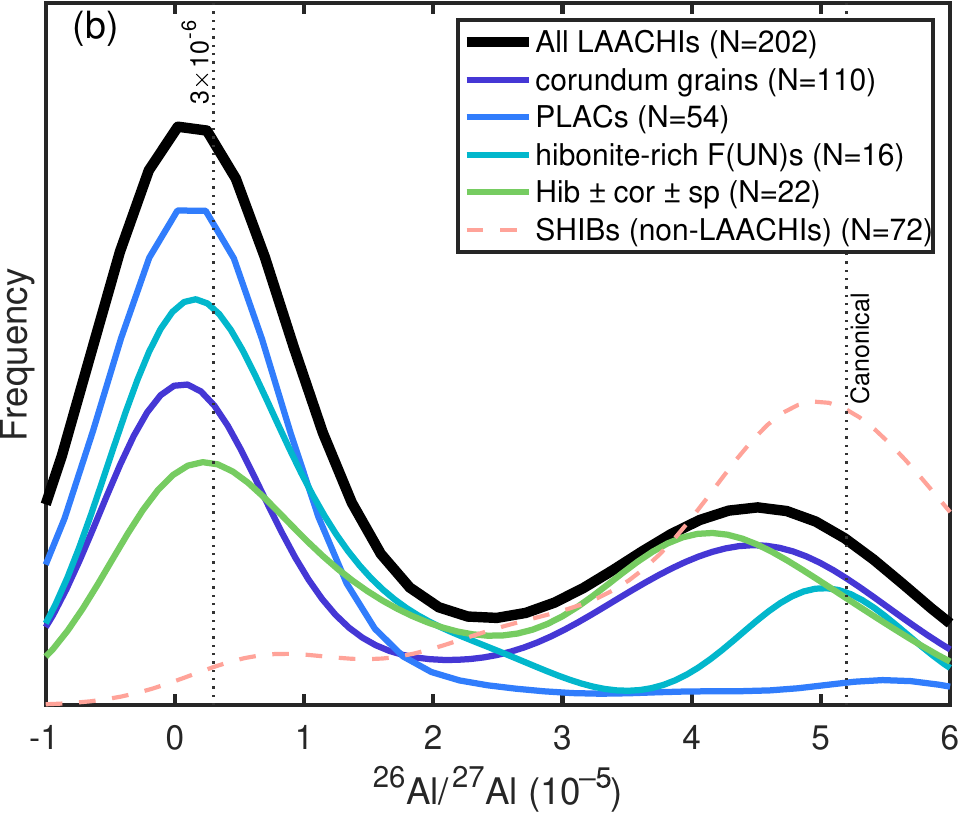}
  %\caption{A subfigure}
  \label{fig:sub2}
\end{subfigure}
\caption{Histograms (a) and kernel density estimation (KDE) (b) of compiled literature values of initial $(\alratio)_0$ ratios in types of inclusions that are considered low in $\altwosix$, including: corundum grains; PLACs; hibonite-rich FUN CAIs; and other hibonite- or corundum-bearing inclusions. 
{\Steve the distributions are strongly bimodal: a minority} have more canonical initial ratios $(\alratio)_0 \approx 5 \times 10^{-5}$, {\Steve but} the majority of these inclusions {\Steve with low $(\alratio)_0$ record modal} $(\alratio)_0 < 2 \times 10^{-6}$ (black dotted line in (b)). All these types of inclusions with low modal $(\alratio)_0$ are dominated by corundum or hibonite (or sometimes grossite), and we categorize such inclusions as Low-$\alratio$ Corundum/Hibonite Inclusions (LAACHIs). Other non-LAACHI CAIs, not dominated by corundum or hibonite, record canonical $(\alratio)_0$ ratios, including: SHIBs, whose KDE is shown (dashed orange curve); and `normal' type A/B/C CAIs (black dotted line "canonical") dominated by melilite or anorthite, too numerous to show, which cluster strongly around $(\alratio)_0 \approx 5 \times 10^{-5}$.}
\label{fig:alratios}
\end{figure}

In {\bf Figure~\ref{fig:alratios}} we present compiled literature values of the $(\alratio)_0$ ratios of hibonite-rich FUN CAIs, corundum grains, PLACs, and other hibonite-bearing inclusions.
These are displayed as histograms and as kernel density estimations we have calculated.
Each population, especially corundum grains and spinel-bearing inclusions, has a minority of grains that show canonical ratios $(\alratio)_0 \approx 5 \times 10^{-5}$; but all types of LAACHIs show modal values $(\alratio)_0 \leq 2 \times 10^{-6}$, meaning the majority of them must have formed from a reservoir with subcanonical $\alratio$. 
The distribution of $(\alratio)_0$ ratios in SHIBs is shown for comparison.
Figure~\ref{fig:alratios} is to be compared to similar plots of normal CAIs, which show a strong preponderance to have values near the canonical ratio \citep[e.g.,][]{MacPhersonEtal2005,Davis2022}.

These probability density functions make clear that among inclusions dominated by corundum or hibonite or grossite, $(\alratio)_0$ ratios are indeed low, with the modal value $\approx 2 \times 10^{-6}$. 
Conversely, all types of inclusions with low modal $(\alratio)_0$ are ones dominated by corundum or hibonite or grossite.
CAIs not dominated by corundum or hibonite or grossite tend to have  $(\alratio)_0$ ratios consistent with formation from a reservoir with canonical $\alratio$.
It is this evident dichotomy that leads us to categorize many CAIs as LAACHIs.
LAACHIs are typically tens to hundreds of microns in size, with many having morphologies suggestive of aggregation, and many resembling fragments of larger objects. 

%-------------------------------------------
% TABLE 1. DATA 
%$\!\!\!\!\!\!\!\!\!\!\!\!\!\!\!\!\!\!\!\!\!\!\!\!\!\!\!\!\!\!\!\!$
\begin{deluxetable*}{p{1.2in}|c|p{1.4in}|p{0.9in}|p{0.9in}|p{0.6in}|p{0.7in}}
% \rotate
% \renewcommand{\thefootnote}{\alph{footnote}}
\label{table:meteoriticdata}
%\tablenum{1}
\tablecaption{Properties of LAACHIs and other CAIs}
\tablewidth{6.5in}
\tablehead{
\colhead{Objects} & 
\colhead{Sizes ($\mu{\rm m}$)} & 
\colhead{Morphology} & 
\colhead{MgO, TiO$_2$} & 
\colhead{$\Delta^{17}{\rm O}$ ($\permil$)} &
\colhead{$\epsilon^{50}{\rm Ti}$ ${}^{\dagger\dagger}$} &
\colhead{$({}^{26}{\rm Al}/{}^{27}{\rm Al})_0$ ${}^{\dagger}$} 
}
\startdata 
Hibonite-rich F(UN) and HAL-like CAIs$^{a}$ & $80 - 300$ & {\footnotesize Large lath-like or needle-like hib. grains with interstitial sp. grains and layers of mel. and diop. Void spaces in hib.} & 0.02-0.6 wt\%, 0.03-6.4 wt\%  & {$\approx -24$} & {$\pm 200$} & $< 2 \times 10^{-6}$ \linebreak 50\% Low  \\
\hline
Corundum grains$^{b}$ & $0.5 - 15$ & {\footnotesize Single cor. grains, sometimes aggregates.} & $\sim 0$ wt\%, \hfill \linebreak $\sim 0.09$ wt\%  & -31 to -14 \hfill \linebreak Avg. -23 & - & $< 1 \times 10^{-6}$ \linebreak 43\% Low \\
\hline
\mbox{PLACs and}\ \linebreak PLAC-like CAIs$^{c}$     & 30 - 150 & {\footnotesize Large  hib. grains (may be aggregates), ranging from platy to stubby and sp.-rimmed to sp.-free. Pv. found as inclusions within hib. and between grains in aggregates. Voids within aggregates.} & 0.2 - 1.4 wt\%, $\,\,\,\,$ 0.7 - 3.1 wt\%  & $-32$ to $-19$ \hfill \linebreak Avg. -22 & -700 to +2700 & $\approx 2 \times 10^{-6}$ \linebreak 72\% Low \\
\hline
Grossite-rich CAIs$^{d}$ & $60 - 200$ & {\footnotesize Gros.-rich, can contain hib., pv., mel., sp. within grossite or in rims.} & 0.03 - 1.5 wt\%, 0.05 - 1.6 wt\% & $-34$ to $-9$ \hfill \linebreak Avg. -19 & {-} & $< 1 \times 10^{-6}$ \linebreak 50\% Low \\
\hline
Other Corundum \linebreak \mbox{$\pm$ Hibonite} \mbox{$\pm$ Spinel} Inclusions$^{d}$  & $30 - 175$ & {\footnotesize Lath-shaped or platy hib. with cor. and/or sp. grains within or surrounding hib. Void spaces within and/or between hib. or cor. grains.} & 0.02- 13 wt\%, \linebreak 0.05 - 11 wt\%& {$\approx -24$} & {$\pm$ tens?} & $< 2 \times 10^{-6}$ \linebreak 27\% Low \\
\hline
\hline
SHIBs$^{f}$           & $30 - 160$ & {\footnotesize Abundant hib. and sp., with pv., FeO-poor silicates, refractory metal blebs. Voids, cracks common.} & 0.4 - 3.1 wt\%, 0.2 - 6.2 wt\% & $-24$ to $-22$ \hfill \linebreak Avg. -23 & $\approx \pm 100$ & $\approx 4 \times 10^{-5}$ \linebreak 0\% Low \\
\hline
`Normal' CAIs$^{g}$  & $5 - 20,000$  & {\footnotesize Variable, with mel.-rich  type A common.} & $\,$  & $-26$ to $-22$, Avg. -24  & +2 to +15 \hfill \linebreak ($\approx +9$) & $\approx 5 \times 10^{-5}$ \linebreak rare low \\
\hline
\enddata
\tablecomments{
{\it Hib.} $=$ hibonite, {\it Cor.} $=$ corundum, {\it Gros.} $=$ grossite, {\it Sp.} $=$ spinel, {\it Pv.} $=$ perovskite, {\it Mel.} $=$ melilite, {\it Diop.} $=$ diopside.
${}^{\dagger}{\rm Modal}$ value (see Fig.~1), and `Low' fraction with  $(\alratio)_0 < 3 \times 10^{-6}$.
${}^{\dagger\dagger}{\rm Using}$ 49/47 normalization.
References: 
$a$ (Hibonite-rich F(UN) CAIs): 
\citet{RussellEtal1998},
\citet{UshikuboEtal2007}, 
\citet{RoutEtal2009},
\citet{KoopEtal2018}, %need to add done
\citet{SimonEtal2019b};
(HAL-like)
\citet{FaheyEtal1987a}, %check that (a) is HAL yep.
\citet{IrelandEtal1992},
\citet{LiuEtal2016metsoc}, %check that this is SHAL yep.
\citet{LiuEtal2021metsoc}; %this is abstract about Hidalgo
\citet{LeeEtal1980};
$b$ (corundum grains): 
\citet{ViragEtal1991}, 
\citet{NakamuraEtal2007},
\citet{MakideEtal2011};
$c$ (PLACs and PLAC-like CAIs): 
\citet{FaheyEtal1987a}; %make sure (a) is about Mur-A1, Mur-H7, and CB-H2 yep.
\citet{TrinquierEtal2009},
\citet{LiuEtal2009}, %about 8 PLACs in 1 isochron,
\citet{LiuEtal2012},
\citet{KoopEtal2016a}, %make sure this is the 2016 paper about PLACs yep.
\citet{LiuEtal2019},
\citet{ShollenbergerEtal2022};
%$c$ (BAGs): \citet{Liu2008thesis};
$d$ (Grossite-rich CAIs): 
\citet{SimonEtal2019a},
\citet{KrotEtal2019} and references therein,
\citet{WeberEtal1995}; %add ref around grs-rich and say "references therein"
$e$ (Other Corundum $\pm$ Hibonite $\pm$ Spinel):
\citet{HintonBischoff1984},
\citet{Bar-MatthewsEtal1982}, %about BB-5
\citet{Fahey1988thesis}, %need to add done, about F5
\citet{SimonEtal2002},
\citet{SugiuraEtal2007}, %need to add done, about Acfer 094 CAI s2
\citet{RoutEtal2009},
\citet{MakideEtal2013}, %need to add done, about Murchison hibcors
\citet{LiuEtal2019},
\citet{BodenanEtal2020};
$f$ (SHIBs):
\citet{LiuEtal2009},
\citet{LiuEtal2012}, %need to add done
\citet{KoopEtal2016b}, %make sure this is 2016 about SHIBs
\citet{Ireland1990};
$g$ (Normal CAIs): 
\citet{MacPhersonEtal1995},
\citet{McKeeganDavis2003},
\citet{MacPhersonEtal2005}, 
\citet{KrotEtal2010}, 
\citet{WilliamsEtal2016},
\citet{DavisEtal2018}.
}
\end{deluxetable*}
%-------------------------------------------

%--------------------
%FIGURE 1
%\begin{figure*}[t!]
%\centering
  %\includegraphics[width=1\linewidth]{Figure1.png}
  %\vspace{-1.2in}
  %\caption{Histograms ({\it A}) and probability density functions ({\it B}) of compiled literature values of initial $(\alratio)_0$ ratios in types of inclusions that are considered low in $\altwosix$, including: corundum grains; PLACs; hibonite-rich FUN CAIs; and other hibonite- or corundum-bearing inclusions. While a fraction of these, particularly corundum grains, have more canonical initial ratios $(\alratio)_0 \approx 5 \times 10^{-5}$, the majority of these inclusions record $(\alratio)_0 < 2 \times 10^{-6}$. All these types of inclusions with low modal $(\alratio)_0$ are dominated by corundum or hibonite (or sometimes grossite), and we categorize such inclusions as Low-$\alratio$ Corundum/Hibonite Inclusions (LAACHIs). Other non-LAACHI CAIs, not dominated by corundum or hibonite, record canonical $(\alratio)_0$ ratios, including: SHIBs, whose probability distribution function is shown (dashed orange curve); and `normal' type A/B/C CAIs (black dashed line) dominated by melilite or anorthite, too numerous to show, which cluster strongly around $(\alratio)_0 \approx 5 \times 10^{-5}$. In (A), for clarity, we have limited the y-axis to 20, but the corundum histogram bar indicated extends to 40.}
  %\label{fig:alratios}
%\end{figure*}
%--------------------
%--------------------
%FIGURE 1

% SECTION 3: MODEL 
\section{Astrophysical Model for Formation of LAACHIs}

\begin{figure*}[t!]
\centering
  \includegraphics[width=1\linewidth]{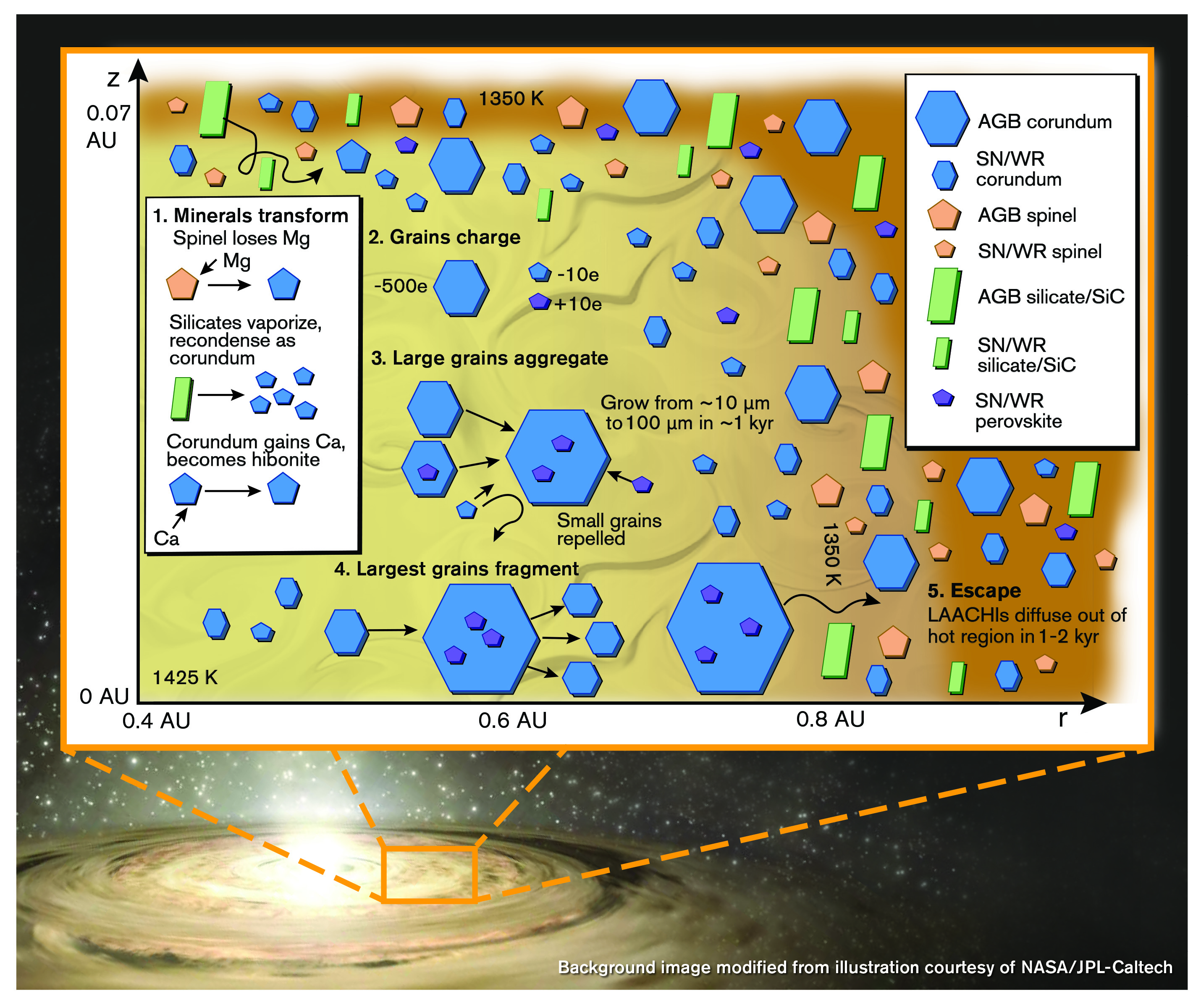}
  \caption{ {\Steve
  Illustration of our model for formation of Low-$\alratio$ Corundum/Hibonite Inclusions (LAACHIs). Very early in solar nebula evolution ($< 10^5$ yr), most grains were presolar. 
  These grains diffused into hot ($T > 1350$ K; light brown), turbulent ($\alpha \sim 10^{-3}$)  regions $\sim 0.6$ AU from the Sun, within a scale height of the disk midplane.
  1.) Corundum, spinel, hibonite, and perovskite were practically the only surviving grains; silicates vaporized, releasing Ca and other vapor, recondensing as corundum. Spinel rapidly converted to corundum by loss of Mg. Corundum reacted with Ca vapor to form hibonite. 
  2.) Grains in this region were charged according to their size and material work function, with most grains negatively charged, except for positively charged perovskite. 
  3.) Their relative velocities were sufficient for the largest grains to coagulate, but small $\altwosix$-bearing  grains generally were repelled. Aggregates of $\altwosix$-free  corundum and hibonite grew to $10 - 10^2 \, \mu{\rm m}$ in  hundreds of years. 
  4.) Growth beyond this size was inhibited by fragmentation.
  5.) LAACHIs escaped this region after thousands of years. }
  (Credit: Sue Selkirk, ASU.)}
  \label{fig:model}
\end{figure*}

{
We develop here a simple but quantitative model to demonstrate that $10-100 \, \mu{\rm m}$ LAACHIs could have formed in the solar nebula, with low $(\alratio)_0$, despite the widespread existence of $\altwosix$.
The key elements of this model are depicted in {\bf Figure~\ref{fig:model}}. 
{\Steve 
As presolar 
%corundum, spinel (and hibonite) 
grains find their way into regions $> 1350$ K, pyroxene, olivine and other grains vaporize; corundum, spinel and hibonite are essentially the only solids in these regions.}
Spinel is likely transformed to corundum by loss of Mg, without gain or loss of Al atoms.
Corundum may partially or completely convert to hibonite by gaining Ca, without gain or loss of Al. 
Hibonite could convert to corundum by loss of Ca (or even gain Ca to convert to grossite), or may remain as hibonite, all depending on what phases are thermodynamically favored. 
As almost the only solids in the hot midplane region, micron-sized grains of corundum or hibonite quickly aggregate and grow into objects tens of microns in size, within hundreds of years.
{\Steve Only large grains ({\EDIT a few to tens of $\mu{\rm m}$} in size) can coagulate, and these large grains derive from AGB stars and contain no live $\altwosix$.}
We assume that all live $\altwosix$ is carried on small (tens of nm) grains that cannot be accreted into the aggregates because they are too negatively charged, and repelled from the negatively charged corundum/hibonite aggregates.
{\Steve These grains include $\altwosix$-bearing refractory grains condensed in the ejecta of a recent, nearby supernova or Wolf-Rayet star.
Because ejecta densities drop so quickly in these environments, dust grains do not grow as large as in AGB outflows. 
These small $\altwosix$-bearing grains also include refractory grains condensed in the environment from Al vapor from vaporized, non-refractory grains.}
We assume that ${}^{48}{\rm Ca}$ and ${}^{50}{\rm Ti}$ anomalies are carried on similarly small grains of perovskite [${\rm CaTiO}_{3}$] that, {\Steve we argue}, are positively charged and can be accreted. 
As such, LAACHIs would be devoid of live $\altwosix$, but would show correlated anomalies in  ${}^{48}{\rm Ca}$ and ${}^{50}{\rm Ti}$.
LAACHIs are likely to diffuse out of the hot midplane region in {\Steve thousands} of years, before they grow larger than $100 \, \mu{\rm m}$. 
Once outside of this region, they may combine with other material to form objects like hibonite-bearing CAIs and SHIBs.
We further hypothesize that spinel in SHIBs formed in other, cooler regions where it was possible to accrete the small grains containing live $\altwosix$, so that although SHIBs carry hibonite, they would show more canonical $(\alratio)_0$ ratios.
}

%--------------------
%FIGURE 2
%--------------------

{
{\Steve In this section} we provide quantitative assessments of the validity of each element of our hypothesis.
We begin in \S 3.1 by describing the temperature structure and other aspects of the hot region of the protoplanetary disk in which LAACHIs form.
{\Steve In \S 3.2 we discuss what phases are thermodynamically favored and how minerals transform from one phase to another. Almost all Al is in the solid phase.
}
{\Steve In \S 3.3 we discuss the likely presolar carriers of Al and $\altwosix$, and their sizes and abundances, both in the solar nebula and in the LAACHI-forming region.}
In \S 3.4 we demonstrate that these {\Steve presolar} grains could exchange oxygen isotopes with the disk gas.
In \S 3.5 we quantify the electrical charging of dust grains in the hot environment and show that micron-sized grains typically would have hundreds of electrons per grain; as a result, we show, the nanospinels would be repulsed by the corundum/hibonite grains (and each other), but that the corundum/hibonite grains could collide with {\Steve and stick to} each other.
Next, in \S 3.6 we quantify the growth rate of {\Steve LAACHIs} and show that objects in the typical size range $10-100 \, \mu{\rm m}$ form before they leave this {\Steve hot region of the disk}.
{\Steve In \S 3.7 we calculate the incorporation of $\altwosix$ into these growing particles, and demonstrate that they would generally form with $(\alratio)_0 < 10^{-7}$.}
{\Steve These various stages are depicted in Figure 2.}
In \S 3.8 we discuss their fate after leaving this region, and their observable characteristics. 
}

\subsection{The LAACHI-Forming Environment}

{\Steve 
We hypothesize that LAACHIs uniquely formed without $\altwosix$, early, and in a region hot enough to vaporize almost but not quite all solids.}
The production of PLACs and other low-$\alratio$ inclusions is widely accepted to have taken place early in the evolution of the solar nebula, before most of its (presolar) material had been thermally processed \citep[e.g.,][]{LiuEtal2009,MakideEtal2011,KrotEtal2012}.
This stage could have occurred even before the Sun had mostly formed.
One of the more widely used treatments of the solar nebula at this early time is that of \citet{YangCiesla2012}, and we adopt their canonical model here. 

We assume formation of LAACHIs at about $r = 0.6 \, {\rm AU}$, at $\approx 0.1$ Myr, when the Sun had a mass of only $0.36 \, M_{\odot}$, yielding an orbital frequency $\Omega = 2.6 \times 10^{-7} \, {\rm s}^{-1}$.
At this point, the mass accretion rate through the disk was $\dot{M} \approx 5.0 \times 10^{-7} \, M_{\odot} \, {\rm yr}^{-1}$ and its surface density $\Sigma \approx 1.0 \times 10^{4} \, {\rm g} \, {\rm cm}^{-2}$.
%, implying an inward radial velocity of the gas $V_{r} = 47 \, {\rm cm} \, {\rm s}^{-1}$, or 98 AU/Myr.
The midplane temperature was $\approx 1425 \, {\rm K}$, so that the isothermal sound speed was $C = 2.25 \, {\rm km} \, {\rm s}^{-1}$. 
The scale height $H = C / \Omega = 8.8 \times 10^{11} \, {\rm cm} = 0.06 \, {\rm AU}$, the midplane density $\rho(0) = 5.5 \times 10^{-9} \, {\rm g} \, {\rm cm}^{-3}$, and the midplane pressure $P = 2.8 \times 10^{-4}$ bar.
The assumed turbulence parameter was $\alpha = 1 \times 10^{-3}$, and the turbulent viscosity was $\nu = \alpha C H = 2.0 \times 10^{14} \, {\rm s}^{-1}$.

{\Steve Under these conditions, the} disk would have had a vertical temperature gradient, with temperatures of a few hundred K near the surface, increasing to $\approx 1425 \, {\rm K}$ at the midplane. 
The surface temperature was dominated by accretional energy, so that $\sigma T_{\rm eff}^{4} = (9/8) \Sigma \, \nu \, \Omega^2$, or $T_{\rm eff} = 235 \, {\rm K}$.
With increasing optical depth $\tau$ into the disk, the temperature increased, as $T \approx T_{\rm eff} \, (3 \tau / 4)^{1/4}$ \citep{Hubeny1990}, and so reached 1350 K, the temperature at which most silicates evaporate at the relevant pressures, at optical depth $\tau \approx 1430$. 
For the Rosseland mean opacity assumed by \citet{YangCiesla2012}, $\kappa = 1 \, {\rm cm}^{2} \, {\rm g}^{-1}$, this entailed a surface density of $1430 \, {\rm g} \, {\rm cm}^{-2}$ between the 1350 K layer and the surface, a small fraction of the total surface density of the disk. 
The $T = 1350 \, {\rm K}$ surface would have been at a height $z = 1.18 \, H$ above the midplane.
We assume an opacity {\EDIT $\kappa \approx 0.1 \, {\rm cm}^{2} \, {\rm g}^{-1}$} in the hotter regions (see {\Steve calculation in} \S 3.6), so that the total optical depth to the midplane was {\EDIT 1790}, and the temperature at the midplane 1425 K.
The conditions at the 1350 K point at $z = 1.18 \, H$, below which silicates were vaporized, were $\rho = 2.7 \times 10^{-9} \, {\rm g} \, {\rm cm}^{-3}$, and $P = 1.3 \times 10^{-4}$ bar.

We note that this hot midplane region would have been sufficiently ionized to support the magnetorotational instability (MRI) and therefore a turbulence parameter $1 \times 10^{-3}$. 
\citet{DeschTurner2015} calculated ionization fractions $\sim 10^{-7}$, and Elsasser numbers $\gg 1$ (for both Ohmic dissipation and ambipolar diffusion) at temperaures 1350 - 1500 K in a similar region. 
Simulations by \citet{SanoEtal2004} show that under a variety of assumptions (zero or non-zero net magnetic flux, isothermal or adiabatic conditions), the MRI saturates at levels equivalent to $\alpha \sim 10^{-3}$, {\Steve justifying the assumption of \citet{YangCiesla2012}, and indicating a self-consistent disk structure.} 

An important quantity is how much time grains would have spent in such a portion of the disk before diffusing out of it.
The average time taken to vertically diffuse a distance $L \approx 1.18 H$ is $L^2 / D$, where $D = \nu / {\rm Sc}$, ${\rm Sc} \approx 0.7$ being the appropriate Schmidt number \citep{DeschEtal2017}.
This yields a median time spent in the $T > 1350 \, {\rm K}$ region of $\approx 120$ years (if remaining at $r = 0.6$ AU).
Only $\sim 10\%$ of grains would have spent less than 20 years in the region, and only $\sim 10\%$ would have stayed longer than 3700 years.
Escape is also possible by radial diffusion.
The radial temperature gradient in the disk was likely 100 K per 0.1 AU, and a typical time to diffuse radially outward $L \sim 0.2 \, {\rm AU}$ (200 K) would be $\sim 1000$ years. 
Particles typically would not have spent more than hundreds of years in this region, {\Steve but those particles that escaped and found their way into chondrites could retain a record of the unique conditions in this environment.}

%The accretional heating liberated at the midplane by the turbulent viscosity (with $\alpha = 10^{-3}$) can be expected to drive convection, with perhaps 10\% of the accretional energy flux $F = 9 \times 10^{5} \, {\rm erg} \, {\rm cm}^{-2} \, {\rm s}^{-1}$ being carried by \textbf{convective} motions \citep{BellEtal1997,MeibomEtal2000}. 
%The convective velocity is given by $0.1 F \approx 4 \rho(0) \, c_{\rm P} \, T \, V_{\rm conv}^3 / (g l_{\rm m})$ \citep{MeibomEtal2000}.
%Assuming $c_{\rm P} = 1.0 \times 10^{7} \, {\rm erg} \, {\rm g}^{-1} \, {\rm K}^{-1}$, $l_{\rm m} = H$, and $g = \Omega^2 z$, we find $V_{\rm conv} = 3 \times 10^{4} \, {\rm cm} \, {\rm s}^{-1}$, about 10\% of the sound speed. 
%This results in convective updrafts and downdrafts, and particles carried in these convective cells will pass from $z = +2.1 \, H$ to $z = -2.1 H$ in about 4 years.  
% NOPE!! CONVECTION IS ABSENT IN THE MIDPLANE REGION, AND WEAK IN THE OTHER REGIONS, AND IN GENERAL NOT IMPORTANT.

\subsection{Stable Phases in the LAACHI-Forming Environment}

\begin{figure*}[t!]
\centering
  \includegraphics[width=1\linewidth]{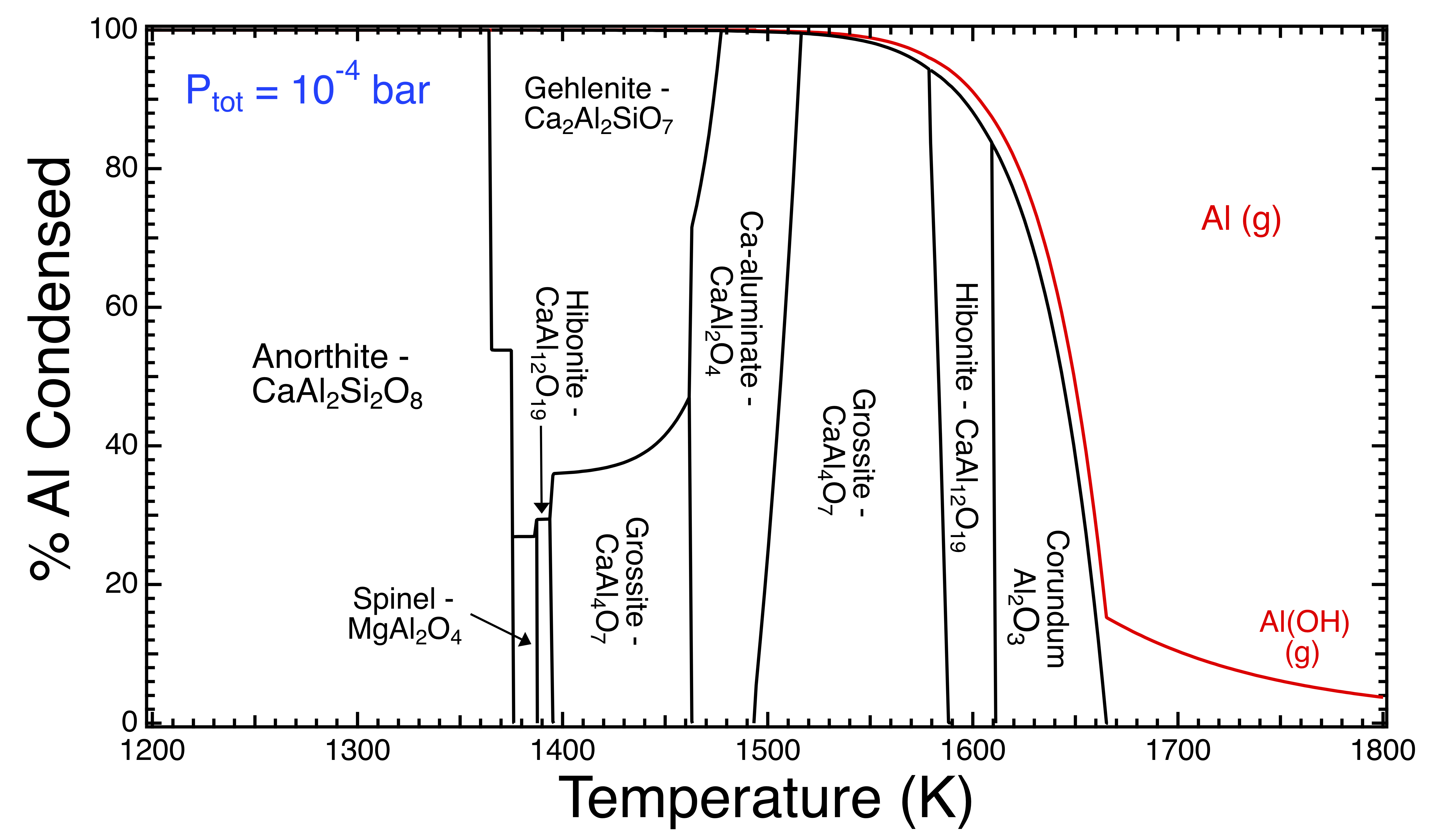}
  \caption{Fraction of all Al atoms existing in various Al-bearing phases at different temperatures in a solar composition gas.  Below 1500 K, essentially all Al exists in solid phases. Above 1650 K it is entirely in gas form. Calculated using ArCCoS code.}
  \label{fig:alphases}
\end{figure*}

{\Steve
Regions like the LAACHI-forming environment are hot enough ($> 1350$ K) to vaporize silicates.
Almost the only minerals that could survive in such a region are the Ca- and Al-bearing minerals that characterize CAIs.}
We have calculated the {\Steve thermodynamically} favored phases using the ArCCoS condensation code \citep{UnterbornPanero2017}, benchmarked against the condensation calculations of \citet{EbelGrossman2000}.
At a pressure $P = 1 \times 10^{-4}$ bar (useful for comparisons to other calculations, and close to the pressure in the LAACHI-forming region; \S 3.1), assuming solar compositions \citep{Lodders2003}, we find that olivine [$({\rm Mg},{\rm Fe})_{2}{\rm SiO}_{4}$] vaporizes fully at 1354 K, and pyroxene [$({\rm Mg},{\rm Fe}){\rm SiO}_{3}$] at 1290 K. 
We adopt 1350 K as the temperature at which silicates vaporize and the opacity drops. 
Above those temperatures, only a handful of minerals are thermodynamically favored:
spinel [${\rm MgAl}_{2}{\rm O}_{4}$], at $\approx \!\! 1375 - 1387 \, {\rm K}$; 
hibonite [${\rm CaAl}_{12}{\rm O}_{19}$], at $\approx \!\! 1386 - 1395 \, {\rm K}$; 
grossite [${\rm CaAl}_{4}{\rm O}_{7}$], at $\approx \!\! 1394 - 1463 \, {\rm K}$; 
krotite [${\rm CaAl}_{2}{\rm O}_{4}$], at $\approx \!\! 1461 - 1516 \, {\rm K}$;
grossite, at $\approx \!\! 1493 - 1588 \, {\rm K}$;
hibonite, at $\approx \!\! 1579 - 1609 \, {\rm K}$;
and corundum [${\rm Al}_{2}{\rm O}_{3}$], at $\approx \!\! 1606 - 1665 \, {\rm K}$.
Gehlenite [${\rm Ca}_{2}{\rm Al}_{2}{\rm SiO}_{7}$] (an end-member of melilite) is also stable from 1364 K to 1478 K.
Among phases without Al, diopside [${\rm CaMgSi}_{2}{\rm O}_{6}$] is unstable at temperatures $> 1360$ K {\Steve although solid solutions with Ti-rich pyroxene may be more stable}.
Perovskite [${\rm CaTiO}_{3}$] is favored from 1316 - 1583 K, but anatase [${\rm TiO}_{2}$] is unstable above 1132 K.
The temperatures of these phase transformations all increase by the  same small amount ($< 10$ K) at pressures of $3 \times 10^{-4}$ bar. 
{\bf Figure~\ref{fig:alphases}} shows the phases into which Al partitions, as a function of temperature, at $P = 1 \times 10^{-4}$ bar.

{\Steve These stability fields determine the fate of the Al in silicate grains that vaporize upon entry into the hot midplane region. 
Depending on the temperature, this Al could condense as any of the phases in Figure~\ref{fig:alphases}. 
For reasons discussed below, we consider it most likely that the Al vapor will condense by self-nucleation into numerous small ($<$ tens of nm) grains.These grains are likely corundum that then transforms into other relevant phases (e.g., hibonite).}
Very little Al would be in the gas phase: 
{\Steve the vapor pressure of Al over corundum is $2.5 \times 10^{-15}$ atm at 1425 K, and $1.5 \times 10^{-16}$ atm at 1350 K \citep{TakigawaEtal2015}.
Equivalently, a fraction $1.6 \times 10^{-6}$ of all Al atoms are in the gas phase at 1425 K, and only $2.0 \times 10^{-7}$ at 1350 K.}
%we calculate $< 2 \times 10^{-4}$ of all Al atoms at 1425 K, and $< 2 \times 10^{-6}$ of all Al atoms at 1325 K.

In contrast to Al, a significant fraction of Ca should be in vapor form at $T > 1360 \, {\rm K}$ above which diopside vaporizes,
{\Steve although we note that more Ca might be found to condense if the ArCCoS code had included solid solutions of Ca-rich pyroxene, Ti-rich pyroxene, and kushiroite [${\rm CaAlAlSiO}_6$].}
The major solids in this {\Steve temperature regime}---presolar spinel and corundum---do not contain Ca.
%Although rare, presolar perovskite [${\rm CaTiO}_{3}$] grains \citep[e.g.,][]{VollmerEtal2007} would exist in this region and carry some Ca.
Although not yet discovered among presolar grains, \citet{DauphasEtal2014} suggested the existence of presolar perovskite [${\rm CaTiO}_{3}$] grains.
These refractory grains would have survived in the hot midplane region, but even if all Ti condensed as perovskite, fewer than 0.03 Ca atoms per Al atom would be sequestered in these solids ($< 4$\% of all Ca atoms), far short of the 0.75 Ca atoms per Al atom in a solar abundance \citep{Lodders2003}.
Even if all Al were in the form of grossite, this would consume only another 33\% of all Ca. 
Throughout much of the hot midplane region, most Ca would have been in vapor form, unless it could have fully reacted with solids, forming gehlenite (melilite), which is unlikely.
%, which would have required the kinetically slow inward diffusion of Si as well.

{\Steve These calculations reveal the equilibrium, thermodynamically favored phases, but kinetic effects matter as well. For example,}
a spinel grain entering the hot midplane region would quickly find itself in regions $> 1387 \, {\rm K}$ where it was unstable.
This does not mean it would be destroyed; as discussed by \citet{ZegaEtal2021}, spinel grains can remain intact by transforming to a different solid phase quickly.
Because Ca and Si vapor existed in the hot midplane region, it would have been possible for corundum and spinel (and hibonite) to transform to more thermodynamically favored minerals such as gehlenite and grossite; but these transformations to thermodynamically favored phases likely were kinetically inhibited, due to the slow diffusion of the tetravalent cation ${\rm Si}^{4+}$ into the minerals.
Instead, as a spinel grain passed through regions with temperatures high enough that spinel was no longer stable, it would have lost Mg (as ${\rm Mg}^{+}$) and O, creating a residue of corundum, as is experimentally observed \citep{RutmanEtal1968}.
This process would have been limited only by diffusion of Mg through spinel, which is rapid.
The diffusion coefficient for Mg in spinel, at about 1600 K, is $D_{\rm Mg} \sim 10^{-16} \, {\rm m}^{2} \, {\rm s}^{-1}$ \citep{ShengEtal1992,LiermannGanguly2002,VogtEtal2015}, about three orders of magnitude faster than self-diffusion of O in spinel \citep{vanOrmanCrispin2010}.
We estimate $D_{\rm Mg} \sim 10^{-19} \, {\rm m}^{2} \, {\rm s}^{-1}$ at 1350 K, in which case a 500 nm-sized spinel grain would have transformed to corundum within $\sim 10$ days. 
{\Steve Nanometer-scale Cr-rich spinel grains would transform even faster, by loss of both Mg and Cr.}
%Nanospinels would of course have transformed even faster, by loss of both Mg and Cr. 
Significantly, the transformation of spinel to corundum would not have required gain or loss of Al atoms or $\altwosix$, and indeed very few Al atoms would have existed in the gas phase.
% Ca is often used by metallurgists to remove spinel inclusions from molten steel, where it substitutes for Mg to produce calcium aluminates \citep[e.g.,][]{DengEtal2021}.

Further transformation of corundum, to hibonite or even grossite, would have occurred upon reaction with Ca vapor, plus diffusion of Ca into the corundum grains.
Ceramicists produce hibonite [${\rm CaO} \cdot 6({\rm Al}_{2}{\rm O}_{3})$, or ``CA6"] by reacting corundum grains with Ca delivered via solid-state diffusion from adjacent grains of grossite [${\rm CaO} \cdot 2({\rm Al}_{2}{\rm O}_{3})$, or ``CA2"] \citep[e.g.,][]{PietaEtal2015}.
{\Steve Single hibonite crystals tens of microns in size formed in these experiments, larger than the grain sizes in the powders.}
Hibonite forms by Ca and O diffusing along the crystal surface perpendicular to the $c$ axis, which is why hibonite forms plates \citep{ChenEtal2016}.
Reaction of Ca vapor with corundum is likely to have created at least rims of hibonite on corundum surfaces, {\Steve such as} is observed in many hibonite-rimmed corundum grains \citep[e.g.,][]{NakamuraEtal2007}. 

Whether corundum could have completely transformed to hibonite or grossite depends on whether Ca could have diffused into grain interiors over relevant timescales.
Transformation of a $\sim \!\! 1 \, \mu{\rm m}$-radius grain in, say, 3 years requires 
a diffusion coefficient $D_{\rm Ca} > 10^{-21} \, {\rm m}^{2} \, {\rm s}^{-1}$ or so at $\approx 1350-1425 \, {\rm K}$ in the relevant minerals.
Unfortunately, the values of $D_{\rm Ca}$ in most of these minerals are not well known.
In general, the diffusion of ${\rm Ca}^{2+}$ in oxides is presumed to be slow, because of its larger ionic radius compared to ${\rm Fe}^{2+}$ or ${\rm Mg}^{2+}$.
\citet{NakamuraEtal2007} estimated the diffusion of Ca in corundum to be no more than 10 times as fast as the diffusion of Al, for which they took $D_{\rm Al}(T) = 1.6 \times 10^{-5} \, \exp(-(510 \, {\rm kJ} \, {\rm mol}^{-1})/RT) \, {\rm m}^{2} \, {\rm s}^{-1}$, following \citet{LeGallEtal1994}.
At $T = 1400 \, {\rm K}$, this would yield $D_{\rm Ca} = 1.5 \times 10^{-23} \, {\rm m}^{2} \, {\rm s}^{-1}$, and a corundum grain would take about 1400 years to transform.
However, we would expect diffusion of the divalent cation ${\rm Ca}^{2+}$ to more than an order of magnitude faster than the diffusion of trivalent ${\rm Al}^{3+}$.
%On the other hand, one measurement in corundum, $D_{\rm Ca} = 3.10 \times 10^{-13} \, {\rm m}^{2} \, {\rm s}^{-1}$ at 1670 K \citep{Frischat1971}, would seem to suggest that Ca diffusion could occur in only minutes.
% I can't find that reference!  I think Doremus (2006) cited it, but it might be for Na diffusion in corundum, not Ca!! 
A more direct estimate of $D_{\rm Ca}$ can be made from the experiments by \citet{PietaEtal2015}, in which CA6 formed via diffusion of Ca through corundum grains of diameter $10 \, \mu{\rm m}$, in only 3 hours at $1600^{\circ}{\rm C}$, implying $D_{\rm Ca} \sim 2 \times 10^{-15} \, {\rm m}^{2} \, {\rm s}^{-1}$ at 1873 K.
As a divalent cation, Ca probably has an activation energy similar to O, which would imply $D_{\rm Ca} \sim 1.5 \times 10^{-21} \, {\rm m}^{2} \, {\rm s}^{-1}$ at 1350 K.
At these rates, a $1 \, \mu{\rm m}$-radius corundum grain could transform to hibonite in about 20 years. 
We view it as likely but not definite that a corundum grain would have fully transformed to hibonite during the hundreds of years it spent in the hot midplane region.

While spinel would have transformed rapidly to corundum, and corundum could have transformed to hibonite, it is not clear that these minerals would have transformed to the other thermodynamically favored phases, including gehlenite and grossite.
Production of gehlenite requires inward diffusion of Si. 
The diffusion coefficients of Si into corundum or spinel are not known, but as a tetravalent cation, it could be expected that $D_{\rm Si} \ll D_{\rm O}$ in these minerals, strongly suggesting that gehlenite production might have been kinetically inhibited.
Production of grossite probably could have proceeded by Ca diffusion into hibonite, but diffusion coefficients in hibonite are not known.
However, \citet{RussellEtal1998} noted that Mg diffuses more slowly in hibonite than in other CAI minerals.
% Measurements of diffusion of Ca in garnet yield $D_{\rm Ca} \sim 10^{-21} - 10^{-20} \, {\rm m} \, {\rm s}^{-1}$ \citep{SchwandtEtal1996,VielzeufEtal2006}

Given that hibonite is thermodynamically favored over a temperature range seen in our modeled hot midplane region, {\Steve and easy to produce by reaction with corundum}, we consider corundum and hibonite to be the most likely phases present.
We emphasize that the preponderance of {\Steve large} corundum and hibonite {\Steve crystals} is {\it not} because {\Steve these} condensed in this region from a solar nebula gas that cooled from above 1665 K, and then cooled below 1605 K to allow transformation to hibonite.
Corundum and hibonite were likely the common phases because presolar grains were corundum, spinel, or hibonite, and at the relevant temperatures ($\approx 1350 - 1425 \, {\rm K}$), spinel would have converted to corundum, and corundum would have converted to hibonite, {\Steve within the $\sim 10^3$ residence time of grains in the hot midplane region, and other phases would have vaporized.}

As spinel loses Mg and O atoms and converts to corundum, it loses 28\% of its mass, and its density increases from $3.6 \, {\rm g} \, {\rm cm}^{-3}$ to $4.0 \, {\rm g} \, {\rm cm}^{-3}$.
A 500 nm-sized spinel grain becomes a 430 nm-sized corundum grain. 
A 50 nm-sized nanospinel grain becomes a 43 nm-sized corundum grain.
Conversion of corundum to hibonite by addition of Ca and O atoms will increase its mass by 9\% and decrease its density to $3.8 \, {\rm g} \, {\rm cm}^{-3}$.
{\Steve We assume corundum grains do not change in size or density as they convert to hibonite.}
%A $1 \, \mu{\rm m}$-radius corundum grain becomes a $1.0 \, \mu{\rm m}$-radius hibonite grain.

%{\Ed We conclude that the presolar grains would have transformed during their residence in the hot midplane region. 
%Spinel grains (and nanospinels) must have rapidly converted to corundum by loss of Mg (and Cr).  
%This new corundum, and presolar corundum, are likely to have converted to hibonite, at least partially.
%Significantly, growth of hibonite by diffusion of Ca into corundum can yield the platy lath morphology associated with PLACs, as seen in the experiments of \citet{PietaEtal2015} (their Figure 4), although we discuss this in more detail in \S 4.4.
%Presumably as corundum grains are accreted onto the surface, reaction with Ca converts them to hibonite with the same crystal orientation.
%More difficult but still possible would have been conversion to grossite. 
%Throughout this process, Al atoms were neither gained nor lost, and $< 10^{-6} - 10^{-4}$ of all Al atoms were in the vapor phase. 
%Nanospinel grains, though converted to corundum, would not have vaporized, and their $\altwosix$ would have been retained in small grains. }

%--------------------
%FIGURE 3
%--------------------

\subsection{Presolar Materials}

A foundational aspect of our model is the interpretation that some of the most primitive ``Solar System" solids are not condensates from a solar-composition gas that cooled from temperatures $> 1600$ K, but are instead partially {\Steve or completely} altered presolar grains that were heated from molecular cloud temperatures $\sim 10$ K.
If it is assumed from the start that PLACs and similar objects derived from condensates from a completely vaporized solar-composition gas, then their lack of $\altwosix$ {\it necessarily} would be because the solar nebula lacked $\altwosix$. 
The only way not to implicitly assume the solar nebula lacked $\altwosix$ early on is to consider the possibility that these objects formed from material that retained some memory of their presolar nature. 
In particular, one cannot explain the existence of Al-bearing minerals lacking in $\altwosix$ unless one considers the presolar carriers of Al and $\altwosix$ in the solar nebula.
The hypothesis we present here requires that a significant fraction of Al resided in {\Steve presolar} micron-sized corundum or hibonite grains (or in minerals that converted easily to corundum and hibonite), but that all live $\altwosix$ resided in a small fraction of tiny (tens of nm in size) grains, of similar mineralogy. 

From studies of the least altered chondrites, a wide variety of presolar grains are known to have existed in the solar nebula, as reviewed by \citet{Zinner2014}: diamonds and graphitic carbon; silicon carbide; silicates like olivine and pyroxene; perovskite; aluminates like corundum, hibonite and spinel; ${}^{54}{\rm Cr}$-rich nanospinels; and many others.
Most are present among chondritic matrix grains at 1-100 ppm levels, but that is just the tiny fraction of materials that have not been thermally or chemically processed in the solar nebula or on the parent bodies.
Originally, of course, 100\% of dust {\Steve in the solar system} was presolar.
The exact abundances in the solar nebula of different types of presolar grains depend on their primordial abundances in the ISM, but also on their ability to survive disk processes, parent body processes, and the process of extracting them from their host chondrites.
These efficiencies are difficult to quantify, although attempts are being made \citep{HoppeEtal2017}.

%, although attempts are being made (Hoppe et al. 2017, Nature Ast ronomy:1, 617–620).
%Here we aim to determine the main carrier grains of the Al and $\altwosix$ in the solar nebula.

The vast majority of grains are expected to derive from asymptotic giant branch (AGB) stars, from progenitors with initial masses $0.5$ to $8 \, M_{\odot}$.
Integration of the {\Steve stellar} initial mass function suggests their outflows should produce more dust than from stars $\geq 8 \, M_{\odot}$ in mass, and this seems to be the case. 
The sources of aluminum oxide presolar grains can be determined based on their oxygen isotope ratios \citep{NittlerEtal1997}---at least for those isolated grains found in chondrites, that started with distinctly non-solar isotopic ratios and that never exchanged oxygen isotopes with the solar nebula.
Presolar grains with oxygen isotopes in Groups 1, 2 or 3 probably derive from AGB stars, while grains from Group 4---with ${}^{18}{\rm O}/{}^{17}{\rm O} > 5.2$---probably
%${}^{18}{\rm O}/{}^{16}{\rm O} > 2.4 \times 10^{-3}$ and ${}^{18}{\rm O}/{}^{17}{\rm O} > 3$---probably 
derive from core-collapse supernovae \citep{NittlerEtal2008,NittlerEtal2020}.
At first only $\approx 10\%$ of \citep{NguyenEtal2007,HynesGyngard2009} presolar oxide grains were classified as coming from core-collapse supernovas, but as the spatial resolution of mass spectrometers has improved and allowed detection of smaller grains, and as magnesium isotope ratios have been measured more precisely, mass fractions $\approx 20\%$ have been inferred \citep{HoppeEtal2015,LiuEtal2022}.
Similarly, \citet{ZegaEtal2011} analyzed five presolar hibonite grains, and concluded that four had Group 1 or 2 oxygen isotopic compositions, but one (KH2) was Group 4.
Thus, about 80\% of Solar System Al came from grains condensed in AGB outflows.
The remaining grains probably came from supernova explosions {\Steve (or Wolf-Rayet winds, also from massive stars $> 20 \, M_{\odot}$)} over Galactic history.
%
% although not necessarily the recent ($< 20$ Myr) past, except for {\Steve massive stars} in the Sun's star-forming region.
%}

{\Steve
\subsubsection{Al-bearing Presolar Grains, their Abundances, and their Sizes} }

We first consider the {\Steve presolar carriers of Al and their sizes.} 
As reviewed by \citet{Zinner2014}, most extant presolar Al-bearing oxide grains are corundum [${\rm Al}_{2}{\rm O}_{3}$] or spinel [${\rm Mg}{\rm Al}_{2}{\rm O}_{4}$], with some also being hibonite [${\rm Ca}{\rm Al}_{12}{\rm O}_{19}$] and chromium spinel [${\rm Mg}({\rm Cr,Al})_{2}{\rm O}_{4}$]. 
In the last decade, presolar silicate grains have increasingly been recognized as the most abundant type of presolar grain \citep{FlossHaenecour2016}.
% \citep{reference}.
Presolar silicate grains themselves often contain Al, with the few measured to date showing Al/Mg $\sim 0.01$ to 0.1 \citep{HoppeEtal2021}.  
Compared to the solar ratio Al/Mg $= 0.08$ \citep{Lodders2003}, this suggests that a non-negligible but poorly known fraction of all Al is carried on presolar silicates. 
{\Steve
Taking an average value 0.03 and comparing to the solar ratio Al/Mg $= 0.08$, we estimate that roughly 60\% of Al in presolar grains resides in aluminum oxides, and 40\% resides in silicates.
We assume similar proportions among grains from AGB stars and massive stars.
}
%{\Ed
%In the context of our model, these silicates will vaporize and release Al into the gas once they enter the hot midplane region (which will likely condense into corundum or onto other grains); but this is unimportant as long as this Al is free of $\altwosix$, and is not the major source of Al.
%We assume the fraction of Al from presolar silicates is low, no more than tens of percent, and that the majority of Al in the Solar System was brought in by presolar aluminum oxides.
%}

Among presolar aluminum oxide grains, roughly half appear to be spinel, and the other half corundum and hibonite, which are not easily distinguished \citep{NguyenEtal2007,NguyenEtal2010,VollmerEtal2009,FlossStadermann2009,FlossStadermann2012,BoseEtal2010,BoseEtal2012,HaenecourEtal2018,NittlerEtal2020lpsc,NittlerEtal2008,ZegaEtal2011}.
The proportion that are hibonite is not well known, but appears to be small: they make up one third of Al-oxide presolar grains in the LL3.2 chondrite Krymka, but a much smaller fraction in other unequilibrated ordinary chondrites (UOCs), and are absent in the H/L3.6 chondrite Tieschitz \citep{NittlerEtal2008}.
Because corundum readily transforms to hibonite in the hot midplane region (\S 3.4), and the two minerals have similar sizes and properties, we do not distinguish between them, and refer to `corundum' to mean a mix of {\Steve mostly} presolar corundum and some hibonite. 
{\Steve Again we assume the proportions of spinel and corundum/hibonite are similar whether grains derive from AGB stars or supernovae/Wolf-Rayet stars.}

The typical sizes of presolar corundum grains are a few microns \citep{MakideEtal2011}, and the typical sizes of presolar hibonite grains are several hundred nm to several microns \citep{NittlerEtal2008,ZegaEtal2011}.
The typical sizes of presolar spinel grains are $\approx 0.5 \, \mu{\rm m}$ \citep[e.g.,][]{ZinnerEtal2005}.
{\Steve These sizes are characteristic of the $\approx 80\%$ of grains that derive from AGB stars, but dust grains from the much faster outflows of supernovae / Wolf-Rayet stars will be smaller.
Theoretical calculations \citep[e.g.,][]{NozawaEtal2003} predict that corundum grains in supernova ejects should grow no larger than about 10 nm in radius, and even the largest silicate grains will not exceed 100 nm in radius ($0.2 \, \mu{\rm m}$ in size). 
Crystal domains also tend to be smaller (10-460 nm) in supernova SiC grains than the $>500$ nm domain sizes in AGB SiC grains \citep{StroudEtal2004,Hynes2010thesis}, strongly indicating faster cooling and condensation in supernova ejecta \citep{Hynes2010thesis}.
From meteoritic studies, higher proportions of Group 4 (supernova) presolar oxides are found among grain size fractions below a few hundred nm \citep{NguyenEtal2007}.

``Nanospinels" are presolar grains composed of Cr-rich spinel [$({\rm Mg},{\rm Fe})({\rm Al},{\rm Cr})_{2}{\rm O}_{4}$] that are $< 200$ nm in size \citep{LiuEtal2022}.
They are remarkable for having ${}^{54}{\rm Cr}/{}^{52}{\rm Cr}$ ratios $3.6 \times$ solar ($\epsilon^{54}{\rm Cr} = +26,000$), and may be the sole carriers of the $\epsilon^{54}{\rm Cr}$ positive anomalies in the Solar System.
Their isotopic anomalies are consistent with formation in a core-collapse type II supernova, probably in its O/Ne or O/C zones \citep{DauphasEtal2010,NittlerEtal2018}.
\citet{DauphasEtal2010} suggested they were typically $< 100$ nm in size but also noted they are concentrated into the colloidal fractions of Murchison and Orgueil separates, indicating sizes $< 30-34$ nm.
If the mass of nanospinels correlates with the $\epsilon^{54}{\rm Cr}$ anomaly, the typical sizes of nanospinels would be in these smallest particles, i.e., $< 50$ nm.}

{\Steve
From the considerations above, including comparing the sizes of these nanospinels with most spinels, formed in AGB outflows, we make the approximation that presolar grains formed in supernova ejecta or Wolf-Rayet winds are a factor of 10 smaller than the similar minerals formed in AGB outflows.}

{\Steve
Once in the hot midplane region, presolar silicate and SiC grains will vaporize, and the 75\% or so of Al atoms in them will enter the gas phase.
These atoms will recondense as corundum or other grains, as described above (\S 3.2). 
The sizes of these condensates will be likely be small; in other contexts, condensation nuclei can be as small as tens of nm \citep{NozawaEtal2003}.
We will assume these condensates have typical radii $a_{\rm c} \sim 22$ nm, like the smallest presolar grains. 
Together these grains are a major reservoir of Al and dominate the surface area of solids in the LAACHI-forming region.}
\vspace{0.1in}

\subsubsection{Live ${}^{26}{\rm Al}$ in Presolar Grains}

Almost none of the grains  carrying Al {\Steve discussed above} should have carried live $\altwosix$.
Because the maximum plausible ratio in a grain formed in a stellar outflow or explosion is $\alratio \sim 1$, decay of just 20 Myr is sufficient to reduce the ratio to $\alratio < 10^{-8}$ within the grain. 
The majority ($\approx 80\%$) of presolar corundum and spinel grains come from AGB stars, which are not at all associated with star-forming regions.
The probability of a molecular cloud being contaminated by a passing AGB star is $\sim 10^{-6} - 10^{-9}$ \citep{KastnerMyers1994,OuelletteEtal2010}, and it can be safely assumed this did not occur.
Those presolar grains in the Sun's molecular cloud from AGB stars therefore would have been from much earlier generations of star formation, after having spent $10^8 - 10^9$ years in the ISM \citep{JonesEtal1996,HirashitaEtal2016,HeckEtal2020}.
AGB grains should contain essentially zero $\altwosix$.
Similar arguments would apply to grains from novas, {\Steve or any stellar source taking more than tens of Myr to evolve}. 

Live $\altwosix$ would be carried exclusively by grains from {\Steve massive stars}.
Unlike other stellar sources, core-collapse supernovae (and Wolf-Rayet winds from their progenitors) {\it are} associated with star-forming regions, and some may have promptly delivered grains to the Sun's molecular cloud \citep{HesterDesch2005,PanEtal2012,DeschEtal2023c}.
Only a small fraction of supernova grains would have been of recent origin, though: like grains from AGB stars, most supernova grains would have been produced over Galactic history, with most recording random ages also $\sim 10^8 - 10^9$ years.
{\Steve The fraction of grains formed in supernovae or Wolf-Rayet stars that contained live $\altwosix$ was probably $\approx (20 \, {\rm Myr}) / (10^9 \, {\rm yr}) \sim 2\%$.}

%It is not known exactly what initial $\alratio$ ratio
%nanospinels might form with in supernova ejecta, but it must be far higher than the canonical Solar System value.
%Simply from a mass balance perspective, if $< 0.4\%$ of Al in the solar nebula was associated with live $\altwosix$, then the component with $\altwosix$ must have had a ratio $> 250 \times$ the canonical value, or $\alratio > 10^{-2}$.

{\Steve 
The initial  ratios of those grains from recent, nearby massive stars was probably $(\alratio)_0 \sim 10^{-2}$.}
\citet{Zinner2014} compiled initial $(\alratio)_0$ ratios for Group 4 oxide grains made of corundum, hibonite or spinel (Figure 15 of that paper).
Corundum grains formed with $(\alratio)_0 \approx 3 \times 10^{-3}$ to $7 \times 10^{-2}$, a hibonite grain formed with $(\alratio)_0 \approx 1 \times 10^{-2}$, and spinel grains formed with $(\alratio)_0 \approx 3 \times 10^{-3}$ to $5 \times 10^{-2}$.  
There is also the example of an extremely large ($25 \, \mu{\rm m}$) presolar SiC grain {\it Bonanza} of supernova origin, with Al-bearing phases within it that apparently record $(\alratio)_0 \sim 0.9$ \citep{GyngardEtal2018}.
Our model does not depend strongly on the exact value, 
{\Steve so based on the above literature, we adopt $\alratio \approx 0.0125$ ($250 \times$ canonical) in the $\altwosix$-bearing presolar grains in the solar nebula.}
This could reflect the value in the supernova ejecta, plus immediate delivery to the Sun's molecular cloud right before its collapse; or this could represent an initial $\alratio \approx 0.9$ in the ejecta of supernovae about 5 Myr before Solar System formation.

{\Steve
\subsubsection{Summary}
}

{\Steve
In Table~\ref{table:grainsbefore} we list the types of Al-bearing presolar grains that would exist in the solar nebula overall,  their stellar sources, the percentage of Al atoms in each phase, their typical sizes (diameters), and the typical initial $(\alratio)_0$ values.
It is assumed that 80\% of Al atoms are in the large grains from AGB stars, and 20\% from supernovae and Wolf-Rayet stars.
It is assumed that 40\% of Al atoms are in silicates / SiC grains, and 60\% in aluminum oxides.
of aluminum oxides, it is assumed that equal amounts are in corundum / hibonite grains and in spinel grains.
In Table~\ref{table:grainsafter} we list the Al-bearing grains that would exist in the LAACHI-forming region, their sources, the percentage of Al atoms in each phase, their typical sizes, their midplane densities (scaled to gas density $5.5 \times 10^{-9} \, {\rm g} \, {\rm cm}^{-3}$), and their typical initial $(\alratio)_0$ values.
It is assumed that the non-refractory grains vaporize in the LAACHI-forming region and recondense as small particles; for simplicity we assume they have the same sizes as the smallest grains, the nanospinels.
We assume spinels grains are transformed to corundum by loss of Mg with some reduction in volume.
}

%----------
% TABLE 2. Grain types
\begin{deluxetable}{l|c|c|c|l}
% \rotate
% \renewcommand{\thefootnote}{\alph{footnote}}
\label{table:grainsbefore}
%\tablenum{2}
\tablecaption{
Al-bearing presolar grains in the solar nebula.}
\tablewidth{0pt}
\tablehead{
\colhead{Mineral} & 
\colhead{Source} &
\colhead{\% of all Al} &
\colhead{Size} & 
\colhead{${}^{26}{\rm Al}/{}^{27}{\rm Al}$}
}
\startdata
Silicate, SiC, etc. & AGB$^a$ & 32\% & microns? & $\approx 0$ \\
{Cor / Hib} & AGB & 24\% & $\sim 2 \, \mu{\rm m}$ & $\approx 0$ \\ 
{Spinel}   & AGB & 24\% & $\sim 0.5 \, \mu{\rm m}$ & $\approx 0$ \\
\hline
Silicate, SiC, etc. & SN/WR$^b$ & 8\% & $\sim 0.1 \, \mu{\rm m}$? & 98\% $\,\,\,$ $\approx 0$ \\
 & & & & \; 2\% $\approx 0.0125$ \\
{Cor / Hib} & SN/WR & 6\% & $\sim 0.2 \, \mu{\rm m}$ & 98\% $\,\,\,$ $\approx 0$ \\
 & & & & \; 2\% $\sim 0.0125$ \\
{Spinel} & SN/WR & 6\% & $\approx 50$ nm & 98\% $\,\,\,$ $\approx 0$ \\
 & & & &  \; 2\% $\sim 0.0125$ \\
\enddata
\tablecomments{
$a$. Asymptotic Giant Branch star. $b$ Core-collapse supernova or Wolf-Rayet star.
}
\end{deluxetable}
%---------------------------------

%----------
% TABLE 3. Grain types
\begin{deluxetable}{l|c|c|c|c|l}
% \rotate
% \renewcommand{\thefootnote}{\alph{footnote}}
\label{table:grainsafter}
%\tablenum{2}
\tablecaption{
Al-bearing presolar grains in the LAACHI-forming region.}
\tablewidth{0pt}
\tablehead{
\colhead{Mineral} & 
\colhead{Source} &
\colhead{\% of all Al} &
\colhead{Size} & 
\colhead{$n ({\rm cm}^{-3})$ $^a$} &
\colhead{${}^{26}{\rm Al}/{}^{27}{\rm Al}$}
}
\startdata
Cor / Hib & Condensates & 40\% & 44 nm & 1540 & $5 \times 10^{-5}$ \\
\hline
Cor / Hib & AGB$^b$     & 24\% & $2 \, \mu{\rm m}$ & 0.0098 & $\approx 0$ \\
Cor / Hib & AGB         & 24\% & $0.44 \, \mu{\rm m}$ & 0.92 & $\approx 0$ \\
\hline
Cor / Hib & SN/WR$^c$  & 6\% & $0.2 \, \mu{\rm m}$ & 2.5 & 98\% $\approx 0$ \\
 & & & & & \; 2\% $\approx 0.0125$ \\
 Cor / Hib & SN/WR & 6\% & 44 nm & 231 &  98\% $\approx 0$ \\
  & & & & & \; 2\% $\approx 0.0125$ \\
\enddata
\tablecomments{
$a$ Particle density, assuming $\rho_{\rm g} = 5.5 \times 10^{-9} \, {\rm g} \, {\rm cm}^{-3}$. $b$. Asymptotic Giant Branch star. $c$ Core-collapse supernova or Wolf-Rayet star.
}
\end{deluxetable}

\subsection{Diffusive Exchange of Oxygen Atoms}

One potential objection to the idea that {\Steve LAACHIs}
%PLACs and low-$\alratio$ inclusions 
derive from presolar grains is that they have oxygen isotopic ratios matching solar nebula reservoirs (\S 2).
For our hypothesis to be viable, presolar grains must have exchanged most of their oxygen atoms with the gas, presumably while in the hot midplane region.
Given the typical excursions of oxygen isotopic ratios in presolar grains ($\sim 10^3 \, \permil$) and the precision of such measurements ($\sim 1\permil$), roughly 99.9\% of all oxygen atoms in a grain must be exchanged.

The fraction of atoms exchanged by a spherical grain of radius $a$ in a time $t$ is determined by the ratio $t / t_{\rm diff}$, where $t_{\rm diff} = a^2 / D(T)$, and $D(T)$ is the temperature-dependent diffusion coefficient.
Direct numerical integration of the diffusion equation {\Steve (or comparison with Equation 6.20 of \citet{Crank1979})} shows that 99.9\% of atoms will exchange after a time $t > 0.69 \, t_{\rm diff}$. 
At a given temperature $T$, we assume a diffusion coefficient governed by an Arrhenius relationship: $D(T) = D_0 \, \exp(-Q/RT)$.
For lattice diffusion of O atoms through spinel, we adopt $D_0 = 2.2 \times 10^{-7} \, {\rm m}^{2} \, {\rm s}^{-1}$ and $Q = 404 \, {\rm kJ} \, {\rm mol}^{-1}$ \citep{RyersonMcKeegan1994}.
For the gas constant $R = 8.3145 \, {\rm kJ} \, {\rm mol}^{-1}$, $Q/R = 48,600$ K. 
Diffusion of O atoms through corundum has been reviewed by \citet{Heuer2008} and \citet{Doremus2006}.
We adopt the value of \citet{Reddy1979thesis}, 
as conveyed by \citet{Heuer2008}: 
$D_0 = 2.5 \, {\rm m}^{2} \, {\rm s}^{-1}$ and {\Steve $Q = 625 \, {\rm kJ} \, {\rm mol}^{-1}$, and $Q/R = 75,200$ K.} 
To our knowledge, oxygen diffusion data do not exist for hibonites or other Al-bearing phases.

While small, these diffusion rates are high enough to allow {\Steve nearly} complete exchange of oxygen atoms with the gas in the hot midplane region.
For $a = 0.25 \, \mu{\rm m}$-radius spinel grains at 1350 K, $D = 5.1 \times 10^{-23} \, {\rm m}^{2} \, {\rm s}^{-1}$,  $t_{\rm diff} = 39$ years, and exchange would have been 99.9\% complete after 27 years.
Nanospinel grains $\sim 25$ nm in radius would have equilibrated much faster. 
{\Steve
Likewise, for corundum grains at 1350 K, $D = 1.6 \times 10^{-24} \, {\rm m}^{2} \, {\rm s}^{-1}$, and at 1425 K, $D = 3.1 \times 10^{-23} \, {\rm m}^2 \, {\rm s}^{-1}$. 
For grains with radius $a = 1 \, \mu{\rm m}$, the values of $t_{\rm diff}$ are $2.0 \times 10^4$ yr and $1.0 \times 10^3$ yr, respectively. 
Isotopic exchange at these temperatures would be 99.9\% complete in 13,000 yr to 710 yr, and would be 90\% complete in 5900 yr to 200 yr. 
Almost all grains at 0.6 AU would have spent hundreds of years near the midplane before diffusing out of the region or being accreted, and therefore would be largely ($> 90\%$) equilibrated with the nebular gas in their oxygen isotopes} with the ${\rm H}_{2}{\rm O}$, CO, and silicate vapor in the gas.
These would have the oxygen isotopic signature near either that of the Sun, at  $\Delta^{17}{\rm O} \approx -29\permil$ \citep{McKeeganEtal2011}; or combined H$_2$O vapor and CO gas, at $\Delta^{17}{\rm O} \approx -35\permil$ \citep{KrotEtal2010}; 
or the reservoir sampled by the majority of CAIs, $\Delta^{17}{\rm O} \approx -23\permil$ \citep{KoopEtal2020}.

\subsection{Charging of Particles}

By virtue of the fact that the region where PLACs and other LAACHIs formed must have been very hot ($T > 1350$ K), they inevitably would have been charged, as quantified by \citet{DeschTurner2015}.
First, all potassium atoms {\Steve (the most easily ionized abundant atom)} would have been in the gas phase, and several percent of them ionized, {\Steve close to as predicted by} the Saha equation, generating free electrons.
Dust grains would have been very negatively charged by the gas, due to adsorption of free electrons on their surfaces, balanced by thermionic emission of electrons.
Thermionic emission is a process whose rate depends on the effective work function $W_{\rm eff}$ of the solid, as $\exp(-W_{\rm eff}/kT)$, where $W_{\rm eff} = W + Z e^2 / a$, $W$ is the work function of the material, and $Z e$ the charge of the grain.
The work function is the energy required to remove an electron from the solid, and larger work functions mean fewer electrons are ejected.
The work function is formally defined for conductors, but thermionic emission from insulators is also governed by a work function-like term, via the Richardson equation.
{\Steve 
Because of this dependence, grains made of materials with high work functions will be negatively charged, and those with low work functions will be positively charged.}

\citet{DeschTurner2015} investigated cases nearly identical to the situation considered here: a single population of dust grains with radius $a = 1 \, \mu{\rm m}$, in a gas of density $n_{\rm H2} = 1.0 \times 10^{14} \, {\rm cm}^{-3}$ and temperatures up to 1500 K.
The solids-to-gas ratio was taken to be a canonical $0.01$, but a ratio $10^{-4}$ was also considered (Figure 11 of Desch and Turner 2015), and found not to affect grain charge or electron density above a temperature of 1100 K.
At high densities, gas-phase electrons come mostly from thermionic emission, and are lost mostly to adsorption to grain surfaces, both effects being proportional to the dust surface area, which therefore is not something grain charges are sensitive to.
%There is a tendency for dust grains to charge until their effective work functions somewhat match the ionization potential of potassium, $4.34 \, {\rm eV}$.
%For example, f
%From Figure 2 of \citet{DeschTurner2015}, if the grains have work function $W = 5.0 \, {\rm eV}$, then they have charge $Z e = -360 e$ at 1500 K, i.e., are charged to a potential of $-0.52$ Volts, so that the effective work function is only $W_{\rm eff} \approx 4.48 \, {\rm eV}$.
% And if the grains have work function $W = 4.5 \, {\rm eV}$, they have charge $Z e = -150 e$, i.e., are charged to a potential of $-0.22$ Volts, so that the effective work function is only $W_{\rm eff} \approx 4.28 \, {\rm eV}$.
%Across the range of plausible work functions $W \approx 2$ to $6 \, {\rm eV}$, the effective work function remains in the range $W_{\rm eff} \approx 3.6$ to $4.9 \, {\rm eV}$.
{\Steve
In practice, \citet{DeschTurner2015} found that  grains made of materials with work functions $< 4$ eV will be positively charged, and those with work functions $> 4$ eV will be negatively charged.}

The LAACHI-forming region considered here differs from the case considered by \citet{DeschTurner2015} in that there are multiple populations of corundum/hibonite grains {\Steve with different radii}.
% : about half the Al in the form of presolar corundum grains, with radius $a = 1 \, \mu{\rm m}$; half the Al in previously spinel grains, with radius $a = 0.25 \, \mu{\rm m}$; and a small fraction of the mass in previously nanospinel grains, with radius $a = 0.025 \, \mu{\rm m}$.
% Corundum grains may have converted to hibonite, and presolar hibonite would have been a non-negligible component as well.
It is straightforward to show that when exposed to the same hot gas and electron density, each population will charge to the same voltage, and therefore the same $W_{\rm eff}$. 
In practice this means the charge on each grain is proportional to its radius, $a$.

Calculation of their exact charges depends on the work functions of corundum and hibonite. 
As discussed by \citet{DeschTurner2015}, a wide variety of astrophysically relevant materials have work functions near 5 eV, including quartz [${\rm SiO}_{2}$] and many other oxides, with $W \approx 5.0 \, {\rm eV}$ \citep{Fomenko1966}, and lunar regolith simulant (mostly silicates), also near $W = 5.0 \, {\rm eV}$ \citep{FeuerbacherEtal1972lpsc}.
%As a general rule, metal oxides tend to have work functions about 0.5 eV greater than the metal, suggesting that since the work function of pure Al is 4.20 eV, that corundum should be near 4.7 eV \citep{Semov1968}.
% I can't find this reference anywhere! 
Corundum's work function is estimated as 4.7 eV \citep{Fomenko1966}.
%, but Density Functional Theory modeling of ${\rm Al}_{2}{\rm O}_{3}$ films overlying Al suggests $W \approx 5.1 \, {\rm eV}$ \citep{CornetteEtal2020}.
Data do not exist for the {\Steve work functions of hibonite and spinel}. 
We therefore adopt $W = 4.7 \, {\rm eV}$ for {\Steve all these minerals.
As long as their work functions are at least 4.5 eV or so, the grains are negatively charged and repel each other, and none of our results are modified.}

At typical temperatures in the hot midplane region of $T = 1350 \, {\rm K}$ and $1425 \, {\rm K}$,  
$a = 1 \, \mu{\rm m}$ grains would have been charged to $-280e$ and $-240e$, and voltages of about -0.35 V \citep{DeschTurner2015}.
The charges on the $0.25 \, \mu{\rm m}$ radius grains at these temperatures would be $-70e$ and $-60e$.
The charges on the $0.025 \, \mu{\rm m}$ former nanospinels would have been $-7e$, and $-6e$. 
The results are summarized in {\bf Table~\ref{table:charges}}.

%----------
% TABLE 2. Charges
\begin{deluxetable}{c|c|c}
% \rotate
% \renewcommand{\thefootnote}{\alph{footnote}}
\label{table:charges}
%\tablenum{2}
\tablecaption{Work functions and charges of refractory grains possibly pertinent to the LAACHI-forming region.}
\tablewidth{0pt}
\tablehead{
\colhead{Material} & 
\colhead{Work Function$^{a}$} &
\colhead{Charge$^{b}$} 
}

\startdata
Perovskite (${\rm CaTiO}_{3}$) & 
3.0 eV & $+400e$ \\
Titanates (e.g., ${\rm SrTiO}_{3}$) & 3.1 eV & $+360e$ \\ 
Hafnia (${\rm HfO}_{2}$) & $\sim$ 3.6 eV & $+160e$ \\
Zirconia (${\rm ZrO}_{2}$) & $\sim$ 4.0 eV & $0e$ \\
Refractory Metal (Mo,Ru) & $\sim 4.5$ eV & $-200e$ \\
Silicon Carbide (SiC) & 4.5 eV & $-200e$ \\
Graphite (C) & 4.7 eV & $-240e$ \\ 
Corundum (${\rm Al}_{2}{\rm O}_{3}$) & 
4.7 eV & 
$-240e$ \\
Hibonite (${\rm CaAl}_{12}{\rm O}_{19}$) & $\sim 4.7$ eV & 
$-240e$ \\
\enddata
\tablecomments{
$a$. See text for details and references. $b$. For $T = 1425$ K and radius $a = 1 \, \mu{\rm m}$. For grains with different radii, charge is multiplied by $(a / 1 \, \mu{\rm m})$.
}
\end{deluxetable}
%---------------------------------

\subsection{Coagulation and Growth} 

In the hot, LAACHI-forming environment, grains would have collided and stuck together, forming larger aggregates.
For practical purposes, all the grains were corundum or hibonite (with presumed identical properties), 
{\Steve 
including former spinel grains, with radii as listed in Table 2: 
$1.0 \, \mu{\rm m}$, 
$0.25 \, \mu{\rm m}$, 
$0.1 \, \mu{\rm m}$, 
$0.025 \, \mu{\rm m}$, or 
$a_{\rm c} \sim 10$ nm.}
A `large' corundum grain of radius $a_1$ would sweep up other grains of radius $a_2$ at a rate $n_2 \, \pi (a_1 + a_2)^2 \, V_{12} \, S_{12}$, where $n_2$ is the number density of other grains, $V_{12}$ the relative velocity between the large grain and the other grains, and $S_{12}$ a factor to account for sticking efficiency and electrical effects on the cross sections.
Here we quantify these terms,
{\Steve using the densities and properties listed in Table 2.}
The densities are estimated assuming a midplane density $\rho = 5.5 \times 10^{-9} \, {\rm g} \, {\rm cm}^{-3}$, so that $n_{\rm H2} = \rho / (1.4 m_{\rm H2}) = 1.2 \times 10^{15} \, {\rm cm}^{-3}$, and the abundance ratios of \citet{Lodders2003}, yielding a density of Al atoms $8.1 \times 10^{9} \, {\rm cm}^{-3}$.
% , with half in $0.25 \, \mu{\rm m}$ radius grains, and half in $0.025 \, \mu{\rm m}$ grains. 
% Taking the density of corundum to be $\rho_{\rm cor} = 4.0 \, {\rm g} \, {\rm cm}^{-3}$, and noting there are 51.0 amu per Al atom in ${\rm Al}_{2}{\rm O}_{3}$, we find $n_2 = 2.1 \times 10^{-2} \, {\rm cm}^{-3}$ for $a = 1 \, \mu{\rm m}$ grains, and $n_2 = 1.3 \, {\rm cm}^{-3}$ for $a = 0.25 \, \mu{\rm m}$ grains. 
% One could estimate the density of former nanospinels by noting that $\altwosix$ atoms in this region must be $5.23 \times 10^{-5}$ times the number density of Al atoms, or $2.1 \times 10^{5} \, {\rm cm}^{-3}$. 
% A nanospinel with radius 22 nm and $\alratio = 0.05$ has $1.0 \times 10^{5}$ $\altwosix$ atoms, and so the number density of nanospinel grains was about $0.14 \, {\rm cm}^{-3}$.
These densities of course would scale linearly with the gas density $\rho_{\rm g}$. 

%{\Steve For the dust surviving in the LAACHI-forming region, the
%opacity $\kappa$ at $\lambda$ $=$ $2 \, \mu{\rm m}$ (the Wien peak at 1500 K) associated with this assemblage is 
%\[
%\kappa \approx \frac{0.13 \, \pi (1 \, \mu{\rm m})^2 }{ 5.5 \times 10^{-9} \, {\rm g} \, {\rm cm}^{-3} } \, \left[ 
%(1540 \, {\rm cm}^{-3}) \, (0.022)^2 \, (0.069)
%+(0.0098 \, {\rm cm}^{-3}) \, (1.0)^2 \, (1) 
%\right.
% \]
% \[
% \left. 
% +( 0.92 \, {\rm cm}^{-3}) \, (0.22)^2 \, (0.69) 
% +( 2.5 \, {\rm cm}^{-3}) \, (0.1)^2 \, (0.31)
% +( 231 \, {\rm cm}^{-3}) \, (0.022)^2 \, (0.069)
% \right].
%  \approx 0.080 \, {\rm cm}^{2} \, %{\rm g}^{-1},
%\] 
%where the factor of 0.13 accounts for the fact that ${\rm Al}_{2}{\rm O}_{3}$ (as sapphire) is about 87\% transparent in the near infrared}
%(e.g., \url{https://www.photonics.com/Articles/Transparent \_Ceramics\_Enabling\_Large\_Durable/a57166}).
{\EDIT
For dust surviving in the LAACHI-forming region, the opacity $\kappa$ at $\lambda = 2 \, \mu{\rm m}$ (the Wien peak at 1500 K) associated with this assemblage is 
\[
\kappa \approx \left( \frac{\rho_{\rm d}}{\rho_{\rm g}} \right) \, \frac{3}{4 \rho_{\rm cor}} \, \frac{Q_{\rm ext}}{a} \approx 0.1 \, {\rm cm}^{2} \, {\rm g}^{-1},
\]
where the dust-to-gas ratio $\rho_{\rm d} / \rho_{\rm g} \approx 1.26 \times 10^{-4}$ if all the Al is in corundum grains, assuming solar abundances \citep{Lodders2003}, and $Q_{\rm ext} / a \approx 4000 \, {\rm cm}^{-1}$ for alumina particles formed by combustion \citep{KoikeEtal1995}.}
%{\Steve and the other factors are $2\pi a/\lambda$.}
This opacity is consistent with the vertical thermal structure described in \S 3.1.

The relative velocities between two grains 1 and 2 are on the order of tens of cm/s, and obey the relation
\begin{equation}
V_{12}^2 = V_{\rm B,1}^2 + V_{\rm B,2}^{2} + V_{\rm T,12}^2,
\end{equation}
where $V_{\rm B} = (k T / m)^{1/2}$ is the Brownian velocity of grains with mass $m$, and $V_{\rm T}$ is a velocity difference due to turbulence
\citep{XiangEtal2020}.
The velocities of former nanospinel grains are dominated by Brownian velocities $V_{\rm B} \approx 27 \, \left( T / 1400 \, {\rm K} \right)^{1/2} \, \left( a / 0.025 \, \mu{\rm m} \right)^{-3/2} \, {\rm cm} \, {\rm s}^{-1}$, but Brownian motion is negligible for larger particles. 
The velocity difference between two particles 1 and 2 ($a_1 > a_2$) is given by \citet{OrmelCuzzi2007} as
\begin{equation}
V_{\rm T,12} = V_{\rm g} \, {\rm Re}^{1/4} \, \left( {\rm St}_{1} - {\rm St}_{2} \right)
\end{equation}
in the limit that ${\rm St}_{1} < {\rm Re}^{-1/2}$, where the Stokes number of particle 1 is  
${\rm St}_{1} = \Omega \, t_{\rm stop,1}$ and its stopping time is $t_{\rm stop,1} = \rho_{\rm cor} a_{1} / (\rho_{\rm g} C)$, where $C = (k T / \bar{m})^{1/2}$. 
The velocity of the largest-scale eddies is $V_{\rm g} = (3\alpha/2)^{1/2} C$, which would have been $8.7 \times 10^3 \, {\rm cm} \, {\rm s}^{-1}$ in our model. 
The Reynolds number is ${\rm Re} = \alpha C^2 \Omega^{-1} / \nu_{\rm m}$, where $\nu_{\rm m} \approx 4 \times 10^{3} \, {\rm cm}^{2} \, {\rm s}^{-1}$ is the molecular viscosity of an ${\rm H}_{2}$-He gas at 1350-1425 K \citep[see][]{Desch2007}, yielding ${\rm Re} = 4-7 \times 10^{10}$.
For conditions at the midplane, this formula for the relative velocity is obeyed by particles up to about 100 microns in size, and the relative velocity tends to approach $0.38 \, (a_1 - a_2) \, {\rm cm} \, {\rm s}^{-1}$ if $a_1$ and $a_2$ are measured in microns.
For larger particles, with ${\rm St} > {\rm Re}^{-1/2}$ (and $a_{1} \gg a_{2}$), the relative velocity is 
\begin{equation}
V_{\rm T,12} = 1.73 \, V_{\rm g} \, {\rm St}_{1}^{1/2}.
\end{equation}
\citep{OrmelCuzzi2007}.
Typical relative velocities would have been tens to hundreds of cm/s, and would have scaled with density $\rho$ and turbulence level $\alpha$ as $\alpha^{3/4} \, \rho^{-3/4}$.

The efficiency of accretion must also be multiplied by a factor $S_{12} = (1 - {\rm PE}/{\rm KE})$, where KE is the kinetic energy and PE the electrical potential energy, to account for the electrical repulsion or attraction between two particles \citep{XiangEtal2020}.
This term is exactly analogous to the gravitational focusing factor $[1 + V_{\rm esc}^2 / V_{\rm rel}^{2}]$ that increases the collision cross section for two masses; except that unlike gravity, for which PE is always negative and the term $1 - {\rm PE}/{\rm KE} > 1$, enhancing the cross section, here PE is negative (enhancing the cross section) only for oppositely charged particles, and PE is positive (reducing the cross section) for similarly charged particles.
It is understood that this term vanishes if ${\rm KE} < {\rm PE}$, i.e., if particles lack sufficient kinetic energy to overcome their mutual repulsion.
As in the case of gravitational focusing, we can write this term as $(1 - V_{\rm crit}^{2} / V_{12}^2)$, where $V_{\rm crit}$ plays the role of escape velocity.
For particles of identical corundum composition, 
\begin{equation}
V_{\rm crit} = \left( \frac{3 Z_1 Z_2 \, e^2 }{2\pi \rho_{\rm cor}} \right)^{1/2} \, \left( \frac{ a_1^3 + a_2^3 }{ a_1 + a_2 } \right)^{1/2} \, \frac{1}{a_1^{3/2} a_2^{3/2}}, 
\end{equation} 
or 
$V_{\rm crit} = 8.3 \, $
$\left[ (a_1^3 +a_2^3)/(a_1+a_2) \right]^{1/2} / (a_1 a_2) \, \times$ 
$\left[ \left| \bar{Z} \right| / 500 \right] \,$ 
${\rm cm} \, {\rm s}^{-1}$, where $a_1$ and $a_2$ are measured in microns, and $\bar{Z} e$ is the charge on a $1 \, \mu{\rm m}$-radius grain.
If $a_1 \gg a_2$, then this expression reduces to 
$V_{\rm crit} = (8.3 / a_2) \, (\bar{Z} / 500) \, {\rm cm} \, {\rm s}^{-1}$, independent of the size of the larger particle.
Large particles must have velocities exceeding tens of cm/s in order to sweep up $\altwosix$-bearing nanospinels.
%\Steve 
{It is because this critical relative velocity is comparable to the expected relative velocities that accretion may be inhibited by electric charge.} 

%---------------------------------
% TABLE 3. Velocities
\begin{deluxetable}{c|c|c|c}
% \rotate
% \renewcommand{\thefootnote}{\alph{footnote}}
\label{table:velocitiesupper}
%\tablenum{3}
\tablecaption{Expected ($V_{12}$) and critical ($V_{\rm crit}$) relative velocities (cm/s) of pairs of corundum particles, at $z = 1.18 H$, where $T = 1350$ K. Bold type denotes cases where $V_{12} > V_{\rm crit}$ and grains can accrete.}
\tablewidth{0pt}
\tablehead{
\colhead{$\,$} & 
\colhead{$a_2 = 1 \, \mu{\rm m}$} & 
\colhead{$a_2 = 0.22 \, \mu{\rm m}$} & 
\colhead{$a_2 = 22 \, {\rm nm}$} 
}
\startdata
$a_1 = 1000 \, \mu{\rm m}$ & 
 $V_{12} = {\bf 193} $ &
 $V_{12} = {\bf 193} $ &
 $V_{12} = 195 $ \\
$\,$ & 
 $V_{\rm crit} =  {\bf 4.7} $ &
 $V_{\rm crit} = {\bf 21.1} $ &
 $V_{\rm crit} = 211  $ \\
  \hline
  $a_1 = 250 \, \mu{\rm m}$ & 
 $V_{12} = {\bf 96.3} $ &
 $V_{12} = {\bf 96.3} $ &
 $V_{12} = 102 $ \\
$\,$ & 
 $V_{\rm crit} =  {\bf 4.6} $ &
 $V_{\rm crit} = {\bf 21.1} $ &
 $V_{\rm crit} = 211  $ \\
  \hline
$a_1 = 100 \, \mu{\rm m}$ & 
 $V_{12} = {\bf 60.9} $ &
 $V_{12} = {\bf 60.9} $ &
 $V_{12} = 68.9 $ \\
$\,$ & 
 $V_{\rm crit} =  {\bf 4.6} $ &
 $V_{\rm crit} = {\bf 21.1} $ &
 $V_{\rm crit} = 211  $ \\
  \hline
$a_1 = 25 \, \mu{\rm m}$ & 
 $V_{12} = {\bf 15.3} $ &
 $V_{12} = 15.8 $ &
 $V_{12} = 36.0 $ \\
$\,$ & 
 $V_{\rm crit} =  {\bf 4.6} $ &
 $V_{\rm crit} = 21.0 $ &
 $V_{\rm crit} = {211}  $ \\
 \hline
$a_1 = 10 \, \mu{\rm m}$ & 
 $V_{12} = {\bf 5.7} $ &
 $V_{12} = 6.3 $ &
 $V_{12} = 32.9 $ \\
$\,$ & 
 $V_{\rm crit} =  {\bf 4.4} $ &
 $V_{\rm crit} = 20.9 $ &
 $V_{\rm crit} = 211  $ \\
 \hline
$a_1 = 2.5 \, \mu{\rm m}$ & 
 $V_{12} = 1.0 $ &
 $V_{12} = 1.8 $ &
 $V_{12} = 32.4 $ \\
$\,$ & 
 $V_{\rm crit} =  4.1 $ &
 $V_{\rm crit} = 20.3 $ &
 $V_{\rm crit} = 210 $ \\
 \hline
$a_1 = 1 \, \mu{\rm m}$ & 
 $V_{12} =  0.15 $ &
 $V_{12} =  1.1 $ &
 $V_{12} = 32.3 $ \\
$\,$ & 
 $V_{\rm crit} =  4.7 $ &
 $V_{\rm crit} = 19.2 $ &
 $V_{\rm crit} = 209  $ \\
 \hline
 $a_1 = 0.22 \, \mu{\rm m}$ & 
 $V_{12} =  1.1 $ &
 $V_{12} =  1.5 $ &
 $V_{12} = 32.3 $ \\
$\,$ & 
 $V_{\rm crit} = 19.2$ &
 $V_{\rm crit} = 21.1 $ &
 $V_{\rm crit} = 202  $ \\
 \hline
 $a_1 = 22 \, {\rm nm}$ & 
 $V_{12} = 32.3 $ &
 $V_{12} = 32.3 $ &
 $V_{12} = 45.7 $ \\
$\,$ & 
 $V_{\rm crit} = 209 $ &
 $V_{\rm crit} = 202 $ &
 $V_{\rm crit} = 211 $ 
\enddata
% \tablecomments{}
\end{deluxetable}
%---------------------------------

%---------------------------------
% TABLE 4. Velocities
\begin{deluxetable}{c|c|c|c}
% \rotate
% \renewcommand{\thefootnote}{\alph{footnote}}
\label{table:velocitieslower}
%\tablenum{4}
\tablecaption{Expected ($V_{12}$) and critical ($V_{\rm crit}$) relative velocities (cm/s) of pairs of corundum particles, at $z = 0 H$, where $T = 1425$ K. Bold type denotes cases where $V_{12} > V_{\rm crit}$ and grains can accrete.}
\tablewidth{0pt}
\tablehead{
\colhead{$\,$} & 
\colhead{$a_2 = 1 \, \mu{\rm m}$} & 
\colhead{$a_2 = 0.25 \, \mu{\rm m}$} & 
\colhead{$a_2 = 25 \, {\rm nm}$} 
}
\startdata
$a_1 = 1000 \, \mu{\rm m}$ & 
 $V_{12} = {\bf 138} $ &
 $V_{12} = {\bf 138} $ &
 $V_{12} = 142 $ \\
$\,$ & 
 $V_{\rm crit} =  {\bf 4.0} $ &
 $V_{\rm crit} = {\bf 18.1} $ &
 $V_{\rm crit} = 181 $ \\
  \hline
  $a_1 = 250 \, \mu{\rm m}$ & 
 $V_{12} = {\bf 68.9} $ &
 $V_{12} = {\bf 68.9} $ &
 $V_{12} = 76.5 $ \\
$\,$ & 
 $V_{\rm crit} =  {\bf 4.0} $ &
 $V_{\rm crit} = {\bf 18.1} $ &
 $V_{\rm crit} = 181 $ \\
  \hline
$a_1 = 100 \, \mu{\rm m}$ & 
 $V_{12} = {\bf 37.6} $ &
 $V_{12} = {\bf 37.9} $ &
 $V_{12} = 50.4 $ \\
$\,$ & 
 $V_{\rm crit} =  {\bf 4.0} $ &
 $V_{\rm crit} = {\bf 18.1} $ &
 $V_{\rm crit} = 181 $ \\
  \hline
$a_1 = 25 \, \mu{\rm m}$ & 
 $V_{12} =  {\bf 9.1} $ &
 $V_{12} =  9.5 $ &
 $V_{12} =  34.5 $ \\
$\,$ & 
 $V_{\rm crit} =  {\bf 3.9} $ &
 $V_{\rm crit} = 18.0 $ &
 $V_{\rm crit} = 181  $ \\
 \hline
$a_1 = 10 \, \mu{\rm m}$ & 
 $V_{12} =  3.4 $ &
 $V_{12} =  3.9 $ &
 $V_{12} =  33.4 $ \\
$\,$ & 
 $V_{\rm crit} =  3.8 $ &
 $V_{\rm crit} = 17.9 $ &
 $V_{\rm crit} = 181 $ \\
 \hline
$a_1 = 2.5 \, \mu{\rm m}$ & 
 $V_{12} =  0.6 $ &
 $V_{12} =  1.4 $ &
 $V_{12} = 33.2 $ \\
$\,$ & 
 $V_{\rm crit} =  3.5 $ &
 $V_{\rm crit} = 17.4 $ &
 $V_{\rm crit} = 180 $ \\
 \hline
$a_1 = 1 \, \mu{\rm m}$ & 
 $V_{12} =  0.2 $ &
 $V_{12} =  1.1 $ &
 $V_{12} = 33.2 $ \\
$\,$ & 
 $V_{\rm crit} =  4.0 $ &
 $V_{\rm crit} = 16.5 $ &
 $V_{\rm crit} = 179  $ \\
 \hline
 $a_1 = 0.22 \, \mu{\rm m}$ & 
 $V_{12} =  1.1 $ &
 $V_{12} =  1.5 $ &
 $V_{12} = 33.2 $ \\
$\,$ & 
 $V_{\rm crit} = 16.5 $ &
 $V_{\rm crit} = 18.1 $ &
 $V_{\rm crit} = 173 $ \\
 \hline
 $a_1 = 22 \, {\rm nm}$ & 
 $V_{12} = 33.2 $ &
 $V_{12} = 33.2 $ &
 $V_{12} = 47.0 $ \\
$\,$ & 
 $V_{\rm crit} = 179 $ &
 $V_{\rm crit} = 173 $ &
 $V_{\rm crit} = 181 $ 
\enddata
% \tablecomments{}
\end{deluxetable}
%---------------------------------

In {\bf Table~\ref{table:velocitiesupper}} we list the relative velocities, $V_{12}$ and critical velocities for collision, $V_{\rm crit}$, between pairs of particles in our model, based on conditions at the $z = 1.18 \, H$ surface ($\rho_{\rm g} = 2.7 \times 10^{-9} \, {\rm g} \, {\rm cm}^{-3}$, $T = 1350 \, {\rm K}$).
%In {\bf Table~\ref{tab:velocitieslower}}, 
In {\bf Table~\ref{table:velocitieslower}} we list the same information for conditions at the $z = 0$ midplane ($\rho_{\rm g} = 5.5 \times 10^{-9} \, {\rm g} \, {\rm cm}^{-3}$, $T = 1425 \, {\rm K}$).
Particles are assumed to be corundum, with work functions 4.7 eV.
Several trends are evident. 

First, throughout the hot midplane region, even particles up to several microns in radius would have been incapable of overcoming their mutual electrical repulsion to collide with each other. 
In contrast, `seed' particles at least $10 \, \mu{\rm m}$ in radius {\it could} sweep up other micron-sized particles (but not $0.25 \, \mu{\rm m}$ particles, unless the seed particles were hundreds of microns in size).
This is because in turbulent eddies, larger particles have greater velocities relative to other particles.
Exceptionally large presolar grains might have served as seed particles: some of the largest presolar hibonite grains studied are several microns in size \citep{ZegaEtal2011}, and other presolar grains approach $20 \, \mu{\rm m}$ in size \citep{Zinner2014}.
A more likely source of seed particles is fragments of previously formed, large, PLAC-like LAACHIs. 
One object that had previously grown to radius $1 \, {\rm mm}$ and then fragmented could produce up to $10^6$ fragments with radius $10 \, \mu{\rm m}$, which could then each accrete $1 \, \mu{\rm m}$-radius grains and grow up to $1 \, {\rm mm}$ in radius themselves, before either fragmenting or diffusing out of the region. 

Another trend evident from Tables~\ref{table:velocitiesupper} and~\ref{table:velocitieslower} is that not even large particles could have accreted the small, $\altwosix$-bearing former nanospinels, unless they were $> 1200 \, \mu{\rm m}$ in radius (at $z = 1.18 H$) or $> 1700 \, \mu{\rm m}$ in radius (at $z\!\!=\!\!0$). 
No LAACHIs less than about 2 or 3 mm in size would have accreted any $\altwosix$, at least for the model parameters explored here.

Assuming perfect sticking efficiencies for accreted grains, the rate at which a large (seed) grain with radius $a$ gained mass by accreting $1 \, \mu{\rm m}$ particles at $z = 0 H$ would have been
\[
\frac{d}{dt} \left( \frac{4\pi}{3} \rho_{\rm cor} a^3 \right) = \frac{4\pi}{3} \rho_{\rm cor} (1 \, \mu{\rm m})^3 \, \times 
\]
\begin{equation}
n_2 \, \pi \left( a + 1 \, \mu{\rm m} \right)^2 \, \left( 0.38 \, {\rm cm} \, {\rm s}^{-1} \right) \, \left( \frac{a}{1 \mu{\rm m}} -1 \right), 
\end{equation}
where {\Steve $n_2 = 1.0 \times 10^{-2} \, {\rm cm}^{-3}$} refers to the density of the $1 \, \mu{\rm m}$-radius corundum grains only.
Defining $y = a / (1 \, \mu{\rm m})$, we can rewrite this as 
\begin{equation}
\frac{d y}{dt} = \frac{1}{t_{\rm grow}} \, \left( y+1 \right) \, \left( 1 - \frac{1}{y^2} \right) \approx \frac{ y }{ t_{\rm grow} },
\label{eq:growth}
\end{equation}
where
{\Steve
\begin{equation}
t_{\rm grow} = 810 \, \left( \frac{\alpha}{1 \times 10^{-3}} \right)^{-3/4} \, \left( \frac{ \rho }{ 5.5 \times 10^{-9} \, {\rm g} \, {\rm cm}^{-3}} \right)^{-1/4} \, {\rm yr}.
\end{equation}
}
The growth timescale would be only $\sim 10\%$ longer at $z=1.18 \, H$.
The solution to Equation~\ref{eq:growth} approaches $y(t) = y(0) \, \exp(+t / t_{\rm grow})$ in the limit that $y(0) \gg 1$.
Because of the exponential nature of the growth, the time to grow from a starting radius $a = 10 \, \mu{\rm m}$, to a particle 1000 times more massive, with $a = 100 \, \mu{\rm m}$, would have been about {\Steve 1860} years; and the time taken to grow to $a = 1000 \, \mu{\rm m}$ would have been {\Steve no more than an additional 1860 years, and possibly no more than an additional 930 years, as larger particles can sweep up the smaller grains in the LAACHI-forming region.}

These growth times of {\Steve thousands} of years are comparable to the time spent by particles in the hot midplane region.
A distribution of particles all starting at $z\!\!=\!\!0$, would have density proportional to $\exp(-z^2 / (4 D t))$ after a time $t$; the fraction still at $z < 1.18 \, H$ {\Steve and remaining in the hot midplane region} would be ${\rm erf}( (1.18 H) / \sqrt(4 D t) )$ $= {\rm erf}( \sqrt{ 30.0 \, {\rm yr} / t} )$.
Above we defined the typical time in the hot region to be $(1.18 H)^2 / D \approx 120 \, {\rm yr}$, and indeed the fraction of particles that would have remained after 120 years would have been ${\rm erf}(0.50) = 52\%$.
{\Steve The fraction of particles that escape to cooler regions of the disk, where they might escape further reprocessing, is ${\rm erfc}(\Delta r/\sqrt{4 D t})/2$ $= {\rm erfc}(\sqrt{255 \, {\rm yr} / t})/2$, where $\Delta r \approx 0.2$ AU (\S 3.1).
After 1000 years, about 25\% of all inclusions in the LAACHI-forming region will have diffused out beyond 0.8 AU. 
The fraction of particles that diffuse to cooler regions, while remaining in the hot midplane region, is the product of these, which peaks at $\approx 5\%$ between 1000 and 2000 years.
In this timespan, particles can grow by factors of 3 to 10 in radius (i.e., from $10 \, \mu{\rm m}$ to $30-100 \, \mu{\rm m}$) and factors of 30 to $10^3$ in mass.}
% That is, about half of all LAACHIs would have stayed in the region about 120 years, long enough to grow from $10 \, \mu{\rm m}$ to $140 \, \mu{\rm m}$ in radius, and increase in mass by a factor of 2.5. 
%About 21\% would have stayed in the region 875 years, long enough to grow  to $100 \, \mu{\rm m}$ in radius; and about 15\% would have stayed in the region 1750 years, long enough to grow  to $100 \, \mu{\rm m}$ in radius.
These sizes and proportions appear consistent with observations of PLACs and corundum grains. 

In principle, some LAACHIs could have remained in the hot midplane region for longer and grown to much larger sizes; but several mechanisms work against growth to scales larger than about 1 mm. 
The most important one is fragmentation: as particles grow to sizes larger than about 1 mm in radius, their velocities relative to other particles exceed $140 \, {\rm cm} \, {\rm s}^{-1}$.
This begins to exceed the fragmentation threshold velocity, estimated from laboratory experiments to be about $1 \, {\rm m} \, {\rm s}^{-1}$ \citep{GuttlerEtal2010}.
Indeed, many PLACs appear to be fragments of larger objects hundreds of microns in size \citep[\S 2.3.3;][]{KoopEtal2016a}.
The survival of the singularly large inclusion {\it HAL} may owe to its apparently melted nature.
To form its three large hibonite laths, it would seem necessary to melt and recrystallize this aggregate somehow; this would be consistent with the known isotopic fractionation of {\it HAL} and the presence of glass \citep{FaheyEtal1987a}.
Other inclusions, not quite as large as {\it HAL} (e.g., {\it M98-8}, {\it 31-2}) do not appear melted; we conjecture that had they been larger they would have fragmented.

Large objects the size of {\it HAL} also may be removed effectively from the LAACHI-forming region.
Objects will settle to the midplane if ${\rm St} > 0.2 \alpha$ \citep{DeschEtal2018}; for this environment that means radius $> 2$ mm.
Once there, they might be removed efficiently by meridional flow \citep{DeschEtal2018}.

{\Steve 
\subsection{Incorporation of $\altwosix$ from the Gas}
}

{\Steve
With the growth of the particles quantitatively described, it is possible to estimate the amount of $\altwosix$ they would incorporate from the gas. 
In the region where $T = 1425 \, {\rm K}$ and $\rho_{\rm gas} = 5.5 \times 10^{-9} \, {\rm g} \, {\rm cm}^{-3}$, only a fraction $\sim 1.6 \times 10^{-6}$ of all Al atoms will be in the gas phase in equilibrium.
The number density of gas-phase Al atoms will be $n_{\rm Al} = 1.3 \times 10^{4} \, {\rm cm}^{-3}$. 
The $\alratio$ ratio of this gas is likely to be close to the solar system average, ${\cal R} = (\alratio) \approx$ ${\cal R}_{\rm SS} = 5 \times 10^{-5}$, as it comprises Al from a variety of presolar grains, including perhaps silicate and SiC grains from a recent, nearby supernova or Wolf-Rayet star, but mostly other presolar grains from AGB stars that have no live $\altwosix$. 
Particles will acquire $\altwosix$ atoms from the gas at a rate proportional to their surface area:
\begin{equation}
\frac{d}{dt} \left[ {}^{26}{\rm Al} \right] = \, 4\pi a^2 \, n_{\rm Al} \, {\cal R} \, \frac{1}{4} \left( \frac{ 8 k T}{ \pi m_{\rm Al} } \right)^{1/2} \, S,
\end{equation}
where $S \approx 0.05$ is the sticking coefficient \citep{TakigawaEtal2015}. 
Assuming $a(t) = a_0 \exp \left( +t / t_{\rm grow} \right)$, the solution to Equation 6, we find the final number of $\altwosix$ atoms in a particle of final radius $a$ is 
\begin{equation}
{}^{26}{\rm Al} = \frac{1}{2} \left( a^2 - a_{0}^2 \right) \, n_{\rm Al} \, {\cal R} \, \frac{1}{4} \left( \frac{ 8 k T}{ \pi m_{\rm Al} } \right)^{1/2} \, S \, t_{\rm grow}.
\end{equation}
The total number of ${}^{27}{\rm Al}$ atoms in the particle is
\begin{equation}
{}^{27}{\rm Al} = \frac{4\pi}{3} a^3 \, \frac{ \rho_{\rm cor} }{ 51 m_{\rm p} },
\end{equation}
where $51 m_{\rm p}$ is the mass of ${\rm Al}_{2}{\rm O}_{3}$ per Al atom).
After growing to a size $a$, the average $(\alratio)$ in a particle is
\begin{equation}
\frac{{}^{26}{\rm Al}}{{}^{27}{\rm Al}} = \frac{3}{2} \, \left( \frac{ a^2 - a_{0}^2 }{ a^3 } \right) \, \frac{ 51 m_{\rm p} }{ \rho_{\rm cor} } \, n_{\rm Al} \, {\cal R} \, \frac{1}{4} \left( \frac{ 8 k T}{ \pi m_{\rm Al} } \right)^{1/2} \, S \, t_{\rm grow}.
\end{equation}
The quantity in parentheses has a maximum of $0.4 / a_0$ at a radius $a = 1.7 a_0$, and decreases as $1/a$ at large $a$.
Because no particle spends all its time at the midplane, the average value of $n_{\rm Al}$ is lower than the midplane value; integrating over $z$, we find the average value is lower by a factor of 0.45.
Assuming this, and adopting $t_{\rm grow} = 810 \, {\rm yr}$ and $S \approx 0.05$ from above, yields 
\begin{equation}
\frac{{}^{26}{\rm Al}}{{}^{27}{\rm Al}} \approx 0.06 \, {\cal R} \, \left( \frac{ a }{ 1 \, \mu{\rm m} } \right)^{-1}.
\end{equation}
Particles like {\it HAL} with an average radius $\approx 70 \, \mu{\rm m}$, would have $(\alratio)_0 \approx 0.002 \, {\cal R}$, or about $4 \times 10^{-8}$.
}

More sophisticated modeling should be undertaken to explore the diversity of particle histories, but we conclude from this simple treatment that growth of aggregates of corundum / hibonite, to tens or hundreds of microns, is predicted to take place during the {\Steve thousands} of years or so particles spent in this environment, and that growth to these sizes would not generally have included live $\altwosix$. 
%\Steve 
{Electrical charge has been considered in previous models of coagulation and growth \citep[e.g.,][]{Okuzumi2009,XiangEtal2020,AkimkinEtal2020}, but to our knowledge, it has not been contemplated how this would lead to chemical or isotopic fractionations.}

\subsection{Subsequent Fate}

After diffusing out of the hot midplane region, newly-formed LAACHIs (PLACs, corundum grains, etc.) would find themselves in cooler environments where other phases are thermodynamically favored, as described in \S 3.4.

One possible outcome is that reaction with the gas could transform hibonite or corundum to the thermodynamically favored phases spinel, melilite, anorthite or diopside, by addition of Mg, Si, and Ca.
This is probably limited, though, as diffusion of these cations into the minerals could have been kinetically inhibited at lower temperatures.
The inclusion {\it A-COR-01} has a rim of diopside about $5 - 10 \,\mu{\rm m}$ thick \citep{BodenanEtal2020}.
Assuming the rate-limiting step is diffusion of Ca into corundum, and estimating $D_{\rm Ca} \approx 1.5 \times 10^{-21} \, {\rm m}^{2} \, {\rm s}^{-1}$ at 1350 K (\S 3.4), implies a rim of thickness $\approx 6 \, \mu{\rm m}$ would grow in $\approx 700 \, {\rm yr}$, roughly the time it would take to diffuse radially to conditions 50 K cooler.

Another possible outcome is the accretion of spinel produced at lower temperatures elsewhere in the disk. 
In adjacent regions of the disk with slightly lower temperatures (e.g., $\approx 1200 \, {\rm K}$, beyond about 0.8 AU), such that olivine and pyroxene would have been stable, the density of solids would be higher by a factor $\sim 10^2$ relative to the LAACHI-forming region. 
While grains would have had similar charges, and the relative velocities would have been similar, a key difference is that silicate particles could have
{\Steve condensed} 
and quickly ($\sim 10^4$ years) grown to sizes $> 1 \, {\rm mm}$ capable of accreting small, $\altwosix$-bearing grains. 
We do not model these regions, but consider it likely that spinel or other minerals formed in these regions would have contained live $\altwosix$. 
Agglomerations of spinel and hibonite would resemble SHIBs, with near-canonical $\alratio$ ratios.
In general, as LAACHIs interact with material outside the hot midplane region where they formed, there would be opportunities for them to incorporate $\altwosix$-bearing material. 

\section{Discussion}

The detailed model presented here provides a framework for interpreting the observations made of PLACs and other LAACHIs, to infer their origins and to make predictions about other potentially observable features. 
{\Steve One potential application is the understanding of Rare Earth Element (REE) patterns. 
Many PLACs are depleted in Yb and Eu \citep{Ireland1990}, with condensation temperatures of 1475 K and 1347 K \citep{Lodders2003}.
Eu, at least, would not fully condense in the LAACHI-forming region. 
However, an understanding of the REE patterns in PLACs and other LAACHIs lays beyond the scope of this paper. }
Here we discuss the implications for refractory metal abundances, stable isotope anomalies, short-lived radionuclides, and detailed mineralogy  of LAACHIs.
These can be used to test the hypothesis presented here.

\subsection{Refractory Metal}

Almost as important to the question of what PLACs and other LAACHIs grew from is what they did {\it not} include during their growth.
Although corundum and hibonite are among the most refractory mineral phases to condense from a solar composition gas, there are other, rarer, minerals that could survive at even higher temperatures. 
Besides perovskite, %[${\rm CaTiO}_{3}$],
zirconia [${\rm ZrO}_{2}$] and hafnia [${\rm HfO}_{2}$] are stable (at $10^{-4}$ bar) up to 1758 K and 1694 K \citep{Lodders2003}.
Refractory metal alloys of Mo, Ru, Rh, W, Re, Os, Ir and Pt condense at temperatures between 1403 K for Pt, and 1817 K for Re \citep{Lodders2003}.
Refractory metal nuggets (RMNs) are a phase rich in these elements; although it has been argued that these primarily form by precipitation from CAI melts \citep{SchwanderEtal2015}, a subset of RMNs appear more consistent with condensation, and certainly a small fraction of RMNs appear to be of presolar origin \citep{SchwanderEtal2015,DalyEtal2017}. 
Therefore it is reasonable to assume that at least some refractory metal grains would have existed in the LAACHI-forming region.
Despite this, and although refractory metal is commonly found in SHIBs, it is relatively rare in PLACs \citep{SchwanderEtal2015}.

%Condensates of refractory metal are probably a constituent of refractory metal nuggets (RMNs), which are FeNi grains sub-micron to micron-sized, enriched in the refractory metals above by $\approx 12-23$ times the abundances in CI chondrites \citep{DalyEtal2017}.
%All of these phases might have existed in the PLAC-forming region.
%RMNs are commonly found in SHIBs \citep{SchwanderEtal2015}.
%Refractory metal has been found in SHIBs---for example, SHIB 1-9-5 contains refractory metal enclosed with spinel \citep{HanEtal2017lpsc}---but to our knowledge, refractory metal enclosed in PLACs has not been observed. 
%Any successful model of PLAC formation must explain their absence. 
%
In the context of the model presented here, refractory metal grains would not have been incorporated into PLACs.
The work functions of the refractory metals above range from 4.3 eV for Mo and 4.5 eV for W, and 4.6 eV for Ru, to 5.3 eV for Ir and Pt \citep{Fomenko1966}.
Grains made of refractory metals would have been micron-sized and negatively charged in the LAACHI-forming environment, although not quite as much as corundum and hibonite grains (being dominated by Mo and Ru, with typical work function $\sim 4.5$ eV).
Their electrical repulsion would have excluded them from the growing PLACs, same as the $\altwosix$-bearing corundum grains.
Interestingly, the work functions of ${\rm HfO}_{2}$ and ${\rm ZrO}_{2}$ are roughly 3.6 eV and 3.8 eV, respectively \citep{Fomenko1966}.
If bare grains of these materials existed, they would be neutral to positively charged, and effectively accreted, although they would be exceedingly rare. 

\subsection{Stable Isotope Anomalies}

\subsubsection{Oxygen}

A potential argument against a mostly presolar origin for the Al in PLACs and other LAACHIs is that their oxygen isotopes do not reflect the wide variety exhibited by presolar grains \citep[e.g.,][]{Zinner2014}.
As outlined in \S 3.3, diffusive exchange of oxygen isotopes with the nebula is rapid in regions that reached 1350 K: roughly 25 years for $0.25 \, \mu{\rm m}$-radius spinel grains  to exchange 99.9\% of O atoms; {\Steve and 3500 years for $1 \, \mu{\rm m}$-radius corundum grains, to exchange 90\% of O atoms.
At 1425 K, a $1 \, \mu{\rm m}$ radius corundum grain exchanges 90\% of oxygen in 190 years.}
These are {\Steve comparable to or} shorter than the typical time a PLAC would have spent in the hot midplane region, roughly hundreds of years.
%; and at higher temperatures the equilibration timescale would have been even shorter.
There is a strong sensitivity to temperature, however, and corundum grains {\Steve would take thousands of years to equilibrate} at 1325 K.
Any PLACs that formed mostly from grains that didn't heat above 1325 K might not have fully exchanged with the solar nebula gas at $\Delta^{17}{\rm O} \approx -35\permil$ to $-23\permil$, and might appear mixed with a presolar composition that presumably was near $\Delta^{17}{\rm O} \approx 0\permil$, with variable $\delta^{18}{\rm O}$
\citep{KrotEtal2010}.
In fact, many PLACs have $\Delta^{17}{\rm O}$ extending up to $\approx -17\permil$, and those that do show a slightly greater spread in $\delta^{18}{\rm O}$ \citep{KoopEtal2016a}.
We conclude that in many PLACs ($\Delta^{17}{\rm O} < -23\permil$), oxygen isotope exchange with the solar nebula gas was very ($> 99.9\%$) complete, but that in some ($\Delta^{17}{\rm O} > -23\permil$), isotopic exchange perhaps was not quite as complete. 
This is consistent with the predicted timescales for oxygen isotopic equilibration in the LAACHI-forming environment.

\subsubsection{Magnesium}

For the most part, the Mg isotopic compositions of PLACs and related inclusions do not appear different from solar \citep{Liu2008thesis}. 
In some inclusions, analysis spots show excesses in ${}^{26}{\rm Mg}/{}^{24}{\rm Mg}$, and isochrons can be constructed, revealing the decay of $\altwosix$.
This is not true in most of the inclusions, and in fact analysis spots in many PLACs show deficits of $3-4\permil$ in the ${}^{26}{\rm Mg}/{}^{24}{\rm Mg}$ ratio relative to the chondritic value
\citep[][and references therein]{Ireland1988,Liu2008thesis}.
{This observation has been difficult to explain, but our model predicts such a nucleosynthetic effect.}

In the context of our model, PLACs and other LAACHIs mostly would have exchanged Mg isotopes with the abundant Mg vapor in the gas, with a solar composition; but some  might have retained Mg from the minor amounts in the Al-bearing phases.
These would not have been spinel or corundum: presolar corundum grains contain very little Mg, and most Mg would have diffused out of spinel as it converted to corundum.
Any Mg remaining in these grains would have equilibrated with solar Mg.
In contrast, Mg can exist as a minor element in hibonite, which contains structural Mg in the form of coupled substitution of ${\rm Mg}^{2+}$ and ${\rm Ti}^{4+}$ for two ${\rm Al}^{3+}$ atoms.
As a tetravalent cation, ${\rm Ti}^{4+}$ may not diffuse out of hibonite easily, and this may allow hibonite to retain the paired ${\rm Mg}$.
PLACs would have mostly exchanged Mg with the gas and isotopically equilibrated; but to the extent they retain a presolar signature, it would be from the Mg in presolar hibonite grains.

Among presolar oxide grains of spinel and corundum, deficits in ${}^{26}{\rm Mg}/{}^{24}{\rm Mg}$ are found almost exclusively on Group 1 and Group 3 grains.
Similarly, the only measured Group 3 presolar hibonite grain exhibited
% $\delta^{26}{\rm Mg} = $
a deficit of $-28\permil$ \citep{NittlerEtal2008}. 
The majority of presolar aluminum oxide grains, those from AGB stars, appear to have deficits of ${}^{26}{\rm Mg}$.
In contrast, supernova-derived Group 4 grains of spinel, hibonite, and corundum show $\delta^{26}{\rm Mg}$ excesses \citep[from the presolar grain database, \url{http://presolar.physics.wustl.edu};][]{HynesGyngard2009}.
As only a small fraction of
%the solar complement of 
Mg existed in solid form in the LAACHI-forming environment, it is difficult to make quantitative predictions of the deviation of Mg isotopic compositions relative to bulk Solar System or terrestrial standards; but of the Mg in the region, LAACHIs would have preferentially accreted Mg from the low-$\delta^{26}{\rm Mg}$ Group 1 and 3 spinel and corundum grains and Group 3 hibonite grains, and excluded the high-$\delta^{26}{\rm Mg}$ Group 4 nanospinels.
This potentially explains the deficits in ${}^{26}{\rm Mg}$ of $\approx -3\permil$, if roughly 10\% of their Mg retained the signature of presolar hibonites.

\subsubsection{Calcium and Aluminum}

LAACHIs would not be expected to show any major deviations from bulk Solar System (``chondritic") 
%solar 
in most Ca or Al isotopic ratios.
Al definitely would have recorded a presolar isotopic signature: it is a major element in corundum, spinel, and hibonite, and Al in these minerals would not be exchanged with the gas.
However, Al is monoisotopic.

Ca is a refractory element with multiple isotopes.  
While a minor phase in presolar corundum and spinel (CaO $\ll 1$wt\%), presolar hibonite contains 8wt\% CaO.
However, this Ca would have mostly exchanged with solar-composition Ca vapor in the gas (\S 3.4)
In contrast to ${\rm Mg}^{2+}$, which is coupled to ${\rm Ti}^{4+}$ in hibonite, ${\rm Ca}^{2+}$ could have diffused through hibonite within hundreds of years or less, although the exact diffusion coefficients are unknown.

The only means of acquiring a Ca isotopic composition recognizable as presolar is through inheritance of isotopic anomalies by accretion of presolar perovskite grains.
Such grains formed in supernovae are the likely source of ${}^{48}{\rm Ca}$ excesses \citep{DauphasEtal2014}.
Being small grains, one might expect them to also exchange Ca atoms with the gas: the diffusion coefficients of Ca and Ti through perovskite are unknown, but the diffusion of similar Ba through ${\rm BaTiO}_{3}$ is $3 \times 10^{-18} \, {\rm m}^{2} \, {\rm s}^{-1}$ at 1350 K \citep{SazinasEtal2017}, implying exchange with the gas in about 4 days, assuming a grain radius of $1 \, \mu{\rm m}$ \citep[e.g.,][]{HanEtal2015}.
As short as this is, it is longer than the average time for a perovskite grain to collide with and stick to a corundum or hibonite grain, about 0.5 days (after accounting for the focusing due to their charge; \S 4.2.4). 
Perovskite grains liberated from evaporating aggregates of silicates therefore might not spend enough time as bare grains to exchange with the gas and lose their signature of ${}^{48}{\rm Ca}$ excess, before being incorporated into LAACHIs. 

Without pursuing detailed modeling, our expectation is that with the exception of ${}^{48}{\rm Ca}$ and other anomalies acquired from perovskite grains, LAACHIs should appear solar in their Ca isotopic compositions. 
Measurements of $\delta^{42}{\rm Ca}$ and $\delta^{43}{\rm Ca}$ in PLACs indeed are not resolved from a solar composition \citep{KoopEtal2016a}.

%If a presolar signature were to persist, it would probably come either from presolar hibonite or perovskite (presolar corundum and spinel probably contain $\ll 1$wt\% CaO).
%Because we invoke Group 3 hibonite as a source of some Mg in LAACHIs, they are the most likely contributors.
%The only measurements of a Group 3 hibonite show slight excesses in Ca isotopes (relative to ${}^{40}{\rm Ca}$): $\delta^{42}{\rm Ca} = +32 \pm 7\permil$, $\delta^{43}{\rm Ca} = +54 \pm 18\permil$, and $\delta^{44}{\rm Ca} = +70 \pm 8\permil$ \citep{NittlerEtal2008}.
%Assuming a few percent of this presolar Ca were retained in a PLAC (a smaller fraction than we infer from Mg), we would estimate very small excesses ($< 1\permil$) in ${}^{42}{\rm Ca}$, ${}^{43}{\rm Ca}$, and ${}^{44}{\rm Ca}$.

\subsubsection{Titanium}

Another well-known feature of PLACs is that they exhibit large and correlated excursions in both $\epsilon^{50}{\rm Ti}$ and $\epsilon^{48}{\rm Ca}$ anomalies \citep{ZinnerEtal1986,ZinnerEtal1987,FaheyEtal1987a,FaheyEtal1987b,Ireland1990,SahijpalEtal2000,MeyerZinner2006,LiuEtal2009,KoopEtal2016a}.
\citet{DauphasEtal2014} have argued compellingly that the carrier of positive anomalies is presolar perovskite [${\rm CaTiO}_{3}$] grains formed in supernova ejecta containing both the neutron-rich isotopes ${}^{48}{\rm Ca}$ and ${}^{50}{\rm Ti}$.
Acquisition of a greater-than-average number of such carriers would lead to ${}^{50}{\rm Ti}$ excesses, while incorporation of a lower-than-average number would lead to deficits, relative to terrestrial (near Solar System average).
These carriers have not been discovered, but having formed in supernova ejecta they are likely to be small, $<$ tens of nm in size.

Like the small nanospinel grains excluded from growing LAACHIs,
presolar perovskite grains would have been small enough to be excluded from PLACs if they were negatively charged, which would mean PLACs could not carry ${}^{48}{\rm Ca}$ or ${}^{50}{\rm Ti}$ excesses;  remarkably, though, perovskite in the LAACHI-forming region is likely to have been {\it positively} charged.
The work functions of titanates are predicted from Density Functional Theory to be much lower than other minerals \citep{JacobsEtal2016}.
The work function of ${\rm SrTiO}_{3}$ is 3.12 eV \citep{WranaEtal2019}, and the work function of ${\rm CaTiO}_{3}$ in particular is 3.0 eV \citep{ZhongHansmann2016}.
As such, these grains would have been positively charged, with $Z \approx +400 \, (a / 1 \, \mu{\rm m})$. 
Not only would small titanate grains not be excluded from the growing PLAC, their rate of capture actually would have been enhanced above the uncharged case, by over an order of magnitude. 
%[Indeed, a large spinel grain has been found encased in perovskite \citep{ZegaEtal2021}, although this has been argued to have condensed there.]
PLACs could have easily incorporated perovskite carriers, perfectly consistent with the notable abundance of perovskite in PLACs, and with them exhibiting correlated excesses/deficits in ${}^{48}{\rm Ca}$ and ${}^{50}{\rm Ti}$.

This model matches observations  not just qualitatively, but quantitatively as well.
PLACs are known to be quite variable in their abundances, with $\epsilon^{50}{\rm Ti}$ ranging from about -750$\epsilon$ to $+1000\epsilon$  \citep{ZinnerEtal1986,HintonEtal1987,Ireland1990,LiuEtal2009,KoopEtal2016a,ShollenbergerEtal2022}.
% FUN CAIs show $\epsilon^{50}{\rm Ti} \approx \pm 350$ \citep{KrotEtal2014,ParkEtal2014lpsc}.
\citet{ShollenbergerEtal2022} recently demonstrated that the smaller spread in $\epsilon^{50}{\rm Ti}$ values among large, `normal' CAIs (typically only a few epsilon units) is attributable to an averaging of $\sim 10^4$ hibonite grains each.
Working in reverse, they concluded that a single hibonite grain $30 \, \mu{\rm m}$ in size might have included $\sim 3 \times 10^6$ presolar grains 100 nm in size, each with $\epsilon^{50}{\rm Ti} \sim 2 \times 10^6$, consistent with the observation that presolar grains exhibit $\epsilon^{50}{\rm Ti}$ up to $10^5 - 10^6$ \citep{JadhavEtal2008,NittlerEtal2018}.
We note that this is exactly consistent with our picture of LAACHIs being aggregates of smaller, quasi presolar grains.
These sizes and numbers of presolar grains are roughly consistent with presolar ${\rm CaTiO}_{3}$ grains $< 50$ nm in size being a major carrier of ${}^{50}{\rm Ti}$ excesses. 

We note that LAACHIs would not be the only samples to show correlated excesses/deficits in ${}^{50}{\rm Ti}$ and ${}^{48}{\rm Ca}$.
Essentially all materials would have accreted perovskite grains and exhibited the same trends. 
Our point is that even though these anomalies would be carried on very small ($< 50$ nm) grains, as small as the nanospinels that are excluded from LAACHIs, perovskite grains would not be excluded, and LAACHIs would show these correlations.

\subsubsection{Other Stable Isotopes}

Because titanates (almost uniquely) would have been positively charged, other isotopes carried by titanates, besides ${}^{48}{\rm Ca}$ and ${}^{50}{\rm Ti}$, could have been accreted into LAACHIs.
{\Steve For example,}
%${}^{84}{\rm Sr}$ %(or ${}^{88}{\rm Sr}$) and 
%${}^{135}{\rm Ba}$ are observed to correlate with ${}^{48}{\rm Ca}$ excesses \citep{BerminghamEtal2022lpsc}, and 
 \citet{ShenLee2003} and \citet{ChenEtal2015} detected excesses/deficits in ${}^{138}{\rm La}$ in CAIs that correlate with ${}^{50}{\rm Ti}$.
It is reasonable to assume these anomalies are also carried by perovskite or titanate grains, as these are the forms into which 
{\Steve La condenses}.
%Sr, Ba, and La condense \citep{Lodders2003}.
If so, then we predict LAACHIs should exhibit correlated excesses or deficits in 
%${}^{84}{\rm Sr}$, ${}^{135}{\rm Ba}$, and 
${}^{138}{\rm La}$ as well.

We note that PLACs would not have incorporated the ${}^{54}{\rm Cr}$-rich nanospinels that likely carried live $\altwosix$, and with a condensation temperature of 1291 K \citep{Lodders2003}, Cr would have evaporated from these grains in the LAACHI-forming region anyway.
It is likely that any Cr in PLACs would be isotopically solar, and not exhibit excesses in ${}^{54}{\rm Cr}$.
This may help to explain the apparent decoupling between ${}^{50}{\rm Ti}$ and ${}^{54}{\rm Cr}$, despite their general concordance \citep{TrinquierEtal2009}.

Finally, if small, refractory metal grains were the carriers of other isotopic anomalies such as ${}^{94}{\rm Mo}$ or ${}^{183}{\rm W}$, it is likely that LAACHIs would not have incorporated these stable isotope anomalies either, as small metal grains would have high work functions and would be negatively charged.

\subsection{Short-lived Radionuclides}

In the context of our model, other short-lived radionuclides (SLRs) would be incorporated into PLACs, or not, depending on the size and charge of the carriers. 
We argue that PLACs did not contain live $\altwosix$ because live $\altwosix$ resided only on small, negatively charged grains (nanospinels that lose Mg and Cr to become corundum, possibly transforming further to hibonite) that cannot be accreted.
Other SLRs would be on similarly small grains, because only those formed in supernovae or Wolf-Rayet star ejecta would be delivered to the Sun's molecular cloud before the SLRs decayed, and these grains are generally much smaller than grains formed in AGB outflows.
In the context of our model, if these grains survived in the LAACHI-forming region and were small and negatively charged, they would be excluded from the inclusions as they grew. 
This picture may help identify the carriers of various SLRs.

\subsubsection{Calcium-41}

The most important example {\Steve of an SLR that can be understood better in the context of our model} is ${}^{41}{\rm Ca}$, with a half-life of only 0.10 Myr \citep{JorgEtal2012}.
This existed in the solar nebula at a level ${}^{41}{\rm Ca}/{}^{40}{\rm Ca} \sim 10^{-8}$ \citep{SrinivasanEtal1994}.
Because of its short half-life, any live ${}^{41}{\rm Ca}$ must have been carried by grains of supernova origin, as is the case for ${}^{26}{\rm Al}$.
Moreover, it is known that PLACs that did not sample live $\altwosix$ also did not sample live ${}^{41}{\rm Ca}$, and vice versa \citep{SahijpalEtal2000,Liu2008thesis}.
%These observations help us to identify the carrier of ${}^{41}{\rm Ca}$. 
Therefore, in the context of our model, to not be incorporated into PLACs, ${}^{41}{\rm Ca}$ must have been carried on small, negatively charged presolar  grains of supernova {\Steve or Wolf-Rayet star} origin, able to survive at temperatures $> 1400$ K.
There are few candidate presolar grains.

Presolar SiC grains are one candidate. 
\citet{LiuEtal2018} extracted several micron-sized SiC presolar grains of supernova origin, measuring one to have formed with ${}^{41}{\rm Ca}/{}^{40}{\rm Ca} = 3.0 \times 10^{-3}$.
SiC has a work function $> 4.5$ eV \citep{Fomenko1966}, so these grains would have been negatively charged, and likely not accreted.
However, these grains likely would have been destroyed in the LAACHI-forming environment.
Although refractory in the sense that SiC forms at high temperature in environments with C/O $> 1$ \citep{Ebel2006}, \citet{MendybaevEtal2002} found that presolar SiC grains $\sim 1 \, \mu{\rm m}$ in radius would survive in the solar nebula for only $\sim 1$ year at $\approx 1400 \, {\rm K}$.
If SiC grains carried a significant fraction of all ${}^{41}{\rm Ca}$, this would have been released as Ca vapor in the LAACHI-forming region, and all corundum and spinel grains would have incorporated ${}^{41}{\rm Ca}$ as they converted to hibonite.

Low-density (LD) graphite presolar grains of supernova origin also contained live ${}^{41}{\rm Ca}$ when they formed, with ${}^{41}{\rm Ca}/{}^{40}{\rm Ca} \sim 10^{-2}$ \citep{AmariEtal1996,TravaglioEtal1999}.
Some trace elements in LD graphite grains appear to have been concentrated in subgrains, but Ca appears to have been distributed uniformly throughout \citep{TravaglioEtal1999}.
Graphite has a work function of 4.7 eV \citep{Fomenko1966} and would be negatively charged, although many LD graphite presolar grains are several microns in size, large enough to be accreted anyway by growing PLACs.
However, graphite also would have been destroyed by oxidation reactions in the LAACHI-forming region, about $10^3$ times faster than SiC grains  \citep{MendybaevEtal1997lpsc}, so we rule it out, too.

Presolar perovskite grains would be a reasonable candidate, as they contain Ca, and such grains are hypothesized to have formed in supernova ejecta \citep{DauphasEtal2014}.
However, the work function of perovskite is 3.0 eV, and such grains would not be excluded from growing PLACs. 
If these were the primary carrier of live ${}^{41}{\rm Ca}$, then most PLACs would not contain ${}^{26}{\rm Al}$ but they would all contain ${}^{41}{\rm Ca}$, contrary to what is observed. 
Also, ${}^{48}{\rm Ca}$ (and ${}^{50}{\rm Ti}$) excesses are not obviously correlated with the one-time presence of ${}^{41}{\rm Ca}$ (and ${}^{26}{\rm Al}$) \citep{Liu2008thesis}.

Presolar hibonite grains of supernova origin, like grain KH2 analyzed by \citet{ZegaEtal2011}, are a reasonable candidate. 
In one presolar hibonite grain of supernova origin measured for it, the ratio when it formed was inferred to be ${}^{41}{\rm Ca}/{}^{40}{\rm Ca} \approx 4.3 \times 10^{-4}$ \citep{Zinner2014}.
However, to account for the solar ${}^{41}{\rm Ca}/{}^{40}{\rm Ca}$ ratio, these grains would have to account  for $\sim 2 \times 10^{-5}$ of all Ca and, given that hibonite has Ca/Al = 0.08, a fraction $\sim 2 \times 10^{-4}$ of all Al.
This is about an order of magnitude higher than might be expected. 
Hibonites comprised a small but unquantified fraction of all presolar Al-bearing grains; we assume this was $\sim 10\%$. 
We assume that 20\% of presolar hibonite grains are of supernova (vs.\ AGB star) origin, and that a fraction $\sim 5 \times 10^{-4}$ formed in $< 0.1$ Myr before the Sun's formation, assuming an average lifetime of $2 \times 10^8$ yr \citep{JonesEtal1994}.
We therefore estimate that presolar hibonite grains of recent supernova origin might have contributed only $\sim 10^{-5}$ of all Al.
This makes them less than likely to be the carriers of live ${}^{41}{\rm Ca}$.

We consider the ${}^{54}{\rm Cr}$-rich nanospinels  that are putatively the carriers of live $\altwosix$ to also be the most likely carriers of live ${}^{41}{\rm Ca}$.
If nanospinels with live ${}^{41}{\rm Ca}$ formed in a supernova {\Steve or Wolf-Rayet star} $< 10^5$ yr before Solar System formation, and were incorporated rapidly into the Sun's molecular cloud, they might retain ${}^{41}{\rm Ca}/{}^{40}{\rm Ca} \sim 10^{-2}$ from the {\Steve massive star} \citep{RauscherEtal2002}.
These would have to make up a fraction $\sim 10^{-6}$, of all ${}^{40}{\rm Ca}$. 
Assuming these same grains also had ${}^{26}{\rm Al}/{}^{27}{\rm Al} \sim 0.05$, they would have had a fraction $\sim 10^{-3}$ of all the Al. 
Therefore in these grains, the ${}^{40}{\rm Ca}/{}^{27}{\rm Al}$ ratio would have had to be $\sim 10^{-3}$ times the solar ratio, or $\sim 7 \times 10^{-4}$.
Converting this to a mass fraction, nanospinels would have had to be $\sim 0.1$wt\% CaO. 
Nanospinels are reported to contain ``negligible" Ca \citep{DauphasEtal2010}, but typical detection limits for energy-dispersive X-ray spectroscopy (EDS) are $\sim 0.1$wt\%.
Although not detected in other grains, some Ca appears to have been incorporated in a presolar spinel grain of AGB origin, at a level of 0.26wt\% \citep{ZegaEtal2014}, suggesting the same reactions might occur in {\Steve massive star} outflows.
The same nanospinel grains we hypothesize carried ${}^{26}{\rm Al}$ therefore are also plausible as carriers of ${}^{41}{\rm Ca}$.

In the context of the model we have presented, the correlated absence or presence of both ${}^{26}{\rm Al}$ and ${}^{41}{\rm Ca}$ \citep{SahijpalEtal2000,Liu2008thesis} requires that the carrier of ${}^{41}{\rm Ca}$, like ${}^{26}{\rm Al}$, be on a small ($< 1 \, \mu{\rm m}$), negatively charged (work function $\geq 4.5$ eV) refractory (surviving up to $> 1400 \, {\rm K}$) grain of recent ($\sim 10^5$ yr) {\Steve massive star} origin.
Presolar graphite and SiC grains contained live ${}^{41}{\rm Ca}$, and hypothetical presolar perovskite might have as well, but these fail to meet all criteria. 
Presolar hibonite grains are a possibility, but we consider nanospinels with $\sim 0.1$wt\% CaO to be the most likely candidate.
%This demonstrates how a model of PLAC formation can help identify SLR carriers.

%Type X presolar SiC grains of supernova origin such as {\it Bonanza} \citep{GyngardEtal2018} may be another possibility, and are known to carry ${}^{50}{\rm Ti}$ excesses \citep{NittlerEtal1996,LinEtal2010}.

\subsubsection{Beryllium-10}

Another very important short-lived radionuclide in the solar nebula was ${}^{10}{\rm Be}$, which decays to ${}^{10}{\rm B}$ with a half-life of 1.39 Myr \citep{ChmeleffEtal2010,KorschinekEtal2010}.
Since its discovery \citep{McKeeganEtal2000}, Be-B systematics have been measured in dozens of CAIs, and over 54 valid isochrons derived \citep{DunhamEtal2022}, including in many objects we would classify as LAACHIs.
Nearly all of these cluster around a canonical value $\beratio \approx (7.1 \pm 0.2) \times 10^{-4}$, with a statistical spread completely consistent with measurement errors alone \citep{DunhamEtal2022}. 
Based on this, \citet{DunhamEtal2022} argued that $\beten$ was inherited from the molecular cloud, where it was produced in gas, ices, and dust grains, by spallation of nuclei such as O and Si, by Galactic Cosmic Rays (GCRs). 
The level $7 \times 10^{-4}$ is consistent with the GCR flux appropriate for the Sun's formation in a spiral arm 4.6 Gyr ago \citep{DunhamEtal2022}. 

Interestingly, some inclusions do not record a canonical value $\beratio \approx 7 \times 10^{-4}$.
Two inclusions (the CAIs {\it Lisa} and {\it B4}) appear to have been irradiated near the surface of their parent bodies \citep{DunhamEtal2022}.
The FUN CAIs {\it KT-1} and {\it CMS-1}  appear from combined Al-Mg and Be-B systematics and $\epsilon^{50}{\rm Ti}$ values to have formed at about 0.8 Myr after most CAIs \citep{DunhamEtal2021metsoc,DeschEtal2023c}; and the hibonite-bearing type A FUN CAI Axtell 2771, which records $\beratio \approx 4 \times 10^{-4}$ \citep{DunhamEtal2022}, also may have formed or been reset around 1 Myr after most CAIs.
All other CAIs with non-canonical $\beratio$ are hibonite-dominated and $\altwosix$-poor, consistent with our definition of LAACHIs. 
These include: nineteen hibonite grains regressed together to yield $\beratio = (5.3 \pm 1.0) \times 10^{-4}$ \citep{LiuEtal2009,LiuEtal2010}; three hibonite grains regressed together to yield $\beratio = (5.2 \pm 2.8) \times 10^{-4}$ \citep{MarhasEtal2002}; 
%hibonite grain {\it CH-C4} with $\beratio = (8.0 \pm 3.3) \times 10^{-4}$, and 
the inclusion {\it HAL}, with $\beratio = (4.4 \pm 1.5) \times 10^{-4}$ \citep{MarhasGoswami2003lpsc}; the inclusion {\it SHAL}, with $\beratio = (3.06 \pm 0.63) \times 10^{-4}$ \citep{LiuKeller2017lpsc}; and the inclusion {\it DOM 31-2}, with $\beratio = (14.8 \pm 6.4) \times 10^{-4}$ \citep{DunhamEtal2022}.
% Oops I see Ushikubo et al. (2006) lpsc abstract with many inclusions. 
% Should I not include abstracts? 
% I decided to not include the Marhas one for CH-C4, or the Ushikubo et al. 2006 abstracts, since better measurements of the same inclusions exist. But the only measurements of HAL and SHAL are in those two abstracts. 
LAACHIs appear to record an average value $\beratio \approx 5.3 \times 10^{-4}$, but with some evident spread.

In the context of our model, we would expect LAACHIs to record a canonical $\beratio = 7.1 \times 10^{-4}$ at the time of their formation, and lower values if the Be-B system were thermally reset later. 
We expect Be to have been absent from presolar corundum and spinel grains at the time of their formation in the outflows of evolved stars (due to astration), and for insignificant Be to be produced in the grains while in the ISM. 
Be and $\beten$ were produced by GCR spallation in the molecular cloud, in gas and all solid phases.
In the LAACHI-forming region, these isotopes would quickly entered the gas phase, but Be would have condensed quickly into the LAACHIs, the only stable solids there; the condensation temperature of Be-bearing melilite is $\approx 1445 \, {\rm K}$ \citep{Lodders2003}, and we presume Be could have entered corundum or hibonite as well, presumably along with Ca as hibonite formed. 
The fact that LAACHIs would incorporate live ${}^{10}{\rm Be}$ but not ${}^{26}{\rm Al}$ immediately explains the observed decoupling between $\altwosix$ and $\beten$ in PLACs \citep{Liu2008thesis,LiuEtal2010}. 
However, explaining the slightly lower $\beratio$ ratios in some imclusions requires resetting of the Be-Be system at a few Myr by transient heating, such as chondrule-forming events. 

The inclusions {\it HAL} and {\it SHAL} do indeed appear heated: they have large (hundreds of microns) crystals of hibonite (presumably formed by melting and recrystallization), and the large isotopic fractionations characteristic of FUN inclusions, due to vaporization. 
Melting at $> 1.5$ Myr after other CAIs, when chondrules were commonly melted \citep{VilleneuveEtal2009}, would yield $\beratio \approx 3.4 \times 10^{-4}$, consistent with the ratios recorded by {\it HAL} and {\it SHAL}.
In contrast, the inclusion {\it DOM 31-2} shows strong evidence it was {\it not} melted, and probably did not experience any thermal alteration later. 
This is consistent with its higher initial $\beratio$, marginally consistent with the canonical value but not at all consistent with lower values like that of {\it HAL}. 
We conjecture that most other hibonite grains experienced heating at about 0.6 Myr, that reset their Be-B systems, probably by loss of about 30\% of their B, yielding $\beratio \approx 5.3 \times 10^{-4}$.
Given boron's low condensation temperature, $\approx 906$ K \citep{Lodders2003}, diffusion and evaporative loss of B may have occurred even after LAACHIs left the hot midplane region. 
However, we expect that at least some LAACHIs might have avoided heating, and may record canonical ${}^{10}{\rm Be}/{}^{9}{\rm Be} \approx 7 \times 10^{-4}$. 
%Indeed, many of the hibonite grains already analyzed, when not regressed together \citep[e.g., PLACs Mur-P6 or Mur-P1 of][]{LiuEtal2009}, appear more consistent with recording a canonical $\beratio$ ratio.
Future work should focus on finding the statistical distribution of $\beratio$ ratios in LAACHIs and correlating against evidence for late-stage thermal resetting of the Be-B system.

\subsubsection{Niobium-92}

Finally, ${}^{92}{\rm Nb}$ is another SLR, with half-life of {\EDIT $34.7 \pm 2.4$} Myr \citep{AudiEtal2003,Kondev2021}, known to have existed in the early Solar System \citep{Harper1996}. 
Niobium is expected to condense into titanates \citep{Lodders2003}.
If the carrier of ${}^{92}{\rm Nb}$ was a titanate grain (e.g., ${\rm BaTiO}_{3}$ or Ba-bearing perovskite) that condensed in a stellar outflow, then these grains were likely incorporated into LAACHIs, such that all PLACs should carry ${}^{92}{\rm Nb}$ at {\Steve whatever is the} solar canonical value.

% it is likely correlated with ${}^{48}{\rm Ca}$ and ${}^{50}{\rm Ti}$ anomalies and taken up into all PLACs. 

\subsection{Mineralogy and Microstructures}

The model presented here hypothesizes that corundum and hibonite are among the last surviving grains as solar nebula material heats up, not the first minerals to condense in a cooling nebula. 
These two scenarios may lead to observable differences. 
In both scenarios, hibonite forms by reaction of corundum with Ca vapor. 
In the cooling nebula scenario, the corundum would be purely condensed from solar nebula gas.
In the scenario presented here, some of the corundum would be presolar in nature, condensed in stellar outlows; {\Steve
although it is recognized that irradiation amorphizes interstellar grains \citep{KemperEtal2004}, we also expect these grains to be annealed by spending thousands of years at temperatures $> 1000$ K before entering the LAACHI-forming region. 
As well,} a significant fraction would be formed from presolar spinel grains that have lost Mg. 
These different formation pathways may lead to different microstructures, perhaps consistent with PLACs.  

The structures of both corundum and hibonite are based on close-packed oxygen sub-lattices. 
The corundum structure is trigonal, with space group {\it R-3c}, and consists of hexagonally close-packed layers (ABA stacking), with two thirds of the octahedral interstices filled with ${\rm Al}^{3+}$.
Corundum commonly forms \{001\} hexagonal prisms terminated by \{001\} pinacoids, that are commonly tabular.
Hibonite is hexagonal with space group {\it P6$_3$/mmc}.
Its structure consists of cubic-close packed (ABC stacking) spinel-like ${\rm Al}_{6}{\rm O}_{8}$ S layers alternating with hexagonal (ABA) ${\rm CaAl}_{6}{\rm O}_{11}$ R layers \citep{NagashimaEtal2010}.
The ${\rm Al}^{3+}$ in hibonite occupies both octahedral and tetrahedral sites in the S layers, and octahedral and trigonal dipyramidal sites in the R layers.
The ${\rm Ca}^{2+}$ occupies a position within the central oxygen sheet in the R layer and has a coordination of 12.
The transformation of corundum to hibonite, by the addition of ${\rm Ca}^{2+}$, requires a partial transformation from hexagonal (ABA) to cubic (ABC) stacking of oxygens, combined with a rearrangement of the ${\rm Al}^{3+}$ cation positions.
The transformation may involve a martensitic-like mechanism similar to that described for the transformation of olivine to ringwoodite \citep{Poirier1991}.
Solid-state synthesis of hibonite from ${\rm CaCO}_{3} + 6 \, {\rm Al}_{2}{\rm O}_{3}$ appears to preserve the general close-packed layers of corundum by diffusion of Ca inward along (001) oxygen layers, resulting in aggregates of platy hibonite crystals \citep{DominguezEtal2001}.
If the transformation of corundum condensed in the solar nebula---or presolar corundum condensed in stellar outflows---to hibonite occurs by a similar mechanism, one would expect formation of platy hibonite crystals and aggregates of sub-parallel hibonite laths like those found in PLACs.
Incomplete reaction of corundum would likely result in corundum cores with hibonite rims, or domains of excess spinel-like layers within hibonite.

Our model uniquely predicts formation of some hibonite by reaction of Ca with corundum formed by loss of Mg from presolar spinel.
In contrast to the above, the transformation of presolar spinel to corundum by Mg and O loss is not likely to produce platy hibonite crystals. 
Spinel, with space group {\it Fd-3m}, has a face-centered structure based on a cubic-close packed arrangement of oxygen atoms.
Transformation of spinel to corundum by Mg and O loss involves a significant loss of oxygen, and would likely involve a complete restructuring of the lattice, without preservation of the oxygen sublattice. 
If the transformation did involve a preservation of the close-packed oxygen, there would be four possible orientations of the resulting corundum (001) corresponding to the four orientations of the \{111\} planes in the spinel structure.
As a result, the transformation of $0.5 \, \mu{\rm m}$ spinel grains would likely result in polycrystalline corundum grains.
Transformation of these to hibonite would likely produce polycrystalline hibonite aggregates.
Hibonite formed from a combination of presolar corundum and presolar spinel would be a mixture of sub-parallel platy crystals like those found in PLACs, and finer randomly oriented hibonites. 
Excess ${\rm Al}_{2}{\rm O}_{3}$ from incomplete transformation would likely include stacking defects consisting of excess spinel-like layers \citep{HanEtal2015,HanEtal2022} as well as corundum inclusions.

The petrologic characteristics of PLACs are varied, and consistent with formation of some hibonite from spinel-derived corundum (as well as from condensed corundum). 
Two thirds of the PLACs analyzed by \citet{KoopEtal2016a} are not actually platy. 
Many hibonite-rich inclusions are single platy crystals, but many are also polycrystalline hibonite aggregates (with or without additional spinel).
Notably, corundum inclusions are common, and perovskite grains are abundant. 
Future investigations of platy vs.\ polycrystalline aggregate structures in PLACs, and searches for stacking defects in corundum, may be a severe test of our model. 

\section{Conclusions} \label{sec:conclusions}

Our Solar System contained live SLRs when it formed, including $\altwosix$.
The abundances of these SLRs were remarkably uniform, and the majority of meteoritic inclusions appear to have formed from a reservoir with $\alratio = (\alratio)_{\rm SS} =  5.23 \times 10^{-5}$, strongly implying that $\altwosix$, like other SLRs, was inherited from the Sun's molecular cloud \citep{DeschEtal2023c}.
Yet a subset of objects we term LAACHIs (Low-${}^{26}{\rm Al}/{}^{27}{\rm Al}$ Corundum/Hibonite Inclusions) appear to have formed with $(\alratio)_0 < 3 \times 10^{-6}$, or even $< 10^{-7}$, too low to have formed or been reset before parent body accretion from the same reservoir.
These presumably early-forming objects, which include corundum grains, PLACs, BAGs, and other hibonite-corundum inclusions, must have formed without as much $\altwosix$, despite its existence in the solar nebula.

We have presented a model here for how this could have happened.
As suggested by \citet{LarsenEtal2020}, we hypothesize that the Al in LAACHIs derived from micron-sized presolar grains (of corundum, spinel, and hibonite) that had resided in the ISM too long for any live $\altwosix$ to remain.
The $\altwosix$ instead derived from a small fraction of Al-bearing grains formed in recent ($<$ few Myr) supernovae within the Sun's star-forming region; these grains necessarily would have been much smaller ($< 50$ nm), with the ${}^{54}{\rm Cr}$-rich ``nanospinels" being a very likely candidate.
At $\sim 10^5$ yr into disk evolution, at 0.6 AU, temperatures near the midplane were $\approx 1350 - 1425 \, {\rm K}$, and corundum and hibonite were essentially the only stable phases.
In this region, micron-sized corundum and hibonite grains would have accumulated into objects $10 - 100 \, \mu{\rm m}$ in size.
Because most grains in this region would have been strongly negatively charged, the small $\altwosix$-bearing grains would have been excluded from the growing PLAC or corundum grain.

Our model explains quantitatively how inclusions rich in corundum and hibonite could have formed without $\altwosix$. 
The model assumes plausible conditions that occurred at one time and place in the solar nebula's history, according to \citet{YangCiesla2012}.
Our model does not address many key questions, such as: what is the lifetime of this region?; How many LAACHIs are produced (e.g., relative to other CAIs)?; and, How sensitive is the result of $\altwosix$ exclusion to the exact conditions in this region?
In future work we plan to model the trajectories of individual growing LAACHIs, across a range of disk properties, to address these issues.
The results of this paper, however, suffice to conclude that PLACs, corundum grains, and other LAACHIs all could have formed and not incorporated live $\altwosix$. 

Our model is similar to, but distinct from, other models that hypothesize that Al was carried on $\altwosix$-free presolar grains and that $\altwosix$ was carried on other presolar grains \citep{TrinquierEtal2009,LarsenEtal2020}.
However, noting that $\altwosix$-free inclusions like PLACs are refractory, these models all considered the ultimate cause of heterogeneity to be differences in the refractory nature of these two types of grains, with $\altwosix$-bearing grains preferentially {\it vaporized}.
This does not explain why the $\altwosix$ did not recondense into {\Steve the relevant} solids.
Our model differs from those in that we consider the carrier of $\altwosix$ to be refractory grains.

Our model provides important contextualization for the formation of PLACs and corundum grains, etc.
These are often argued to be the first solids formed in the Solar System, based on three lines of evidence: corundum and hibonite condense at the highest temperatures, and therefore first in a cooling gas; they show a wide spread in $\epsilon^{50}{\rm Ti}$ anomalies indicating incomplete ``homogenization"; and they lacked $\altwosix$, which must have been injected ``late" into the solar nebula, after their formation.
We concur with the consensus view that PLACs and other LAACHIs formed early, but not for any of these stated reasons.

We do not interpret corundum and hibonite inclusions as the first condensates from a solar composition gas that cooled from $> 1600 \, {\rm K}$.
Rather, these are {\Steve formed from} the last surviving presolar grains from the Sun's molecular cloud in a disk region with gas that heated above 1400 K. 
The distinction matters in that some aspects of the grains' presolar nature---such as ${}^{50}{\rm Ti}$ anomalies and lack of live $\altwosix$---will be retained.
Since much of the hibonite would have originated as spinel that lost Mg before acquiring Ca, it is possible that this history would be recorded in crystallographic defects or minor element abundance patterns.
Not all presolar properties would be retained, though; in particular, these grains would have nearly fully exchanged oxygen isotopes with the gas, and their Mg and Ca isotopes would have mostly reflected a solar composition as well.  

We concur that the large variations in $\epsilon^{50}{\rm Ti}$ reflect incomplete homogenization, but disagree that this requires PLACs and related inclusions to form before ${}^{50}{\rm Ti}$ was ``mixed in the solar nebula", which often carries the implication that the mixing was spatial, i.e., varied with position in the protoplanetary disk (\S 1.2).
We instead agree with \citet{ShollenbergerEtal2022} that the mixing was due to aggregation of smaller presolar grains, which did advance with time; but which was primarily a function of the mass of the particle, not a spatial mixing in the nebula over time. 

Most importantly, we do not interpret their lack of $\altwosix$ as meaning they formed before injection of $\altwosix$ into the solar nebula \citep{SahijpalGoswami1998,SahijpalEtal2000}, as we consider $\altwosix$ to have been inherited from the molecular cloud and never absent from the solar nebula \citep{DeschEtal2023c}.

Our model makes detailed predictions that can be tested. 
PLACs and other LAACHIs should have formed with canonical ${}^{10}{\rm Be}/{}^{9}{\rm Be} \approx 7 \times 10^{-4}$, and even if thermally reset in the nebula at 3 Myr (e.g., in chondrule-forming transient heating events) before incorporation into carbonaceous chondrite parent bodies, should record 
${}^{10}{\rm Be}/{}^{9}{\rm Be} > 2 \times 10^{-4}$.
Our model strongly implicates small ($< 50$ nm) ${}^{54}{\rm Cr}$-rich nanospinels as the carriers of live ${}^{41}{\rm Ca}$, although presolar hibonite is another possibility.
Our model therefore allows for ${}^{41}{\rm Ca}$ to be excluded from inclusions when  ${}^{26}{\rm Al}$ is, but predicts they would contain live ${}^{92}{\rm Nb}$ even if they didn't contain live $\altwosix$.
Because small perovskite grains hypothesized to be the carriers of positive ${}^{48}{\rm Ca}$ and ${}^{50}{\rm Ti}$ anomalies could be accreted, PLACs and other LAACHIs should show correlated excesses or depletions in these two isotopes.
Notably, although both are associated with {\Steve explosions of massive stars \citep{MeyerBermingham2022lpsc}}, we predict the carriers of live $\altwosix$ were {\it not} the carriers of ${}^{50}{\rm Ti}$ anomalies.
%as suggested by \citet{LarsenEtal2022}.
We predict correlations of ${}^{48}{\rm Ca}$ and ${}^{50}{\rm Ti}$ with ${}^{84}{\rm Sr}$, ${}^{135}{\rm Ba}$ and ${}^{138}{\rm La}$ (assuming these anomalies are carried by presolar titanate grains), but not ${}^{54}{\rm Cr}$ (which are carried on nanospinels).

Our model has important implications for chronometry. 
Despite forming at $t \approx 0$ in a solar nebula with spatially uniform $\alratio$ ratio, the reservoir sampled by LAACHIs (BAGs, PLACs, corundum grains) would have been the bulk Solar System, {\it minus} the $\altwosix$-rich supernova presolar grains, which therefore would not have had the canonical $\alratio$ ratio of the bulk Solar System reservoir. 
Any inclusion with hibonite or corundum (or possibly grossite) plausibly derived from LAACHIs, such as hibonite-bearing CAIs or FUN CAIs, or even SHIBs {\Steve (to a lesser degree)}, also might not have sampled a reservoir with a canonical $\alratio$ ratio. 
In these inclusions, the Al-Mg may not be a reliable chronometer.
CAIs dominated by other phases, chondrules and bulk meteorites all probably sampled the bulk Solar System reservoir, and in these objects the ${}^{26}{\rm Al}-{}^{26}{\rm Mg}$ chronometer can be used.
This one observation largely reconciles discrepancies between the Al-Mg chronometer and others, e.g., the Be-B chronometer \citep{DeschEtal2023c}.

Finally, our model results have strong implications for the Sun's astrophysical birth environment. 
PLACs and other low-$\alratio$ inclusions are consistent with all $\altwosix$ in the solar nebula having been inherited from a well-mixed molecular cloud.
There is no need to invoke late injection of $\altwosix$-bearing grains \citep{SahijpalGoswami1998,SahijpalEtal2000,OuelletteEtal2007}, nor models in which the gas infalling from the molecular cloud has variable amounts of supernova-derived inputs \citep[e.g.,][]{NanneEtal2019,LichtenbergEtal2021,PignataleEtal2019,LiuBEtal2022}.
The addition of recently formed supernova grains to the Sun's molecular cloud could have happened up to several Myr before the Sun's birth \citep[e.g.,][]{GoodsonEtal2016}. 
Mixing in molecular clouds is very thorough over these timescales \citep{PanEtal2012}.
Although formation of the Sun in a high-mass star-forming region is required, a nearby supernova explosion during evolution of our protoplanetary disk is not demanded. 

\bigskip
{\bf Acknowledgments}: 
{\EDIT We thank an anonymous referee and especially Andy Davis for a very thorough review that led to substantial improvements in the paper.}
We thank Zachary Torrano for useful discussions.
The work herein benefitted from collaborations and/or information exchange within NASA's Nexus for Exoplanetary System Science research coordination network sponsored by NASA's Space Mission Directorate (grant NNX15AD53G, PI Steve Desch).
Emilie Dunham gratefully acknowledges support from a 51 Pegasi b Fellowship, grant \#2020-1829.

\bigskip

We include as Research Data 
%spreadsheet with information on the meteoritic inclusions discussed in \S 2, that forms the basis of Table 1 and Figure 1. 
%A 
a spreadsheet to calculate the relative velocities in Tables 3 and 4.

%We also need to decide whether to include a spreadsheet of the 26Al data and inclusion morphologies from literature. Ashley has a spreadsheet that'd be ready as a supplemental sheet.
%is also included as Research Data. 


%\appendix

\bibliography{laachis}{}
\bibliographystyle{aasjournal}



\end{document}